\documentstyle[12pt]{article}

\begin{document}

\baselineskip=7mm

\begin{center}

{Regularities of Many-body Systems Interacting by a 
Two-body Random Ensemble }

{ Y. M. Zhao$^{a,b}$
\footnote{Corresponding author. Cyclotron Center, RIKEN,
Hirosawa 2-1, Wako-shi, Saitama  351-0198, Japan.  Tel: +81-48-467-9697; 
fax: +81-48-461-5301. {\it E-mail address}: ymzhao@riken.jp. },
A. Arima$^{c}$, and N. Yoshinaga$^d$}

\vspace{0.2in}
{ $^a$ Cyclotron Center, the Institute of Physical and Chemical Research
(RIKEN), \\
 Hirosawa 2-1, Wako-shi, Saitama 351-0198 Japan \\
$^b$ Department of Physics,  Southeast University, Nanjing 210018 China \\
$^c$ The House of Councilors, 2-1-1 Nagatacho, 
Chiyodaku, Tokyo 100-8962, Japan \\
$^d$ Department of Physics, Saitama University, Saitama 338-0825 Japan }

\tableofcontents

\date{\today}

\end{center}

\newpage

\vspace{0.2in}
{\bf Abstract}
The ground states of all even-even nuclei  have
angular momentum, $I$, equal to zero, $I=0$, and 
positive parity, $\pi=+$. 
This feature  was believed to be  a
consequence of the attractive short-range
interaction between nucleons. However,  in the
 presence of   two-body random interactions, 
the predominance of  $I^{\pi}=0^+$ ground states (0 g.s.)
was found to be robust 
 both  for bosons and for an even number of  fermions. 
For simple systems, such as $d$ bosons, $sp$ bosons, $sd$ bosons, and 
a few fermions in    
single-$j$ shells for small $j$, there are a few 
approaches to predict and/or explain 
  spin $I$ ground state ($I$ g.s.) probabilities. 
An empirical approach to predict   $I$ g.s. probabilities is 
available for general cases, such as 
fermions in a single-$j$ ($j>7/2$) or many-$j$ shells and 
various boson systems,  but a more fundamental understanding
of the robustness of 0 g.s. dominance is still  out of reach. 
Further interesting results are also reviewed  concerning  
other robust  phenomena 
of many-body systems in the presence 
of random two-body interactions,
 such as the odd-even staggering of binding energies, 
generic collectivity, the behavior of average energies, correlations,
and regularities of many-body systems
interacting by a displaced two-body random ensemble. 

\vspace{0.2in}
      
{\it \bf PACS}:   05.30.Fk, 05.45.-a, 21.60Cs, 24.60.Lz 

\vspace{0.2in}

{\it key words}: $I$ g.s. probabilities,
0 g.s. dominance, random interactions,
correlation,  collective motion, average energies. 

\newpage

\section{Introduction }

Atomic nuclei provide  ideal  laboratories to study  
the features of  microscopic  many-body systems with 
finite number of   constituents less than $\sim 300$. They 
are  complex objects with many degrees of freedom and   
exhibit  almost all features  found in other many-body systems. 
The  regularities   in  atomic nuclei  in the presence of 
 random two-body interactions  therefore   provide  an excellent window to 
study general features of low-lying states of many-body systems.  
Therefore, although  discussions  in this article are originated from
 nuclear structure studies, the results can 
have many implications for other fields as well.

The Gaussian orthogonal ensemble of random matrices was first
proposed by Eugene Wigner in Ref. \cite{Wigner},
which was a revolutionary thought in
understanding the spacings of levels observed in resonances
in slow-neutron scattering on heavy nuclei. The two-body random 
ensemble (TBRE), which will be  used in most examples 
in this article,  was introduced
to study statistical properties of
spectra of many-body systems, 
by French and Wong in Ref. \cite{French}, and
by Bohigas and Flores in Ref. \cite{Bohigas}. 
Ref. \cite{Brody} presented
a self-contained account of random matrix physics in quantum
systems concerning spectrum and strength fluctuations. 
The latest review along  similar lines is presented in Ref. 
\cite{Guhr}. 

We also  note that there recently appeared other review articles on  
random matrix ensembles for finite particle systems but with different 
focuses. In Ref. \cite{Dean}, the focus was on the links between many-body
pairing,  as it evolves from
the original nucleon-nucleon force, the manifestations 
of superfluity in nuclear matter (say, neutron stars), and  
pairing in atomic nuclei. In  Refs. \cite{Kota,Zelevinsky2}
the discussion was focused on 
statistical mechanics and onsets of chaos in finite many-body
systems.
In Ref. \cite{Zele-review}, the focus was on the 
geometric chaoticity of angular momentum couplings, its possible 
implications on the energy centroids, and the multipole collectivity 
in the presence of random two-body interactions. 
In this article we shall focus on
the features (particularly, orders and 
correlations) of low-lying levels of many-body systems
in the presence of random two-body interactions, explaining both 
the observations and the present 
status towards ``understanding"  these features.

In many-body systems such as molecules and atomic nuclei,
the interactions by themselves have no trace of symmetry 
groups for vibrational  or rotational modes. However, 
the low-lying states often exhibit a pattern suggestive of
symmetries for these modes. One may   ask to what 
extent the low-lying states acquire order from the basic properties
of  interactions 
such as rotational invariance and possibly
other symmetries such as isospin invariance. 
In other  words, some properties 
such as vibration   or rotation 
might  dominantly  occur in the low-lying states of many-body systems while 
the others might occur only with  small probabilities.

Atomic nuclei with an even proton number Z and an even neutron number N
are examples which follow these lines.   
The  angular momenta and  parities ($I^{\pi}$) 
 of the ground states   of even-even nuclei are always 0$^+$,
  and   the structure of the low-lying states  
is characterized by  a tripartite classification \cite{Casten,Zamfir}, i.e.,  
seniority region, anharmonic vibrational region and rotational  
region. Also there exists an odd-even staggering of binding energies, etc. 
One can ask whether these   features   are robust for
general many-body systems.  This can be studied by permitting   
interactions to be more and more arbitrary. 
  
This question of robustness was first studied by
 Johnson, Bertsch and Dean in Ref. \cite{Johnson1},  
where a  dominance of $I^{\pi}=0^+$ ground states (0 g.s.)
was obtained  by using the TBRE Hamiltonian.
In Ref. \cite{Johnson2}, the 0 g.s. dominance was found to  
be insensitive to the 
monopole pairing interaction,  and to be  related to  
a reminiscence of generalized seniority suggested  
in Refs. \cite{Talmi,Talmi-text}.  
In Refs. \cite{Bijker2,Bijker3},  it was found 
that the $sd$ bosons produce  both vibrational and rotational spectra 
as well as the 0 g.s. dominance in presence of the TBRE Hamiltonian. 
In Ref. \cite{Papenbrock}, it was found that odd-even 
staggering of binding energies in finite metallic grains
and metal clusters \cite{Satula}  
arises from purely random two-body interactions.
These interesting
results suggest that the above features are not only the  
consequences of  
attractive pairing interactions, but are much more general than they 
were previously assumed. Many authors made efforts to
seek the origin of these observations and to look for more
robust features along these lines. This   article aims at
reviewing these achievements.

As was emphasized by Feshbach \cite{Feshbach},
the studies of atomic nuclei   provide us with many 
``universals" of the physical world.
In this article  we will  show that 
 microscopic many-body systems
in the presence of  random interactions provide
us with a new method to discover and to study  universal features 
of microscopic  systems in nature. Discoveries and 
 understandings of these interesting  patterns  
 are very exciting  topics in physics.

In Sec. 2 we shall define the Hamiltonians for systems such as
nucleons and/or other fermions in a single-$j$ or many-$j$ shells,
bosons with single spin $l$ or many $l$'s. In the process of
doing this, we shall also give a   brief introduction to the
nuclear models at a basic level, which will be helpful 
to those who are not familiar with the nuclear structure theory.
We then define  the two-body random ensemble that we take in this paper.

In Sec. 3  we shall  concentrate on  statistics of  
the  distribution of spin $I$ 
in the ground states in the presence of the TBRE Hamiltonian. 
We shall not restrict the discussions  to the
0 g.s. probability (denoted as  $P(0)$) which is
 obtained by diagonalizing the TBRE Hamiltonian, 
 but also study 
 other $I$ g.s. probabilities (denoted as $P(I)$). 
 We shall   go to  systems with odd numbers of fermions  
 to study  other $I$ g.s. probabilities and 
 parity distributions  of the ground states as well.

In Sec. 4 we shall review the efforts  to understand
the 0 g.s. dominance of simple systems suggested in earlier works,
where one can evaluate  $I$ g.s. probabilities.
The systems are restricted to 
 $d$-, $sp$- and $sd$-boson systems, and 
fermions in a single-$j$ shell with $j\le 7/2$.

In Sec. 5 we shall go to more complicated systems,
such as fermions in a  single-$j$ (for large $j$) or many-$j$ shells and 
$sdg$ bosons, etc.,  for which 
the understanding of  $P(I)$'s is not yet available. 
However,  an empirical approach will be suggested to predict  $P(I)$'s 
of these systems. This empirical approach also shows that
the 0 g.s. dominance is related to certain interactions with
specific features. Some properties of $I_{\rm max}$ g.s.
probabilities will be  found and explained, and some
features of the energy gaps for 0 g.s. and 
the $n$-body matrix elements of $I=0$ states
will be pointed out.

In Sec. 6 we shall discuss other features
of many-body systems interacting by random interactions, including
  average energies,   collectivity,
and  normal ordering of spin $I$ for the yrast states. 
We shall show that the energy centroids of spin $I$ states (denoted by 
$\overline{E_I}$)
have large probability to be the lowest energy only if 
$I\simeq I_{\rm min}$ or $I \simeq I_{\rm max}$. 
Compact trajectories  of  $\overline{E_I}$'s can be  found.  Bijker and 
Frank \cite{Bijker2,Bijker3}  discovered  
that  both the vibration  and the rotation 
are  robust in the low-lying states of
 $sp$- and $sd$-boson space. 
For fermionic degrees of freedom, say,  nucleons in the 
$sd$ shell,   rotational spectra  do not appear if one uses  
a general two-body Hamiltonian. We will show  
that the quadrupole-quadrupole correlation is  
very important to get a collective rotational spectrum.
We will also see in Sec. 6  that spin $I$  in the yrast bands obtained by  
the TBRE Hamiltonian has a large possibility to be normally ordered
(i.e.,  the sequence   0, 2, 4 $\cdots$). 
In the fourth part of  Sec. 6 we shall
review results of constraints of the TBRE Hamiltonian in order to 
reproduce some global features of atomic nuclei. 

In Sec. 7 we summarize this paper and conclude 
on this    interdisciplinary field.

In the Appendix we listed mathematical notations of this paper.

\newpage

\section{Space and ensembles of the Hamiltonian}

In this Section we define the Hamiltonians 
to establish notations and conventions used  in  this 
paper. Towards this goal, we  present a brief introduction to the 
nuclear shell model \cite{Mayer,Jensen,Haxel}
and its boson approximation, the interacting boson 
model  (IBM)  \cite{Arima1,Arima2,Arima3,Iachello1}. 
This  helps those who are not experts at
the nuclear structure theory, which is frequently
involved in the studies of the present subject, 
 to catch easily  the essential point of physics.
More detailed discussions can be found in Ref. 
\cite{Talmi-text}. We also explain the random samplings 
and the two-body random ensemble. These are very basic 
concepts to study, e.g., the question how
large a probability is for a certain 
spin $I$ to be the spin of the ground state.

\subsection{Hamiltonian of the shell model and its approximations}

The simple picture that  particles move independently in a
one-body potential well, which well describes the atoms, metals  etc., 
is also a key to the nuclear structure theory.  Here the
one-body potential is given by the mean field which is produced by 
all nucleons except the ``single" nucleon in consideration.
The Hartree-Fock method provides us with an 
approach to derive a single-particle potential from the two-body  
interactions, with the requirement that the energy 
for   $A$-particle   Slater determinants 
should be a minimum. Numerical 
results of Hartree-Fock calculations indicate that the harmonic-oscillator  
wave functions are   good approximations to the 
self-consistent wave functions. A breakthrough was made by  
Mayer and Jensen \cite{Mayer,Jensen,Haxel} through  introducing the 
strong spin-orbit coupling in the nuclear force. This nuclear shell model
 (SM) makes understandable  a vast amount of experimental data of  
spins, magnetic moments,
isomeric states, and the ``magic numbers" for proton number Z or 
neutron number N  equals   2, 8, 20, 28, 50, 82, 126. The SM has 
been playing a central role in understanding the
nuclear structure since it was established in 1949.
It  is defined by a set of spin-orbit coupled single-particle 
states  with quantum numbers $n_rljm$,   corresponding to the radial quantum
number $(n_r)$, orbital angular
momentum $(l)$,  total angular momentum $(j)$ and its $z$-component $(m)$,
respectively. The parity for each orbit is determined by $(-)^l$. 
Because the Hamiltonian is rotationally invariant, the single particle
energies are independent of $m$. For all single-particle states within
one major shell, the value of  each  $j$ is unique, 
 so one usually  suppresses  
the quantum numbers $nl$ and 
uses only the total angular momentum $j$ to label
a certain single-particle level. When we define 
the  Hamiltonian below, we shall use
$j_1, j_2, j_3, j_4$ to label different  single-particle levels. 
In atomic nuclei there are two types of particles, neutrons and 
protons.  One introduces the  isospin quantum number $t$, which 
is mathematically very similar to   spin $s$, with 
$t=1/2$ and $m_t=\pm 1/2$  corresponding to protons and neutrons, 
respectively. Because $t$ for both protons and neutrons is the same, 
one can keep only $m_t$ to label
the single-particles states. 
 The single-particle energies
for protons and neutrons are different, with the
proton ones higher mainly
due to the Coulomb force.
It is usually believed that  at least one major shell
is necessary to adequately describe low-lying states of a given nucleus.

One property  of nuclear systems is that the residual
interaction  between the valence nucleons,
i.e., the part of the interaction which can not  
be covered by the mean-field part, 
is very strong and leads to   strong configuration mixings. 
Therefore, one has to diagonalize the Hamiltonian
in the shell model space  to obtain the ``true" eigenvalues 
and wave functions.

The shell model Hamiltonian that is usually taken includes a one-body term
\begin{equation}
 H_1 = \sum_{j m m_t} e_{j  m_t} a_{jm, m_t}^{\dag} a_{jm, m_t}, 
\end{equation}
and a two-body term 
\begin{eqnarray}    
H_2= &  \frac{1}{4} & \sum_{j_1j_2 j_3j_4,JT} 
\sqrt{(1+\delta_{j_1 j_2})(1+\delta_{j_3j_4})}   ~G_{JT} (j_1j_2, j_3j_4)
  \nonumber \\
   &&  \sum_{M_J M_T}  
A^{\dagger} \left(j_1 j_2\right)^{(JT)}_{M_J M_T}
A \left(  {j_3} {j_4}\right)^{(JT)}_{M_J M_T} ~ , \label{V}
\end{eqnarray}
where
\begin{eqnarray}
&& A^{\dagger} \left(j_1 j_2\right)^{(JT)}_{M_J M_T}
= \sum_{m_1 m_2  m_{t_1} m_{t_2}} (j_1 m_1, j_2 m_2|J M_J) 
(\frac{1}{2} m_{t_1}, \frac{1}{2} m_{t_2} | T M_T)
 a_{j_1 m_1, m_{t_1} }^{\dagger}
 a_{j_2 m_2, m_{t_2} }^{\dagger} , \nonumber \\ 
&& A \left(  {j_3} {j_4}\right)^{(JT)}_{M_J M_T}  =
\left( A^{\dagger} \left(j_3 j_4\right)^{(JT)}_{M_J M_T} \right)^{\dagger} ~
 .  \label{Clebsch}
\end{eqnarray}
In Eq. (\ref{Clebsch}),    $(j_1 m_1, j_2 m_2|J M_J) $ denotes  
the Clebsch-Gordan coefficient.  In Eqs. (\ref{V}-\ref{Clebsch}) 
$J$ ($T$) is the total spin (isospin) resulting from 
one particle  in the $j_1$ orbit and another one in the $j_2$
orbit, or one particle in  the $j_3$ orbit and 
another one in the $j_4$ orbit.
$M_J$ and $M_T$ are the $z$ components of
$J$ and $T$, respectively.
The $a_{j_1}^{\dagger}$  and $a_{j_2}^{\dagger}$ are single-particle 
creation operators.  
$  G_{JT} (j_1 j_2, j_3 j_4) $'s are   two-body matrix elements
defined by 
\begin{equation}
G_{JT} (j_1j_2, j_3 j_4) = 
\langle \left( j_1  j_2 \right)^{JT}_{M M_T} | V |
\left( j_3   j_4\right)^{JT}_{M M_T} \rangle, 
\label{rot-invariance}
\end{equation}
for a two-body interaction, V.
The $G_{JT} (j_1j_2, j_3 j_4)$'s are independent of $M$ and $M_T$. 
The Hamiltonian of 
Eq. (\ref{V})  respects parity conservation, which means that 
the parity product for the four orbits $j_1 j_2 j_3 j_4$   is positive. 
The Hamiltonian Eq. (\ref{V}) also respects rotational and 
time-reversal invariance, which means that 
 $V$ in Eq. ~(\ref{rot-invariance}) 
is a scalar and takes  real values.

In this paper we  shall also consider systems in which there are
only one type of valence particles (i.e., neutrons or protons).
For a single-$j$ shell   
 one then can  suppress the labels $j_1j_2 j_3 j_4$ 
($j_1=j_2= j_3= j_4 \equiv j$) and $T$ $(T\equiv 1)$ 
and simply denote the two-body matrix elements by
$G_J$, where the allowed values of $J$ are $J$=0, 2, $\cdots$, $2j-1$.
The number of $G_J$'s 
of fermions in a single-$j$ shell is $ j+\frac{1}{2}$. 
It is noted that   the ``one type of particle  
systems"  here should be understood in 
a broader sense: the identical particles are not necessarily
identical valence nucleons, they can be also   
other types of identical particles  (say, electrons). Therefore, we 
refer to these systems as  ``fermions" in a single-$j$ shell 
or many-$j$ shells.

In this paper we shall also use 
a restricted separable Hamiltonian  for 
fermions in  many-$j$ shells:
\begin{equation}
 H=H_1 + H_{\rm P} +   V_{\rm ph} ~,     \label{sep1}
\end{equation}
where $H_1$, $H_{\rm P}$ and $V_{\rm ph}$ are
the spherical single-particle energy term,
 generalized pairing,  and
particle-hole type interactions, respectively. The
definition of $H_{\rm P}$ is as follows, 
\begin{equation}
  H_{\rm P} = V_0 + V_2 + \cdots ~~.   \label{sep2}
\end{equation}
Here
\begin{eqnarray}
&& V_0=  G_0  {A}^{\dagger (0)}  {A}^{(0)},   ~~
{A}^{\dagger (0)}  =   \sum_{j} 
\frac{\sqrt{2j+1}}{2}(a_{j}^{\dagger}  
a_{j}^{\dagger})_0^{(0)} ~,
 \nonumber \\
&&
V_2 =   G_2 \sum_M {A}^{ \dagger (2)  }_{M}  {A}^{(2)}_M,   ~~
{A}^{\dagger (2)}_M = \sum_{j_1 j_2} q(j_1 j_2) \left(
a^{\dagger}_{j_1}   a^{\dagger}_{j_2} \right)^{(2)}_M ~,
\end{eqnarray}
with 
\begin{equation}
{A}^{(0)} = \left( {A}^{ \dagger (0)} \right)^{\dagger} ~, ~
{A}^{(2)}_M = \left( {A}^{\dagger (2)}_M \right)^{\dagger} ~, ~
q(j_1 j_2) =
 - \frac{ \langle j_1 ||
r^2 Y^{(2)} || j_2 \rangle }{\sqrt{5}}   ~.  \nonumber
\end{equation} 
The particle-hole interaction $V_{\rm ph}$ takes the form 
\begin{equation}
 V_{\rm ph} =\kappa \sum_M Q_M    Q_M   +   \cdots, 
\end{equation}
where
\begin{equation}
 Q_{M} = \sum_{j_1 j_2 m_1 m_2} q(j_1 j_2) (-)^{j_2 - m_2}
 (j_1 m_1, j_2 m_2 | 2 M) 
 a^{\dagger}_{j_1 m_1} 
a_{j_2 ~ -m_2} ~.  \label{qab2}
\end{equation}
The operator $Q$ is called the quadrupole operator, and
the interaction $\sum_M Q_M    Q_M$ is called
the quadrupole-quadrupole interaction.  
The validity of  truncating  two-body interactions to  the  
 pairing and the quadrupole-quadrupole interaction  
in   nuclear physics was studied 
by Kumar and Baranger \cite{Kumar1,Kumar2}.

The difficulty   to apply the nuclear shell model is that the dimension 
of the configuration space involved in a  major shell increases 
very rapidly  with the number of valence nucleons.
For  medium and heavy nuclei, the number of states is usually
too huge to handle even for a very good computer. The status of 
diagonalization of the shell model Hamiltonian was reviewed in 
Ref.  \cite{Koonin}. 
Because of this difficulty,   many efforts were made to 
simplify the shell model approach. Important milestones include 
the Bohr-Mottelson-Rainwater collective model (or the
so-called geometric model) 
\cite{Bohr,Rainwater} and  the interacting 
boson model (IBM) \cite{Arima1,Arima2,Arima3,Iachello1}.
The  geometric model was introduced  
by Bohr, Mottelson, and Rainwater  in 1950-1953 as an alternative
to the shell model. The link between the geometric
model and the shell model was established by Elliott \cite{Elliott} in 1958.  
Intrinsic deformation is introduced in the geometric collective model. 
In the IBM the building blocks of the Hamiltonian and the
model space are $s$ bosons with spin zero and $d$ bosons with spin two,
which are  interpreted as correlated $S$ nucleon pairs with spin zero
and $D$ nucleon pairs with spin two. Namely,
the IBM is a phenomenological model which first truncates
the full shell model space to the $S$ and $D$ pair subspace and
next maps the $S$ and $D$ pair subspace to the 
$s$ and $d$ boson subspace. The total boson number is  
 conserved in the Hamiltonian and equal to half of the valence nucleon  
 number for an even-even nucleus. The parity of $s$   and $d$ bosons 
 is positive.  The relationships between the
 IBM and the geometric description were discussed 
in Refs. \cite{Iachello1,Ginox,Dieperink,Scripta}.

The IBM in which the distinction between
protons and neutrons is left out is called the IBM-1. 
The   IBM-1 Hamiltonian \cite{Iachello1}
 that we use in this paper
is as follows, 
\begin{eqnarray}
H &=& E_0 + e_d \sum_m d^{\dagger}_m  d_m
+ \sum_{L=0,2,4} \frac{1}{2}  
c_L \sum_M  \left( d^{\dagger}   d^{\dagger} \right)^{(L)}_M            
           \left (d d \right)^{(L)}_M   
           \nonumber \\
  &+& \frac{1}{\sqrt{2}} \epsilon_{ddds}  \sum_M \left[ 
     \left( d^{\dagger}   d^{\dagger} \right)^{(2)}_M  
            d_M s 
+   h. c.    \right]
          \nonumber \\
  &+& \frac{1}{2} \epsilon_{ddss}  \left[
    \left( d^{\dagger}  d^{\dagger} \right)^{(0)}   
          \left( ss \right)^{(0)}  
+ h. c.      \right] +
  \epsilon_{ssss}    s^{\dagger}   s^{\dagger}     s s   ~,  \label{sd-boson1}
\end{eqnarray}
where
\begin{eqnarray}
&&   \left(  d^{\dagger}   d^{\dagger} \right)^{(L)}_M
 = \sum_{m_1 m_2} ( 2 m_1 2 m_2 | LM ) d^{\dagger}_{m_1}   d^{\dagger}_{m_2} ~,
~~  \left(dd \right)^{(L)}_M  = \left( \left(
 d^{\dagger}   d^{\dagger} \right)^{(L)}_M \right)^{\dagger},  \nonumber 
\end{eqnarray}
and we take this notation for bosons hereafter.

Similarly,  we shall also go to the $sdg$ and $sp$  systems.
The spin$^{\rm parity}$ for a $g$ boson and a $p$ boson is 
$4^+$ and $1^-$, respectively.  The $sdg$ boson systems 
have been  studied  by including $g$ bosons  
as well as $sd$ bosons in describing deformed nuclei. 
The $sdg$ Hamiltonian is very similar to that of the $sd$ IBM-1 Hamiltonian 
except that there is another   one-body term for  the $g$ bosons  
and there are many more two-body terms (totally 32 two-body terms)
in which $g$ bosons are involved. 
One  sees  Ref. \cite{Iachello1}  for details of the $sdg$ IBM. 
The $sp$ interacting boson model is also called  the ``vibron" model 
\cite{vibron}, which is used to describe both rotations and vibrations of 
the diatomic molecules.
The total parity for a $sp$-boson system is given by $(-)^I$, where $I$ 
is the total spin of a certain  state for this $sp$-boson system. 
Like the $sd$ IBM, the total number of bosons 
is also conserved in the $sp$  IBM (or the vibron model).
The  Hamiltonian for the $sp$ bosons is as follows, 
\begin{eqnarray}
H &=& E_0 + e_p \sum_m p^{\dagger}_m  p_m
+ \sum_{L=0,2}  \sum_{M} \frac{1}{2}  
c_L \sum_M  \left( p^{\dagger}   p^{\dagger} \right)^{(L)}_M            
           \left (p p \right)^{(L)}_M   
           \nonumber \\
  &+& \frac{1}{2} \epsilon_{ppss}  \left[
    \left( p^{\dagger}  p^{\dagger} \right)^{(0)}   
          \left( ss \right)^{(0)}  
+  h. c.   \right]
          \nonumber \\
  &+&   
  \epsilon_{ssss}    s^{\dagger}   s^{\dagger}  ss
 + \epsilon_{spps} \sum_m
             s^{\dagger}   p^{\dagger}_m 
           p_m s  ~   .  \label{sp-boson1}
\end{eqnarray}
We shall also  use a schematic $sp$-boson
Hamiltonian
\begin{equation}
H = -{\rm cos} \chi \sum_m p^{\dagger}_m  p_m +
\frac{ {\rm sin} \chi}{4(n-1)} \left( s^{\dagger} s^{\dagger} -
p^{\dagger}  p^{\dagger}  \right)^{(0)} 
\left(  ss - pp \right)^{(0)} ~,   \label{vibron-H}
\end{equation}
which contains the basic
features of the model,  to  exemplify the so-called 
mean-field approach to spin $I$ ground state probabilities.

The success of the $sd$ IBM in describing the low-lying excitations 
of atomic nuclei  stimulated  calculations 
 within the subspace constructed by
 correlated $S$ and $D$ pairs. 
Similar to the IBM, Ginocchio constructed symmetry-dictated
$SD$ pairs \cite{toy-model}. 
This approach was further studied by Wu {\it et al.} and 
called  the Fermion dynamical symmetry model \cite{Feng}.
To overcome the restriction in  Ginocchio's model  that the structure of 
 the $SD$ pairs is very specific, Chen \cite{Chen0} developed 
 recursion formulas which are applicable for arbitrary
pairs. Based on this technique, a nucleon pair approximation (NPA) 
of the shell model was suggested in Refs. \cite{Chen,unified}. The 
 NPA Hamiltonian in this paper is defined by Eq. ~(\ref{sep1}).

Parallel to fermions in a single-$j$ shell, we shall also study  
boson systems with spin $l$. The  Hamiltonian for
a boson system with spin $l$  is written as  
\begin{equation}
H =  \sum_{{\rm even} L} \sum_{M_L}  \frac{1}{2} 
G_L   \left( b_l^{\dagger}   b_l^{\dagger} \right)^{(L)}_{M_L}
           \left (b_l b_l \right)^{(L)}_{M_L} ~,  \label{single-boson}
\end{equation}                                                 
where   $ b_l^{\dagger}$ is the creation operator for bosons with spin $l$.
The number of two-body matrix elements for
a boson system with spin $l$ is $l+1$, and
$ \left (b_l b_l \right)^{(L)}_{M_L} = \left(
\left( b_l^{\dagger}   b_l^{\dagger} \right)^{(L)}_{M_L}
\right)^{\dagger} $.

In the numerical calculations of this paper,
 the one-body terms in the above Hamiltonian will be  neglected. 
It is   noted that the inclusion of one-body terms does not change  
the statistics substantially.

\subsection{Monte Carlo samplings}

 The name ``Monte Carlo" arises from the ``random" character
 of the method and the famous casino in Monaco.
The Monte Carlo   method is
well known and  a  powerful method in almost all  
fields of physics. For instance, there are 10 review papers with  
 ``Monte Carlo"  in their titles in the journal   {\it Physics Reports}.  
This method has been applied to many problems, such as the evaluation 
 of   high-dimensional  integrals, 
the shell model Monte Carlo method  in nuclear physics \cite{Koonin},
 the quantum  Monte Carlo simulation in 
 solid state physics \cite{Linden},
 etc.

It is usually  difficult to investigate questions such as  
how large the probability is for  a system to have a  
rotational spectrum or to have a certain spin $I$ ground state  
for the full space of   two-body interactions.  In particular,  
when the number of 
two-body interactions is very large, it is impossible to  
evaluate those probabilities by summing over the results obtained from 
the usual quadrature points throughout the full parameter space.

The approach by random samplings of the ensemble is a very efficient and
simple way to study the  regularities of low-lying states 
of many-body systems   {\it throughout} the 
possible parameterizations. The  advantage  
  by Monte Carlo samplings can be seen, for example, 
in the  calculation of integrals with many variables. 
Suppose one calculates an integral with $D$ variables, then the accuracy
by the usual quadrature method is $p^{-2/D}$ while  
that of the Monte Carlo calculation is $p^{-1/2}$. Here  
$p$ is the number of   quadrature points  of the 
conventional quadrature method, or the number of  samplings  
of the Monte Carlo method  (obviously, $p >> 1$). 
When the integral is of very high dimension (i.e., $D >> 1$), the  
Monte Carlo sampling is an efficient way to evaluate the integral, while  
the usual quadrature method becomes completely impossible.

There are many computer programs to generate random numbers
with specific distributions.  Of course, a sequence of numbers generated
by a deterministic program can never be truly random. Nevertheless, 
the sequence looks ``random" (called
 pseudorandom) enough for practical purposes, and thus can be  
used as if they were truly random.  A good discussion of uniform 
random number generators  and of tests whether or not 
these ``random" numbers work properly can be found in Ref. \cite{Koonin2}.

In studying the  regularities 
of many-body systems in the presence of
random interactions,  one thousand samplings
present   very stable statistics and thus 
are deemed to be  ``reliable" samplings. 
One then says that a ``reliable" statistics is obtained for 
  aspects such as  spin $I$ ground state probabilities.

\subsection{Two-body random ensemble}

The two-body random ensemble is defined as follows. 
The two-body matrix elements are independent of each other, and take
random values which follow Gaussian distribution  with  
an average being zero. The distribution width 
is set to be 1 for diagonal two-body  
 matrix elements  and $1/\sqrt{2}$ for off-diagonal 
 two-body matrix elements. More specifically, 
 if  $G_{JT} (j_1j_2, j_3 j_4)$ and
$G_{J'T'} (j'_1j'_2, j'_3 j'_4)$
are two arbitrary (different)  two-body matrix elements
for valence nucleons in an open shell,  one has        
\begin{eqnarray}
&&  \rho(G_{JT} (j_1j_2, j_3 j_4)) = \frac{1}{\sqrt{2 \pi x}}
  {\rm exp}(- \frac{ \left[ G_{JT} (j_1j_2, j_3 j_4) \right]^2 }{2 x} ),
  \nonumber \\
&& \rho(G_{J'T'} (j'_1j'_2, j'_3 j'_4)) = \frac{1}{\sqrt{2 \pi x'}}
  {\rm exp}(- \frac{ \left[ G_{J'T'} (j'_1j'_2, j'_3 j'_4) \right]^2 }{2 x'}), 
  \label{tbre}
\end{eqnarray}
where
\begin{equation}
 x = 
 \left\{ \begin{array}{ll}
 1              & {\rm if ~} |(j_1 j_2)JT \rangle = |(j_3j_4)JT \rangle \\
 \frac{1}{2}    & {\rm otherwise} 
\end{array} \right. ~ , ~
 x' = 
 \left\{ \begin{array}{ll}
 1              & {\rm if ~} |(j'_1 j'_2)J'T' \rangle = |(j'_3j'_4)J'T' \rangle \\
 \frac{1}{2}    & {\rm otherwise}  
\end{array} \right. ~.   \nonumber 
\end{equation}
One sees for the ensemble average that 
\begin{eqnarray}
&& \langle \left[G_{JT} (j_1j_2, j_3 j_4)\right]^2  \rangle = x, \nonumber \\
&& \langle G_{JT} (j_1j_2, j_3 j_4) ~ G_{J'T'} (j'_1j'_2, j'_3 j'_4) 
 \rangle = 0 ~. \nonumber
\end{eqnarray}

A two-body random ensemble such defined is called the ``TBRE". 
In this paper  we shall discuss  fermions in a single-$j$ shell
or many-$j$ shells, and bosons with spin $l$ or more than one spin. 
The  Hamiltonians for these systems are defined in  subsection 2.1. 
The TBRE Hamiltonian  for each of these systems is given by  
an ensemble in which  two-body matrix elements, such as  
$G_J$'s for fermions in a single-$j$ shell or $G_L$'s in 
Eq. (\ref{single-boson}) for bosons with spin $l$, 
or $\epsilon$'s  and $c_L$'s in Eqs. (\ref{sd-boson1})-(\ref{sp-boson1}), 
are taken to be random numbers following the distribution 
of Eq. (\ref{tbre}).  The one-body  interaction parameters are set to be
zero.

In Ref. \cite{Johnson1}, Johnson {\it et al.}  also used an ensemble for which 
the width $x$ has a $J$- and $T$-dependence:  
the width  equals
to the above $x$ multiplied by a factor of $\frac{1}{(2J+1)(2T+1)}$.
This was  called the 
random quasiparticle ensemble (RQE). 
In Refs. \cite{Zele-review,Mulhall1,Mulhall2} the MSU group took 
two-body matrix elements to be
random numbers which are uniformly distributed between $-1$ and 1.

It is noted that the  properties  obtained 
by using the TBRE, the RQE and the uniformly distributed random 
two-body matrix elements are essentially  similar  to  each other, 
although the statistics might be somewhat different. 
This was already noted in the early paper by French et 
al. \cite{French} for the spectral statistics.

The results of this paper are  based on  1000 samplings
of the TBRE Hamiltonian. 
                                                          
\subsection{Summary of this Section}

In this Section we have defined the Hamiltonian corresponding
to  valence nucleons in many-$j$ shells, fermions in a single-$j$ shell
or many-$j$ shells, $sp$ bosons,   $sd$ bosons, $sdg$ bosons, and bosons 
with spin $l$.    We have   presented the connection
between the Hamiltonians of these different systems by
reviewing the nuclear shell model and its various approximations:
 the Bohr-Mottelson geometric model, 
 the Arima-Iachello interacting boson model, 
 etc.  This knowledge is helpful in establishing  
notations and conventions in the paper.  The separable
Hamiltonian is taken for the $SD$ pair approximation of the shell model.
The $sp$ boson model was introduced to describe the rotation and vibration 
of diatomic molecules, but the spirit is very similar to the 
interacting boson model.

We have introduced very briefly the concept of Monte Carlo samplings. The
Monte Carlo approach is widely used in various subjects.
The term ``Monte Carlo" appeared in titles of hundreds of monographs,  and 
Monte Carlo techniques were used in numerous papers.
Here we have explained why the Monte Carlo samplings are 
very efficient  in studying the regularities of many-body
systems in the presence of random interactions.
 We point out that 1000 sets of random interactions
produce statistics  which is ``accurate" 
enough for our discussions.

We have also defined the ``two-body random ensemble" (TBRE).
One-body interaction parameters
in the Hamiltonian are set to be zero,  and two-body 
interaction parameters are independent of each other and chosen to 
follow a Gaussian distribution of  width $\frac{1}{\sqrt{2}}$
for the off-diagonal terms and 1 for the diagonal terms. We 
 note that some authors took non-Gaussian distributions, but  
the pattern exhibited by the statistics such obtained is very similar.

\newpage

\section{ Regularities of the ground states}

In this Section we shall go first to two simple cases:
fermions in a single-$j$ shell and bosons with spin $l$.
We take four particle systems for these two cases and
change $j$ and $l$ values, and focus
on  spin $I$  ground state probabilities  based on
the TBRE Hamiltonian. Then we review the
results for more complicated systems based  on both 
the TBRE Hamiltonian and  displaced TBRE Hamiltonian.
We also study the regularity of parity distribution
in the ground states for  nuclei with random interactions 
and the effect from higher rank   Hamiltonians.

\subsection{Fermions in a single-$j$ shell}
 
Before going to complicated cases,  let us first come to a few simple
examples.   A system with four fermions in a single-$j$ shell
is the simplest but non-trivial  case for  fermionic 
degrees of freedom.
This case was treated by using both 
the TBRE Hamiltonian in Ref. \cite{Zhao1}, and  a uniformly distributed 
random two-body Hamiltonian assuming values between $-1$ and 1  
in Ref.  \cite{Mulhall1}. The results obtained by using these two ensembles 
are essentially the same although the distributions of the
ensembles are very different. Fig. 1  shows  the results of a few
important $I$ g.s. probabilities calculated by the
TBRE Hamiltonian,  with
$j$ ranging from $j=7/2$ to $j=33/2$.
Some of the principal outcomes are:

1) The probability of  ground states with odd  $I$ 
 is  much smaller than that of their neighboring  even  
values of spin,  even if the corresponding numbers of states 
  in the $j^n$ configuration 
 are   comparably large. 

2) The unique $I_{\rm max}$ state has a large probability  
to be the ground state, although 
 this probability  decreases  with $j$
\footnote{This holds also for the case of
 random interactions which  distribute 
 uniformly   between $-1$ and 1. 
 The results of  Ref. \cite{Mulhall1} which claim that 
  this probability    staggers rapidly and  becomes  0 for several
   single $j$ are erroneous.}. 

3) The $I=2^+$ and $4^+$ states have  large probabilities
 to be  the ground state.   This  indicates   
that  small and even angular
momentum  states are favored as the ground states for an even
number of fermions in single-$j$ shells.

4) The 0 g.s.  dominance obtained by using  the TBRE  Hamiltonian 
is not a ``rule" without exceptions. 
In Fig. 1, $P(0)$'s of  four nucleons in a 
$j=\frac{7}{2}$   and $j=\frac{13}{2}$ shell   are smaller
than the corresponding $P(2)$'s.

5)  An interesting oscillation of
$P(0)$'s with respect to $j$  was noticed in Refs. \cite{Mulhall2,Zhao1}. 
This oscillation is synchronous to an increase of the  
 number of $I=0$ states. Namely,
$P(0)$  staggers when $(j-\frac{3}{2})=3k$ ($k$ is a positive integer),
 coinciding  with an increase of the  number of $I$=0 states which is given by  
the largest integer not exceeding $(j-\frac{3}{2})/3$ 
\cite{Zhao9,Ginocchio1}.

Because  the  0 g.s. is dominant for  four fermions in a single-$j$ shell
with the only two exceptions  $j=$7/2 and 13/2, 
one easily gets the intuition that 
 the $P(j)$'s might be  large for  five fermions  in a single-$j$ shell. 
This was  found to be indeed the case in Refs. \cite{Mulhall1,Zhao1}, 
although the $P(j)$'s are not as large as the $P(0)$'s. Fig. 2
presents  a few
examples for $n=5$ in which  $P(j)$'s are indeed very large
in comparison with other $P(I)$'s.

One connection between the dominant $P(0)$
for $n=4$ and the large $P(j)$ for  $n=5$  
  is given by the monopole 
pairing interaction (to be discussed in Sec. 5). 
One should be aware, however, that 
the large $P(j)$'s for an odd number of fermions are not necessarily
related to $P(0)$'s of its even $n$ neighbors,   
  except for  the connection given by the   
monopole pairing interaction. 
In many cases $P(0)$'s of systems with even numbers of 
fermions are much larger than $P(j)$'s of those with odd particle numbers.

Figure 3  plots the $P(0)$'s for 
 $n=4$ up to $j=$33/2, for  $n=6$ up to $j=27/2$,  
and the $P(j)$'s for   $n=5$ and $n=7$ up to $j=27/2$. It is seen that 
 $0$ g.s. probabilities
for   even numbers of fermions in a single-$j$ shell, and  
 $j$ g.s. probabilities for odd numbers of fermions in a 
single-$j$ shell, stagger  synchronously
  at an interval of  $\delta_j=3$  when $j$ is small. 
When $j$ is large,  the  $P(0)$'s and $P(j)$'s seem to
saturate.

\subsection{Bosons with spin $l$}

Figure 4 shows a few important  $P(I)$'s  versus $l$ for four   
  bosons with  spin $l$ \cite{Zhaoym}. 
One sees that  the pattern of $I$ g.s. probabilities for 
 four bosons with  spin $l$ is very similar 
to that of four fermions in a single-$j$ shell. 
For instance,  the $P(0)$'s vs. $l$ stagger  at an interval of
$\delta_l = 3$; the $P(0)$'s are dominant over other $P(I)$'s except
two $l$ shells, $l$=2 or 8; and $P(I_{\rm max})$'s decrease with $l$.  A  
new feature of   $I$ g.s. probabilities for four bosons 
with  spin $l$ is that  the $P(l)$'s
are  considerably large (small) when $l$ is even (odd), i.e., the 
$P(l)$'s exhibit an odd-even staggering behavior.

Next let us come to boson systems with  spin $l$ and particle number $n$. 
Here one should ask about the feature of 
 $P(I)$'s for an odd number of
particles: 
 which angular momentum $I$  g.s.  dominance will appear, $I=0$  or $I=l$?
 In systems with an odd number of 
fermions in a single-$j$ shell, there are no $I$=0 states, and one  
expects  $I=j$ g.s. probabilities  to be   large, 
as shown in Fig. 2.  For bosons with spin $l$ and odd $n$, 
 however, one may have $I$=0 states and thus 
 it is  not known  {\it a priori} whether the $I=0$ or $I=l$ g.s. dominance  occurs
 in these systems.

Figure 5 shows  the $P(l)$'s and  $P(0)$'s in 
boson systems with $l=$4 and  6, and 
$n$ running as large as possible.  
For odd $n$ and $l=1$,  or for $n=3$ and any odd $l$ \cite{Zhao9}, 
 there are no $I=0$ states; 
for odd $n$ and $l=3$, 5, 7 and 9, the $I=0$ states do not exist
unless $n\ge 15$, 
9, 7, and 5, respectively. For fifteen bosons with $l=3$, 
nine bosons with $l=5$, seven bosons with $l=7$ and 
five bosons with $l=9$ or 11,  $P(0) \sim 0 \%$  
according to calculations by using 1000 sets of the TBRE Hamiltonian 
\cite{Zhaoym}.  From Fig. 2 and these odd-$l$ cases it is concluded that 
the $P(0)$'s are usually 
much less than the corresponding $P(l)$'s 
when $n$ is   odd. On the other hand, the $P(0)$'s are  
mostly larger than the  $P(l)$'s when $n$ is an  even number.

   These results  indicate  that
the 0 g.s. dominance is robust
for systems with an even number of $n$, but  not true generally
if $n$ is   odd. In the latter case,  it is observed that 
the 0 g.s. dominance is  easily lost.  
One therefore expects that the 0 g.s. dominance 
is partly connected to an even number of particles.

When one applies the TBRE Hamiltonian to 
$sp$  bosons \cite{Bijker6,Kusnezov2}, $sd$ bosons 
\cite{Bijker2,Bijker3,Bijker6,Bijker4,Bijker5,Bijker-conf,Zhao2,Arima5},
and $sdg$ \cite{Zhao4} bosons 
(we shall come to these cases in subsections 4.6 and 5.2), the 
0 g.s. dominance is  found also for an odd number  of bosons. 
However, to a large extent,  the dominant  
$P(0)$ therein is associated with the $s$ boson condensation, which  
contributes around $40\%$ to the 0 ground state  probability \cite{Bijker4}. 
In other words, one should be aware that $sp$ and $sd$ systems 
are very special systems in which $s$ boson condensation 
produces the 0 g.s. dominance  when the boson number   is odd.  
Without $s$ bosons, the $P(0)$'s of those systems with odd $n$ would be  
drastically smaller and other $I$ g.s. probabilities, such
as those of $I=l$ or $I=I_{\rm max}$,  would
be much larger.  It would be very interesting to carry out
systematic calculations of $P(I)$'s for both even and odd numbers  
of bosons  with various mixtures of spins. 

\subsection{Many-$j$ shells}

Let us exemplify the cases of  fermions in  many-$j$ shells   by using 
the pioneering works of Refs.  \cite{Johnson1,Johnson2}. 
These  authors calculated a few even-even nuclei in the
$sd$ shell (63 independent two-body matrix elements)
and the $pf$ shell (195 independent two-body matrix elements) 
with both neutrons and protons, with 1000 runs of a  Hamiltonian  by
using the following ensembles: 
the  TBRE,  the random quasiparticle
ensemble (RQE) which has an additional $J$ dependence of the width
\cite{Johnson2},    the RQE without monopole pairing (RQE-NP), and the 
RQE with splittings of single-particle energies 
(RQE-SPE). The results of four, six and eight 
neutrons in the $sd$ shell, four protons
and four neutrons in the $sd$-shell
show that there are around $40\% \sim 70\%$ 0 g.s. 
in these cases, although 
in these systems  the   $I=0$ states occupy a very  small 
portion in the full shell model space.
This result was  a surprise to nuclear structure theorists,
and has been attracting much attention since then. 
Table I shows  0 g.s. probabilities by using
different ensembles, and the percentage of $I=0$ states 
in the full shell model space.

In Ref.  \cite{Horoi2}, Horoi, Volya, and Zelevinsky
extended the investigations of 0 g.s. dominance 
to both even-even  and odd-odd nuclei. They 
took  random two-body interactions uniformly distributed between $-1$ and 1. 
They checked the even-even nucleus $^{24}$Mg and the
odd-odd nucleus $^{26}$Al. They found that  
 for  $^{24}$Mg  the  ground states are dominated by  $I=0$ and $T=0$ 
($T$ is the total isospin of the state), and that 
states which  satisfy  the relation $(-)^{I+T} = 1$   are favored. 
For the odd-odd nucleus  $^{26}$Al the  ground 
states are dominated by  $I=1$ and $T=0$,  and the  favored states 
have $(-)^{I+T} = -1$. In both cases the lowest $T$=0 is dominant. 
They also studied a simpler model, i.e.,  four nucleons
in a single-$j$ ($j=15/2$) shell  with both protons and neutrons.
The situation is quite similar to that of $^{24}$Mg:
there is a similar predominance of the lowest
$I, T$ states, and a preference of  the ground states with
$(-)^{I+T}$=1 is easily noticed, while in the case of six protons and 
neutrons in a $j=9/2$ shell the ground states with $(-)^{I+T}$=$-1$
are favored.  Based on these examples,
the authors  concluded that the ground states  
are dominated by the states with $I$ and $T$ as low as possible, and  
quantum numbers satisfying a ``selection" rule  
$(-)^{I+T}$=$(-)^{n/2}$ are favored, where $n$ is the
number of particles. It is interesting to
investigate more cases to see whether or not this observation
is applicable for other systems.

Now let us come to  $P(I)$'s of boson systems with 
spins more than  one. 
The $sd$ and $sp$ boson systems were studied by
Bijker and Frank \cite{Bijker2,Bijker3,Bijker4,Bijker5,Bijker6,Bijker-conf},
and Kusnezov \cite{Kusnezov2}.
In Refs. \cite{Bijker2,Bijker3}, $(sd)^n$ boson systems 
with  $n$ ranging from 3 to 16 
were considered by using random one- and two-body interactions, and 
 a dominance of 0 g.s. ($\sim 60 \%$) was obtained for these cases.
For $sd$ bosons,  
the $I=2$ (spin $l$ of $d$ boson)  g.s  probability 
and the $I_{\rm max}$ g.s. probability are also large, while other
$I$ g.s. probabilities are nearly zero; for $sp$ bosons,   $P(I)$'s 
are very similar to  the $sd$ boson case except that $I=1$ ground states  
(spin $l$ of $p$ boson) are favored instead of $I=2$ ground states.

\subsection{Parity distribution in the ground states} 

Another relevant quantity is    parity distribution in 
the ground states in the presence of the TBRE Hamiltonian. 
For atomic nuclei, all the ground states
of  even-even nuclei are observed to have positive parity,  
while those of odd-odd nuclei have
both positive and negative parity with slightly more 
positive ones.  The nuclei with odd mass numbers
have almost equal number of cases with positive or negative parity.
Table II presents the statistics of parity 
in the  ground states of nuclei with mass numbers larger than 120.
This table is based on the experimental data
compiled in Ref. \cite{Firestone}. 

It is interesting to see whether 
a similar pattern appears in the presence of the TBRE Hamiltonian. 
A series of calculations was carried out in Ref. \cite{Zhaoym2} 
for  four model spaces: 
\begin{itemize}
\item[A] Both protons and neutrons are in the
$f_{\frac{5}{2}} p_{\frac{1}{2}} g_{\frac{9}{2}}$ shell which corresponds to
nuclei with both proton number Z and neutron number N $\sim 40$;
\item[B] Protons in the
$f_{\frac{5}{2}} p_{\frac{1}{2}} g_{\frac{9}{2}}$ shell and neutrons in the
$g_{\frac{7}{2}} d_{\frac{5}{2}}$ shell which corresponds to nuclei with
Z $\sim40$ and N $\sim50$;
\item[C] Both protons and neutrons are in the $h_{\frac{11}{2}}
s_{\frac{1}{2}} d_{\frac{3}{2}}$ shell which corresponds to nuclei with
both Z and N $\sim82$;
\item[D] Protons in the $g_{\frac{7}{2}} d_{\frac{5}{2}}$ shell and
neutrons in the $h_{\frac{11}{2}} s_{\frac{1}{2}} d_{\frac{3}{2}}$ shell which
corresponds to nuclei with Z $\sim50$ and N $\sim82$.
\end{itemize}
These four model spaces do
not correspond to a complete major shell but have been truncated in order to
make the calculations feasible. These truncations are based on the subshell
structure of  the involved single-particle levels.  
It is noted that the number of states (denoted as $D(I)$)
 for positive and negative parity are very {\it close} to each other 
 for all these examples. 
One thus expects that  the probability of  the ground states with
positive parity is around 50$\%$, if one assumes that  each state of the full
shell model space is
 equally probable in the ground state.

We denote valence proton number and valence neutrons 
number by using $N_p$ and $N_n$, respectively.
Because  N and Z  for  closed shells are always even, 
   nuclei with  even values for both $N_p$ and $N_n$ 
correspond to even-even type; nuclei with
odd values for $(N_p+N_n)$ correspond to odd-$A$ type; and
nuclei with odd values for both  $N_p$ and $N_n$ correspond to
odd-odd type. As discussed above, the statistics for these
three types are quite different.

The calculated statistics of parity in the ground states obtained
by using a TBRE Hamiltonian is given in Table III.
For even-even nuclei (even values for both $N_p$ and $N_n$),
it was  noticed that positive parity  
is dominant in the ground states. 
For odd-$A$ nuclei  and doubly-odd nuclei, it is
found  that  probabilities to have positive or negative parity
in the ground states are
almost equal with some exceptions.
In general, there is no favoring for either
positive parity or negative parity in the
ground states of odd mass nuclei and
doubly odd nuclei in the presence of
random interactions. It is noted that
these calculations are done for the
beginning of the shell. For the end of the
shell the results show a similar pattern.

It was also found that the above regularities for parity distributions   hold for
very simple cases: single-closed two-$j$ shells, one with
positive  parity and 
one with negative parity. The following shells
have been checked: $(2j_1, 
2j_2)$ = $(9,7)$, $(11,9)$, $(13,9)$, $(11,3)$, $(13,5)$, $(19,15)$, $(7,5)$,
$(15,1)$. The statistics is very similar to the above results: The probability
of ground states with positive parity is about $85\%$ for an even number of
nucleons, and about $50\%$ for an odd number of nucleons. 

It is very interesting to note that for all even-even nuclei the $P(0^+)$ is usually 
two orders of  magnitude larger than $P(0^-)$. It would be very interesting to
investigate the origin of the large difference in $P(0)$ for positive and
negative parity states, i.e., why the $0^-$ is not favored in the ground
states. As is the case for an odd number of bosons  with spin $l$
 (refer to Sec. 3.2),
spin $I=0$ is {\it not} a sufficient condition to be favored  
in the ground states of a 
many-body system in the presence of  random interactions, i.e.,
in order that the state is favored as a realistic ground state, not only
$I=0$ but also positive parity is required.

A simple system to study the parity distribution of the ground  
states in the presence of random interactions is the $sp$-boson system. First,
it is noted that a $sp$-boson system with an odd number  of 
particles $n$ has the same number
of states with positive and negative parity; while for an even number
 of particles $n$ 
there are {\it slightly} more states with positive parity (the difference is 
only $n+1$).  
The parity for $sp$ bosons is given by $(-)^I$.  The calculated 
results of Ref.~\cite{Bijker6} showed that when the number $n$ of $sp$ bosons
is even, $P(0)+P(n) \sim 99\%$, which leads to 
positive parity  ground states dominance.  When
the number of $n$ is odd, only about 50$\%$ of the ground states in the 
ensemble have $I=0$, and  about $50\%$ have $I=1$ or $I=n$. 
  This leads to about equal percentages 
for positive and negative parity ground
states. This pattern is very similar to that observed for fermion systems.

\subsection{Many-body systems interacting by a displaced TBRE}
 
 While a TBRE is distributed 
symmetrically with respect to  zero, an interesting question is   what  happens 
 if one  uses random interactions with only 
   positive  or   negative signs, or random interactions which are not
distributed symmetrically around  zero.  This  issue 
 is both interesting and
important because interactions in realistic systems, such as
nuclei, atoms etc, are not symmetric around  zero. Below we
present  results calculated by using a displaced TBRE.

Let us  firstly consider two arbitrary ensembles
$\{ G'_{JT} (j_1 j_2 j_3 j_4) \}$ and
$\{ G_{JT} (j_1 j_2 j_3 j_4) \}$,  which 
are related by a shift $ c$:
\begin{equation}
  G'_{JT} (j_1 j_2 j_3 j_4) = G_{JT} (j_1 j_2 j_3 j_4) +   c,  
 \label{tbrex}
\end{equation}
where $ c$ is a constant. If 
 $\{ G_{JT} (j_1 j_2 j_3 j_4) \}$ is symmetric around zero and is described 
 by Eq. (\ref{tbre}), we call $\{ G_{JT} (j_1 j_2 j_3 j_4) \}$ the  
TBRE, and  $\{ G'_{JT} (j_1 j_2 j_3 j_4) \}$ will be called a displaced TBRE
with a displacement $c$.

For fermions in a single-$j$ shell,   the results by using the  
ensemble $\{ G_J' \}$ are exactly the same as those
obtained by  using  $\{ G_J \}$,  
except for a shift $\frac{n(n-1)}{2}  c$  of  the  
 eigenvalue of the ground state. Therefore, a 
 displacement of the TBRE is trivial in a single-$j$ shell.

In Ref. \cite{Zuker}, the authors showed  that
for the $f_{7/2} p_{3/2}$ shell  the displaced TBRE 
with an attractive average leads 
to the 0 g.s. dominance  and
rotational spectra with strongly enhanced B(E2) transitions for a certain
class of model space. We shall discuss the results of 
Ref. \cite{Zuker} in Sec. 6.2.3.

In general, the role played by $c$ in 
the displaced TBRE  of Eq. (\ref{tbrex})
is very complicated \cite{Zhao-comment}. 
For instance, both negative  and positive  
displacements for an even number of fermions in
 many-$j$ shells may  favor (or quench) the 
0 g.s. probability, or produce a  minor change.

Recently, Johnson  \cite{Johnson-comment} also
revisited the consequences of a  
displaced TBRE for $^{44}$Te and $^{48}$Ca. He pointed out
that one should be aware of a fact that the wave function
obtained by the TBRE with an attractive average (i.e., $c < 0$)
is quite close to that obtained by a negative constant value $c$ for  
all the two-body matrix elements, if the width of TBRE Hamiltonian  
is 1 and $c=-3$ or $-2$.

\subsection{Effect of higher rank interactions}

As for the effect of interactions with rank higher 
than two, the only work was done by Bijker and Frank in  
Ref. \cite{Bijker3}. These authors  studied the case of $sd$ bosons  
by adding three-body interactions to the one-body and two-body 
 Hamiltonian (two one-body, seven two-body, and seventeen
three-body interactions), where
they used a scaling which depends on  boson number $n$:  
For one-body terms  this scaling is $1/n$; for two-body terms
this scaling is $1/(n(n-1))$; and 
for three body terms  it is $1/(n(n-1)(n-2))$.
Their calculations showed  that
the inclusion  of  three-body interactions does 
not change the results in a significant way.
When the boson number $n$ is  sufficiently large in comparison to 
the rank of  interactions, the results are essentially
similar. A study by using random ensembles of one- and two-body
interactions also showed similar results to the case of
pure three-body interactions.
In other words,  the basic features  
of $sd$ boson systems do not change significantly  
due to  the inclusion of  three-body interactions.

\subsection{Odd-even staggering of binding energies}

The odd-even staggering of binding energies  is well 
known in nuclear physics.  It is an evidence of  
the pairing interaction between like particles.  
Similar features were found in Ref. \cite{Johnson2} 
for  the angular momentum zero ground states by
using the RQE Hamiltonian for  four to ten neutrons in the $pf$ shell.
Johnson and collaborators made a least square fit of the  
binding energies, which was used to simulate 
  Talmi's formula of binding energies within the
framework of the generalized seniority scheme
for atomic nuclei \cite{Talmi,Talmi-text}, 
for the case of  even numbers of neutrons  
in the presence of the RQE Hamiltonian.
Then they calculated   binding energies of systems with
five and seven neutrons in the same shell by using  the  same sets 
of the RQE Hamiltonian.  Statistics of deviation of
binding energies  from   Talmi's formula showed that  
deviations for an even number of neutrons are 
usually small and symmetric around zero, while those for
an odd number of neutrons are large and positive.
This is very similar to the situation
of binding energies for atomic  nuclei.  

In Ref. \cite{Papenbrock}, Papenbrock, Kaplan and 
Bertsch considered the random  Hamiltonian for
quantum dots or small metallic grains which
conserve total spin.
The results of Ref. \cite{Papenbrock} 
showed that even a purely random two-body
Hamiltonian can give rise to the odd-even staggering of
binding energies.

\subsection{Summary of this Section}

In this Section we showed  
the   robustness of the 
0 g.s. dominance for  the case of even numbers of 
fermions in a  single-$j$ or many-$j$ shells, 
and also for the cases of 
(both even and odd numbers of)  $sd$ or $sp$ bosons. 
There are  only very few counter examples for the case of four fermions 
in a single-$j$ shell ($j=7/2$ and $13/2$).
For neutron-proton systems,
it was found that the $P(0)$ is not dominant for odd-odd systems
\cite{Horoi2,Zhaoym2}. 
For bosons with  spin $l$,
there are many cases  in which the $P(0)$'s are very small
in comparison to  other $P(I)$'s when the boson number $n$ is  odd, 
which suggests that  the 0 g.s. dominance is 
 partly connected to an even number of particles. 
In other words,   the 
 0 g.s. dominance in  many-body systems is 
a robust feature associated with even numbers of particles.

The parity distribution in the ground states
 calculated by using the TBRE Hamiltonian is  found to be robust: 
for the case of even numbers of both valence
protons and neutrons the positive parity is  always  dominant (around
80$\%$), while positive parity and negative parity are
almost equally probable for other cases.
For atomic nuclei, all even-even nuclei have positive parity g.s. 
and the other cases have either positive or negative
parity g.s. with almost equal probabilities (refer to Table II).

A  displaced TBRE Hamiltonian may give a very different pattern 
of $P(I)$'s for the case of fermions in many-$j$ shells 
in comparison to those obtained by a TBRE Hamiltonian.
This means that the $I$ g.s. distribution for
a displaced TBRE Hamiltonian is a much more complicate issue. 
                                                             
The effect of the rank of the Hamiltonian was studied in Ref. \cite{Bijker3} 
for $sd$-boson systems. The calculated results of  Ref. \cite{Bijker3} 
showed that  basic features obtained 
by using the random Hamiltonian do  not change significantly even  if   
one includes  three-body interactions.

 Systematic odd-even binding energy differences 
were first  discussed in Ref. \cite{Johnson2} for a few nucleons
in a $sd$ shell, and revisited in Ref. \cite{Papenbrock} for finite  
metallic clusters by using the TBRE Hamiltonian. 
These calculations concluded that an odd-even staggering
arises  from the TBRE Hamiltonian: stronger
binding energies for systems with  
even numbers of particles   are typically obtained 
in numerical simulations. However,  the 0 g.s. dominance 
and odd-even staggering are not necessarily two facets of the 
same thing.

\newpage

\section{Spin $I$ ground state probabilities of simple systems} 

In this Section we discuss  $P(I)$'s
of  simple systems.  Here the appellation
``simple systems" means  either the eigenvalues of the systems are 
 linear in terms of two-body matrix elements or
one can classify the ranges of two-body matrix elements (because of
some specific features of the systems) in a simple way (see Sec. 4.4).

In this Section
we shall first come to a few   systems   in which
the eigenvalues are linear combinations of two-body matrix elements. 
One feature
for a state to have a large g.s. probability,   
and  an empirical approach to predict  $P(I)$'s  which were 
introduced in Refs. \cite{Zhao1,Zhao2,Arima5,Zhao4}, 
will be discussed.  For 
three and four fermions in a single-$j$ shell ($j \le 7/2$) and
$d$ boson systems, one can 
calculate  $P(I)$'s exactly by using the  
geometry of eigenvalues in the two-body matrix elements. This recipe  
was introduced in Ref. \cite{Chau}, and will be discussed in this Section. 
The evaluation of $P(I)$'s of $sp$ and $sd$ bosons 
by  a mean-field approach  introduced in Refs. \cite{Bijker6,Bijker4,Bijker5}  
will be also discussed.

Due to the simplicity and specific properties of the systems  discussed  
in this Section,   $P(I)$'s of these systems can be predicted very well.  
Thus one can say that  $P(I)$'s  of these systems in the presence  
of random interactions are satisfactorily described.

\subsection{Fermions in a single-$j$ shell  with $j \le$ 7/2}

The eigenvalues of states of  fermions in a single-$j$ shell
with $j \le \frac{7}{2}$ can be written in terms of
linear combinations of the two-body matrix elements.
Here we discuss only the case of four fermions in a $j=\frac{7}{2}$ shell. 
Discussions of systems with $n=3$ and $j \le \frac{7}{2}$ 
 can be found in Ref.  \cite{Zhao4}.

For a $j=7/2$ shell with four fermions,  
 all  the states are labeled by  
 their total angular momenta $I$ and their seniority  
quantum numbers ($v$). The eigenvalues $E_{I(v)}$ are as follows (see Ref. 
\cite{Lawson}):
\begin{eqnarray}
\begin{array}{crcrcrcr}
E_{0(0)}=& {\bf\frac{3}{2}}G_0&+&       \frac {5}{6}G_2&+&       \frac{3}{2}G_4&+&  \frac{13}{6}G_6, \\ \\
E_{2(2)}=&      \frac{1}{2}G_0&+&       \frac{11}{6}G_2&+&       \frac{3}{2}G_4&+&  \frac{13}{6}G_6, \\ \\
E_{2(4)}=&                    & &                   G_2&+&{\bf\frac{42}{11}}G_4&+& {\it \frac{13}{11}}G_6, \\ \\
E_{4(2)}=&      \frac{1}{2}G_0&+&        \frac{5}{6}G_2&+&       \frac{5}{2}G_4&+&  \frac{13}{6}G_6, \\ \\
E_{4(4)}=&                    & &   {\bf\frac{7}{3}}G_2&+&        {\it 1} ~ G_4&+&   \frac{8}{3}G_6, \\ \\
E_{5(4)}=&                    & &        \frac{8}{7}G_2&+&    \frac{192}{77}G_4&+& \frac{26}{11}G_6, \\ \\
E_{6(2)}=&      \frac{1}{2}G_0&+&        \frac{5}{6}G_2&+&       \frac{3}{2}G_4&+&  \frac{19}{6}G_6, \\ \\
E_{8(4)}=&                    & &{\it \frac{10}{21}}G_2&+&    \frac{129}{77}G_4&+& {\bf \frac{127}{33}}G_6. 
\end{array}
\label{7/2-4}
\end{eqnarray}
In Eq. (\ref{7/2-4}), {\bf bold} font is used for the
largest and {\it italic} for the smallest amplitudes in an expansion
in terms of $G_J$. Eq.~(\ref{7/2-4}) can be rewritten as  follows, 
\begin{equation} 
E_{I(\beta)} = \sum_J \alpha^J_{I(\beta)} G_J,    
  \label{alpha}
\end{equation}
where   $\beta$ representing all the 
necessary additional quantum numbers to label the state. 
By using the  TBRE Hamiltonian described  by  
  Eq.~(\ref{tbre})
and the eigenvalues given by Eq.~(\ref{7/2-4}), 
it is easy to obtain the
 probability, $P(I)$, for each $I$ ground state. 
$I$ g.s. probabilities
for four fermions in a $j=7/2$ shell are shown in the row ``TBRE" of 
Table IV, and are obtained by 1000 runs of the TBRE Hamiltonian.

One can also predict   the  $I$ g.s. probability 
 without running the TBRE Hamiltonian. For
 example,  the  $P(0)$    is determined by 
\begin{equation}
 \int dG_0 \int dG_2  \int dG_4
\int dG_{6} ~~
 \rho(G_0)  \rho(G_2)  \rho(G_4) \rho(G_6)~
 |_{\left(\sum_J \alpha^J_{0(0)} G_J <
\sum_J \alpha^J_{I(v)} G_J \right)  },  \label{exact}  
\end{equation}
where $I(v) \neq 0(0)$, and the subscript
``${\left(\sum_J \alpha_{0(0)} G_J <
\sum_J \alpha_{I(v)} G_J \right)  }$"
 is the requirement
for  $G_J$'s which take values
from $-\infty$ to $\infty$. 

The  row  ``pred1" of Table IV  corresponds to  probabilities predicted
by an  integral for each $I^+$ state similar to Eq.~(\ref{exact}) for the
 $0^+$ state.
Probabilities  calculated by using the TBRE  and those predicted 
 by using  integrals like Eq.~(\ref{exact})  are   consistent
  within statistical fluctuations.  

One easily sees from the present  example  that
a state with one or more largest 
 (or   smallest) $\alpha_{I(v)}^J$, for which we used {\bf bold}
 (or {\it italic}) font in Eq. (\ref{7/2-4}), 
  has a very large probability to be the ground state (or the
highest state).
The quantum numbers $I(v)$ of these states are: 
 $I(v)$=0(0), 2(4), 4(4), 8(4). 
$P(I)$'s of
states without the largest and/or the smallest 
$\alpha_{I}^J$ for a given $J$ are very small.
As we shall see from the discussions of next subsection,
this regularity is the feature for a certain state
to have a large probability to be the ground state.

The   regularity  that  large  
$P(I)$'s are related to  the largest or the smallest 
 coefficients among the fixed two-body matrix elements,  
 discussed above,  
is actually very general. 
Now let us look at  linear combinations of 
the random numbers which follow the Gaussian distribution with 
the average being zero and the width being one. 
Suppose that  $F(k)$ be a set of linear combinations  
of $G_J$:
\begin{equation}
F(k) = \sum_J \alpha_k^J G_J, ~~~ k = 1, 2, \cdots {\cal K},    \label{fk}
\end{equation}
where $k$ is used to specify each $F$, and 
 ${\cal K}$ is the total number of different $F$'s.  
One can prove  that distribution functions  of random $F(k)$ are  
\begin{equation}
  \rho(F(k)) = \frac{1}{\sqrt{2\pi} g_k} {\rm exp}
  \left(- \frac{ \left( F(k) \right)^2}{2g_k^2} \right),
 ~~ g_k^2 = \sum_J \left(\alpha_k^J\right)^2. 
   \label{randomx}
\end{equation}
If $\alpha_m^{J'}$ in Eq.~(\ref{fk})
is the largest (or the smallest) among all the 
$\alpha_k^{J'}$ ($k =1, \cdots   {\cal K}$), the  probability of 
 $F(m)$ being either the smallest or the largest number  is large. 
 To show this, let us look at  
\begin{equation} 
 {\cal F}(k) =  F(k) - F(m)  = 
\left( \alpha_k^{J'}  - \alpha_m^{J'} \right)   G_{J'} + \left( \sum_{J \neq J'} 
\left( \alpha_k^{J} - \alpha_m^{J}  \right) G_J \right),   \label{shift}
\end{equation}
where $k \neq m$. The right hand side of  Eq.~(\ref{shift}) has two
 terms, both of which are   random numbers which
 follow the Gaussian distribution. The value of  
 $\left( \alpha_k^{J'}  - \alpha_m^{J'} \right)  G_{J'}  $ 
 is  negative or positive for all 
 $ {\cal F}(k)$'s, and thus effectively produces 
 a ``simultaneous"  shift to either negative side or 
 positive side
 for all $ {\cal F}(k)$'s,
 depending on the sign
 of $\left( \alpha_k^{J'}  - \alpha_m^{J'} \right)  G_{J'}$. 
Therefore,  all of the functions ${\cal F}(k)$'s    
have large probabilities to be both negative and positive,  i.e., 
 $F(m)$ has a large probability to be either the 
smallest or the largest. 

  If there are two or more   
  coefficients $\alpha_m^{J'}$ ($J'=0,2,\cdots 2j-1$)
  which are the largest or smallest  for different functions $F(k)$, 
   the probability of  finding  $F(m)$ as either   
   the smallest or the largest is expected to increase.

By using Eq.~(\ref{7/2-4}) and Eq. ~(\ref{randomx}), 
    the distribution  width,  $g_{I}$ \footnote{We shall define
    another width, $\sigma_I$, in Sec. 5.6. },  
for  each state  of four fermions in a $j=7/2$ shell 
  are listed in the last row of  Table IV.
It is easily noticed that there is no correlation between
the  $P(I)$ values and their corresponding $g_{I(v)}'s$. 
For example, $P(0)$ ($\sim 20\%$) is much larger than $P(5)$ ($=0$) 
and $P(6)$ ($\sim 0$) although  $g_{I(v)=0(0)}$
is much smaller than $g_{I(v)=5(4)}$ and $g_{I(v)=6(2)}$.

\subsection{An empirical approach}

Because  $P(I)$'s in
Eq. ~(\ref{exact}) cannot yet be determined by a simple analytic procedure, 
one has to evaluate this integral  numerically. It is therefore  
desirable  to find  a simple alternative
method to evaluate  $I$ g.s. probabilities.

In Refs. \cite{Zhao2,Zhao4}, such a substitute was given. The idea is based
on the  observation discussed above: The state with a sizable $I$ g.s. probability
involves the largest and/or smallest $\alpha_{I(v)}^J$ with $J$ fixed. 
Thus the $I$ g.s. probability  might be proportional to the 
number of the  largest and/or smallest $\alpha_{I(v)}^J$. 
Let ${\cal N}'_I$   be the  sum of   numbers   
 of  the smallest and the largest $\alpha_{I(v)}^J$ 
with a fixed $J$ for a certain $I$. Then 
the $I$ g.s. probability is approximately 
given by
\begin{equation}
P(I) = {\cal N}'_I /N_m, \label{empirical-I}
\end{equation}
where $N_m=2N-1$ with $N$ 
the number of two-body matrix elements. 
Note that   $N_m=2N-1$ is used 
instead of $2N$,  because
all  $\alpha_{I(v)}^{J=0} (I \neq 0)$'s are 0 (there is
no smallest $\alpha_{I(v)}^{J=0}$), and that 
 $\sum_I {\cal N}'_I = N_m$.

Now we exemplify this empirical approach
by using four fermions in a $j=7/2$ shell.
Here $N=j+\frac{1}{2} = 4$, and $N_m = 2 \times 4$ -1 = 7. 
From Eq. (\ref{7/2-4}), it is easy to find the
largest (or the smallest) $\alpha_{I(v)}^{J}$   for  different  $I(v)$ 
states but fixed $G_J$: 
\begin{eqnarray}
{\rm for ~} G_0 ~ & &  \alpha_{0(0)}^{0} = {\bf \frac{3}{2}} ~ {\rm is ~
the ~ largest} ~; \nonumber \\
{\rm for ~} G_2 ~ & &  \alpha_{4(4)}^{2}~ ({\rm or ~} \alpha_{8(4)}^{2})
= {\bf \frac{7}{3}} ~ ({\it \frac{10}{21}}) ~  {\rm is ~
the ~ largest ~ (smallest) } ~; \nonumber \\
{\rm for ~} G_4 ~ & &  \alpha_{2(4)}^{4}~ ({\rm or ~} \alpha_{4(4)}^{4})
= {\bf \frac{42}{11}} ~ (~ {\it 1}~ ) ~  {\rm is ~
the ~ largest ~ (smallest) } ~; \nonumber \\
{\rm for ~} G_6 ~ & &  \alpha_{8(4)}^{6}~ ({\rm or ~} \alpha_{2(4)}^{6})
= {\bf \frac{127}{33}} ~ ({\it \frac{13}{11}}) ~  {\rm is ~
the ~ largest ~ (smallest) } ~. \nonumber 
\end{eqnarray}
We thus find that
\begin{eqnarray}
{\cal N}'_0 = 1 && {\rm given ~ by ~ the ~ largest} ~ \alpha_{0(0)}^{0};
\nonumber \\
{\cal N}'_2 = 2 && {\rm given ~ by ~ the ~ largest} ~ \alpha_{2(4)}^{4}
 {\rm ~ and ~ the ~ smallest} ~   \alpha_{2(4)}^{6}; \nonumber \\
{\cal N}'_4 = 2 && {\rm given ~ by ~ the ~ largest} ~ \alpha_{4(4)}^{2}
 {\rm ~ and ~ the ~ smallest} ~   \alpha_{4(4)}^{4}; \nonumber \\
{\cal N}'_8 = 2 && {\rm given ~ by ~ the ~ largest} ~ \alpha_{8(4)}^{6}
 {\rm ~ and ~ the ~ smallest} ~   \alpha_{8(4)}^{2} ~.  \nonumber 
\end{eqnarray}
According to the above empirical approach, we therefore predict that
$P(0)$ is 1/7, and $P(I)$ with $I=2, 4, 8$ and $v=4$ is 2/7.                             
These predicted $I$ g.s. probabilities 
are  given in the row ``pred2" of Table IV.
A very reasonable agreement is easily noticed between   
 the results  obtained by running the  TBRE Hamiltonian,  
those obtained by  multiple integrals such as Eq. (\ref{exact}), 
 the  solutions by using geometry method which will be 
discussed later,  and the predicted values by 
the present empirical approach.

The largest (smallest) $\alpha_{I(v)}^J$'s with fixed $J$
correspond to the ground (highest) state  
when $G_J=-1$ and other $G_{J' \neq J}$'s are zero.  Below 
 the ground (highest) state for   
one of $G_J=-1$ and other $G_{J' \neq J}=0$ will be discussed 
instead of the largest (smallest) $\alpha_{I(v)}^J$'s, for the sake of 
convenience.

Now let us study another case: $d$ boson systems. 
Similar to fermions in a   single-$j$ shell  for small $j$ 
 ($j=\frac{5}{2}$ or  $j=\frac{7}{2}$), the  
 relation between the two-body matrix elements and the
 eigenvalues for $d$-boson systems is also linear.  
The two-body Hamiltonian of a $d$-boson system is given by  
\begin{equation}
  H_d =  \sum_{L,M}  
  \frac{1}{2} c_L 
\left(d^{\dagger} d^{\dagger}  \right)^{(L)}_M 
\left( d  d  \right)^{(L)}_M  .   \label{d-boson} 
\end{equation}
From Eq.~(2.79) of Ref. \cite{Iachello1}, one obtains 
\begin{equation}
E=E_0  + \alpha' \frac{1}{2} n_d (n_d-1) + \beta' \left[n_d (n_d+3) - v(v+3) \right]
+ \gamma' \left[ I (I+1) - 6n_d \right],       \label{eigen0}
\end{equation} 
where $E_0$ contributes only to binding energies, and 
 $n_d$ is the number of $d$ bosons.
Eq.~(\ref{eigen0}) can  be rewritten as follows, 
\begin{equation}
E(v, n_{\Delta}, I)=E_0'(n_d) - \beta'  v(v+3) +  \gamma'   I (I+1).   \label{eigen}
\end{equation} 
The $\alpha', \beta'$ and $\gamma'$ in Eq. (\ref{eigen0})
are linear combinations of $c_0$, $c_2$ and $c_4$.  
From Eq.~(2.82) of Ref. \cite{Iachello1}, one obtains 
\begin{eqnarray}
&&  \alpha' = \frac{1}{7} (4c_2+3c_4), \nonumber  \\
&&  \beta'  =\frac{1}{70} (7c_0-10c_2+3c_4), \nonumber  \\
&&  \gamma' = \frac{1}{14} (-c_2 + c_4).   \label{coef}
\end{eqnarray}

Substituting these coefficients $\beta', \gamma'$ into
Eq.  (\ref{eigen}), and taking the two-body matrix elements
   $c_0, c_2$ and $c_4$ to be
the TBRE defined by Eq. (\ref{tbre}),
one easily calculates   $I$ g.s.
probabilities which are shown in Fig. 6.
It is easy to notice that 

1. The $P(I_{\rm max})$'s are almost constant (around 40$\%$)  for all
$n_d$ ($\le 4$); 

2. The $P(0)$'s and $P(2)$'s are periodic, with a period  $\delta(n_d)$=6.

3. All the $P(I_{\rm max})$, $P(0)$  and $P(2)$  are near
to 0,  20$\%$, 40$\%$, or $60\%$.  Other $P(I)$'s  are
always zero. 

Now let us explain these observations by using
the empirical approach introduced in Refs. \cite{Zhao2,Zhao4}. 
From  Eq.~(\ref{eigen}) and Eq.~(\ref{coef}),  
\begin{eqnarray}
  c_0=-1, c_2=c_4=0: &&
E(v, n_{\Delta}, I)=E_0'(n_d) + \frac{1}{10}  v(v+3);
\nonumber  \\
  c_2=-1, c_0=c_4=0: &&
E(v, n_{\Delta}, I)=E_0'(n_d) - \frac{1}{7}  v(v+3) + \frac{1}{14} I(I+1);
\nonumber  \\
  c_4=-1, c_0=c_2=0: &&
E(v, n_{\Delta}, I)=E_0'(n_d) + \frac{3}{70}  v(v+3) - \frac{1}{14} I(I+1), 
\label{final}
\end{eqnarray}
where  $E_0'(n_d)$ is a constant for all states.
 Based on Eq. (\ref{final}), one obtains Table  V,
 which presents the angular momenta
giving the largest (smallest) eigenvalues
when $c_L=-1$ ($L=$0, 2, 4) and other parameters are   0 for
$d$ boson systems. In Table V, 
$\kappa$ is a  natural number, and $n_d \ge 3$. 
These angular momenta  appear periodically,  originating  from  the
reduction rule of U(5)$\rightarrow$SO(3). 

One can use Table V to predict the $P(I)$ for $d$-boson systems. 
For example, ${\cal N}'_0=3$ and ${\cal N}'_{I_{\rm max}}=2$ for $n=6k$. 
According to  the empirical approach of  
Eq. ~(\ref{empirical-I})
\footnote{Note that when  one searches for
the smallest eigenvalue with $c_0=-1$ and
$c_2=c_4=0$  in case A of Eq.~(\ref{final}), one finds that 
many states with different $I$ are degenerate at the lowest value. 
Therefore, again,   $N_m=2N-1=5$ is used in  predicting
 $P(I)$'s by the formula $P(I)={\cal N}'_{I}/N_m$.}
, the predicted $P(I)={\cal N}'_I/N_m$,
 where $N_m=5$. 
 Thus the predicted   $P(0)=60\%$ and $P(I_{\rm max})=40\%$, and
all other  $P(I)$'s are predicted to be zero for $n=6k$.

A comparison between the  $P(I)$
values predicted by Table V and those in Fig. 6 shows that
a certain $P(I)$ is large if 
one state with angular momentum $I$
 involves the largest and/or
 the smallest $\alpha_{I\beta}^l$ (Eq.~(\ref{alpha}))
for a given $L$ ($L=0,2,4$).

For $d$-boson systems  one also finds that 
 0 g.s. probabilities 
 are very close to zero periodically when 
$n=6 \kappa \pm 1$ ($\kappa$ is a natural number).
These  counter examples of the 0 g.s. dominance are also  
predicted  by   Table V: if $n=6 \kappa \pm 1$, 
the $I=0$ states do not produce the largest and/or smallest 
eigenvalues when one of the parameters $c_l$ is $-1$
and   others  $c_{L'} ~ (L' \neq L)$ are zero.
This feature is  consistent with 
the discussion of Sec. 3.2, where it was 
seen that the $P(0)$ for an odd number of bosons with  
spin $l$ is usually not dominant.

\subsection{Predicted $P(I)$'s  based on geometry of the eigenvalues} 

Chau, Frank, Smirnova, and Isacker \cite{Chau}   showed
that  the simple systems discussed in last subsection 
can be  projected  to a  polyhedron on the axes defined by two-body
matrix elements.  These authors related the above   
$\alpha_{I\beta}^l$ with the largest possible convex
in $W$ dimensions, where $W=N-1$.  In this subsection we shall discuss 
predicted $P(I)$'s  based on geometry of the eigenvalues,
which was suggested in Ref. \cite{Chau}. 

Chau {\it et al.}  discussed   $d$ bosons   and 
four fermions in a  $j=7/2$ shell.  Here let us 
take five $d$ bosons as an example,
where $W=3-1=2$. The procedure  suggested by
Chau {\it et al.} is as follows, 

1) ~ One rewrites the eigenvalues of Eq. (\ref{eigen0}) as 
\begin{equation}
{\cal E}_{n_{\delta}, v, I}  = c_4 + \sum_{L} C_{n_{\delta}, v, I} (c_L - c_4),
\end{equation}
where ${\cal E}_{n_{\delta}, v, I} = E_{n_{\delta}, v, I} \frac{2}{n(n-1)} =
\frac{1}{10}  E_{n_{\delta}, v, I}$, is called the  ``scaled energy" 
 \cite{Chau}. 

2) ~ The scaled energy of an arbitrary eigenstate is  represented as a point 
in a plane spanned by $(c_2 - c_0)$ and $(c_4-c_0)$. 
All points (corresponding
eigenvalues) were found to be confined to a 
compact region with the size of one unit 
in each direction.  In the case of $d$ bosons,
the $I$ g.s. probability is related to each
angle at the corresponding vertex $i$ by  
\begin{equation}
p(I)_i  = \frac{1}{2} - \frac{\theta_i}{2 \pi}, \label{vertex} 
\end{equation}
where $p(I)_i$ is the probability of $I$ 
 to be the ground state contributed from the vertex $i$. The 
eigenstates for which corresponding points are not vertices
can not be the ground states for the TBRE Hamiltonian.
In Fig. 7   the twelve solid circles have one-to-one correspondence to the 
twelve  states with different $I$ and $v$ for five $d$ bosons.  
The angles $\theta_i$ ($i=1, \cdots 4)$ at vertex are also labeled
in Fig. 7.

3) ~ One finally sums $p(I)_i$
over all the $i$ vertices to get the total $P(I)$.
                                                    
This method is appropriate to  discuss  $P(I)$'s of systems
in which the number of two-body matrix elements is not large 
(3 or 4) (where the angle is relatively easy to
evaluate),  and meanwhile the Hamiltonian is diagonal. 
For more complicated cases this approach should be generalized.
Similar to  Eq. (\ref{exact}), the approach of Ref. \cite{Chau} presents
 exact $P(I)$'s without using random interactions.

\subsection{Mean-field method}

 $P(I)$'s  of $sp$ bosons in the presence of the TBRE Hamiltonian  
were first addressed   by Kusnezov in Ref. 
\cite{Kusnezov2}. He  addressed  $P(I)$'s by using random polynomials. 
The procedure is as follows:  First, choose $sp$ 
a system with a large boson number for which 
 the dimension of the Hilbert space is large. 
The Hamiltonian is then reduced to a tri-diagonal form with 
the trial Lanczos   state for which  
the number of $p$ bosons is equal to $I$. Next one constructs  
the off-diagonal and diagonal matrix elements in
terms of  $n_p/n$ and $I/I_{\rm max}$.

In the  limit of large boson number $n$, the lowest 
eigenvalues for each $I$ states can be 
found in terms of these off-diagonal and diagonal matrix elements
of the tri-diagonal matrix, 
and they can furthermore be written in the form  
of a  parabolic function of $n_p/n$, 
with the coefficients 
 determined by  matrix elements of  interactions
and $I/I_{\rm max}$. Then one is able to analyze the properties
of each term to evaluate  $I$ g.s. probabilities.
The advantage of the approach  in Ref.  \cite{Kusnezov2} is that  it avoids 
the diagonalization  of matrices. 
In general,  the lowest eigenvalues are not
quadratic polynomials, but have a more 
complicated form and should be studied
more carefully, as pointed out in Ref. \cite{Bijker5}.
The results of  Ref.  \cite{Kusnezov2}
is consistent with the mean-field approach of Refs. 
\cite{Bijker4,Bijker6,Bijker-conf}, 
which will be introduced   in this subsection.

In Refs. \cite{Bijker6,Bijker4}, Bijker and
Frank suggested   a mean-field analysis to predict 
$I$ g.s. probabilities in  the vibron model  and 
the $sd$ IBM. They  used the connection between 
potential energy surfaces of the  Hamiltonian and  
geometric shapes. Let us 
discuss below only the case of $sp$ bosons  with 
a schematic Hamiltonian, because the philosophy 
for a general vibron Hamiltonian and 
 the $sd$ IBM Hamiltonian is similar.

The schematic vibron Hamiltonian  of Ref. \cite{Bijker6} 
 is given by Eq. (\ref{vibron-H}) in Sec. 2.1. 
The range of $\chi$ in Eq. (\ref{vibron-H}) is from $-\frac{\pi}{2}$  to
$\frac{3\pi}{2}$. 
The coherent state of the vibron model is given by 
\begin{equation}
|n, \alpha \rangle = \frac{1}{\sqrt{n!}} \left(
{\rm cos} \alpha ~ s^{\dagger} +
{\rm sin} \alpha ~ p_0^{\dagger} \right)^n |0 \rangle,
\end{equation}
where $\alpha \in \left[0, \frac{\pi}{2}\right]$.
The potential surface is given by the expectation
value of  the vibron Hamiltonian
in the coherent state, i.e., 
\begin{equation}
E(\alpha) =  \frac{1}{4} {\rm sin} \chi ~ 
{\rm sin}^4 \alpha + {\rm cos} \chi ~ {\rm sin}^2 \alpha. 
\end{equation}

The equilibrium shape is obtained by calculating
one- and two-order derivatives of $E(\alpha)$ with
respect to $\alpha$, and the results can be classified
into the following three classes: 
\begin{eqnarray}
& {\rm (i)}   & \alpha_0 = 0, -\frac{\pi}{2} < \chi \le \frac{\pi}{2} ; \nonumber \\
& {\rm (ii)}  & {\rm cos} 2 \alpha_0 = {\rm cot} \chi,
\frac{\pi}{4} \le \chi \le \frac{3\pi}{4}; 
 \nonumber \\
& {\rm (iii)} & \alpha_0 = \frac{\pi}{2},
\frac{3\pi}{4} \le \chi \le \frac{3\pi}{2}.  
\end{eqnarray}

The   case  (i),
which corresponds to spherical symmetry and produces only 0 ground states, 
  occupies a portion of $(3 \pi/4)/(2 \pi)$ and contributes
 $3/8$=$37.5\%$   to 
$P(0)$; the   case (ii), which gives 0 ground states
when $\frac{\pi}{4} \le \chi \le \frac{\pi}{2}$ and
 $I_{\rm max}=n$ ground states when 
 $\frac{\pi}{2} \le \chi \le \frac{3\pi}{4}$,   contributes
$(1 \pi/4)/(2 \pi)=12.5\%$ to both $P(0)$ and $P(n)$;  the last case
(iii), which produces 0 g.s. (if $n$ is an even number) or 1 g.s.
(if $n$ is an odd number) when ${\pi} \le \chi \le \frac{3\pi}{2}$, 
and produces $I_{\rm max} = n$ g.s. when 
$\frac{3\pi}{4} \le \chi \le \pi$,  therefore the case (iii) contributes 
$\frac{\pi}{2}/(2\pi)$=25$\%$ to  $P(0)$ for even   $n$ or $25\%$ to 
$P(1)$ for   odd  $n$,   and contributes 
 $\frac{\pi}{4}/(2\pi)$=12.5$\%$ to  $P(n)$. 
To sum over these three cases, one obtains 
$P(0) = 75\%$ and $P(n) = 25\%$ for  even  $n$ and
that $P(0) = 50\%$, $P(1)=25\%$, and $P(n) = 25\%$ for   odd  $n$. 

 $P(I)$'s obtained by using the TBRE Hamiltonian for
vibrons, shown in Fig. 8, are very close to this simple prediction.
A detailed mean-field analysis for $sp$  bosons can be found   
in Refs. \cite{Bijker6,Bijker4,Bijker5}.

This mean-field approach was also applied to predict  
 $P(I)$'s of $sd$ boson systems  \cite{Bijker4,Bijker-conf}.  
Recently, Kota  applied this approach  \cite{Dalian} to 
analyze  probabilities of different  
irreducible representations in the ground states obtained
by using the TBRE Hamiltonian of the IBM.

\subsection{Summary of this Section}

In this Section we  first showed  that for systems in which  
eigenvalues are   linear combinations  
of two-body matrix elements, one can apply an 
empirical approach suggested in Refs. \cite{Zhao1,Zhao2,Zhao4} or  
an  approach based on geometry of eigenvalues suggested in Ref. \cite{Chau}
in order to predict  $P(I)$'s   without using  random  
interactions. The examples  include  
$d$ boson systems, three and four fermions in a $j$=$5/2$ or  $7/2$
shell.

We also showed that one can  explain  $P(I)$'s
of systems (such as
 $sp$- and $sd$-boson systems) in which 
one can classify two-body  matrix elements in a simple way. 
 $P(I)$'s for $sp$ and $sd$ bosons in the presence  
of the TBRE Hamiltonian 
were  evaluated  in Refs. \cite{Bijker5,Bijker6} based on 
a mean-field method.
The 0 g.s. dominance  for $sp$ bosons were also discussed in   
Ref. \cite{Kusnezov2} by Kusnezov based on random polynomials.  
The results of these two approaches are consistent, but 
the mean-field approach has a more transparent  picture.

On the other hand, one is unable to predict $P(I)$'s of
the systems discussed in this Section 
without {\it a priori}  knowing  the relations between the eigenvalues 
and the two-body matrix elements 
or the specific features of their eigenvalues.
In this  sense, the behavior of $P(I)$'s
for such  examples discussed in this Section 
has not been understood at  a  fundamental level. 
The   mean-field approach for the $sp$ and $sd$ bosons \cite{Bijker5,Bijker6} 
has not been successfully generalized to other cases.

Because the problem of the 0 g.s. dominance is very difficult, 
one may  leave this  embarrassing situation to the future  and  proceed 
by describing  $P(I)$'s for more complicated systems in which
the eigenvalues are not linear combinations of two-body matrix elements
and one is unable to classify the two-body matrix elements in
the way discussed in Sec. 4.4. Fortunately, 
the empirical approach of Sec. 4.2 was found to be applicable 
to these cases, after very slight modifications.  This  
will be discussed in the next Section.

\newpage 

\section{Spin $I$ ground state probabilities 
 of complicated systems}

In this Section we shall discuss  $P(I)$'s of complicated
systems. 
Here the appellation ``complicated systems" means 
 the eigenvalues of the systems are  not 
 linear in terms of two-body matrix elements and 
one cannot classify two-body matrix elements 
in a simple way (as done for $sp$ or $sd$ bosons in
last Section).

Although there have been no  simple understandings 
of the 0 g.s. dominance in complicated systems so far,  
 the empirical approach discussed in last Section 
was found to be reasonably applicable to prediction of $P(I)$'s, 
after slight modifications. 
In this Section we shall first discuss  results along this line.

We shall first consider 
 fermions in a single-$j$ $(j>7/2$) shell, next go to  
fermions in  many-$j$ shells and boson systems.
An argument of the empirical approach   will be discussed  schematically. 
Then we shall  discuss   $P(I_{\rm max})$'s  which 
were found to be considerably large for fermions in a single-$j$ shell
and boson systems.

In this Section we shall also discuss 
an argument of  the 0  g.s. dominance 
for  four fermions in a single-$j$ shell or four bosons with spin $l$ 
\cite{Zhao8}, studies of the 0 g.s. dominance based on
the distribution width
of the eigenvalues for states with different $I$,
and those based on time reversal invariance of the  Hamiltonian,
 large overlaps  between 
0 g.s. wavefunctions of systems with mass number
differing by two,  results of ground states for spin-$1/2$ fermions, etc.

\subsection{Fermions in a single-$j$ shell}

Let us take the case of four fermions in a single-$j$ shell again. This  
case was first studied 
 by Mulhall, Volya and Zelevinsky \cite{Mulhall1,Mulhall2} by
 using two-body interactions  which  distribute uniformly between $-1$
 and 1. However,  ground state probabilities
 discussed by these authors  are not $P(I)$'s, but actually those 
 of centroids of states  with given $I$ 
 (denoted as ${\cal P}(I)$ in Sec. 6). For instance, they predicted 
$50\%$ of $P(I_{\rm min})$ and another $50\%$ of $P(I_{\rm max})$, which 
is far from the observation but is the behavior of  energy centroids 
with spin $I$.

In  Refs. \cite{Zhao2,Zhao4} the 
empirical approach was generalized as follows:
 First,  one sets  one of the two-body matrix elements $G_J$  
to   $-1$ and  all  others  to  zero.
Then one  finds which angular momentum $I$ gives
the lowest eigenvalue among  all  the  
eigenvalues of the  shell model diagonalization.  
Suppose that the number of independent two-body matrix elements 
is $N$, then the above procedure is repeated $N$ times. 
Next, among the  $N$ runs   one counts  how many times
(denoted as ${\cal N}_I$)   a certain angular momentum $I$ 
gives the lowest eigenvalue  among all the possible 
eigenvalues. Finally, we predict that the probability of $I$ g.s. is  
  given by
\begin{equation}
 P(I) = {\cal N}_I/N ~. \label{empirical-II}
\end{equation}
In Eq. (\ref{empirical-I}) of
Sec. 4.2 we used ${\cal N}'_I$ which is the number of
both the largest and the smallest eigenvalues for $G_J = - \delta_{JJ'}$, 
and $N_m$ which is equal to $(2N-1)$  for four fermions in a $j=7/2$ shell 
\footnote{One can also use  Eq. (\ref{empirical-II}) for 
four fermions in a $j=7/2$ shell. The 
predicted $P(0)=P(2)=P(4)=P(8)=25\%$ if one takes only the largest 
$\alpha$'s. These predicted $P(I)$'s are also reasonably
consistent with other results.}, in order to predict 
the values of $P(I)$ empirically. 

Below let us use Eq. (\ref{empirical-II}) to predict  
 $P(I)$'s of  complicated systems, namely,
we do not use the largest eigenvalues for $G_{J}=-\delta_{JJ'}$.  
The reason of this modification is that the 
largest eigenvalues are usually (exactly or nearly) zero 
for  matrices corresponding many spin $I$'s, 
especially for  many-$j$ shells or a
 single-$j$ shell with large $j$. 
 To have the empirical approach  as simple  as possible, we shall 
use only the lowest eigenvalues  with 
 one of  the $G_J$'s being   $-1$ and the others  being switched off,  
for the case of  fermion systems in  a   single-$j$ shell for large $j$, 
 many-$j$ shells, $sd$- and $sdg$-boson systems.

For fermions in a single-$j$ shell,   tables of the  
angular momenta $I$  which give the lowest eigenvalues
for $n=3$-7 were presented in Refs. \cite{Zhao2,Zhao4}. 
The cases of $n=4$ is given in Table VI as an example. 
Here we also mention two systematic features. 
The first concerns  the quadrupole matrix element  ($G_2$) term. 
It has been well known that the monopole pairing 
interaction always gives  an  
$I=0$ ground state for an even number of fermions in a single-$j$ shell,   
and  an $I=j$ ground state for an odd number of fermions in a single-$j$ shell  
 when $G_0$ is set to be $-1$ and others 0. 
However, little was known   about  the role of 
  the $G_2$  matrix element  in a single-$j$ shell.
It was found in Ref. \cite{Zhao4} that 
the $J=2$ pairing interaction  always gives  an 
$I=n$ ground state for an even number of fermions,  
and an $I=j-(n-1)/2$ ground state for an odd number of fermions, 
 when $G_2$ is set to be $-1$ and others 0.
Another regularity   is that
interactions $G_J =-1$ (all others are zero) with 
$J=2j-3, 2j-5 \cdots$ ($j > 9/2$)  produce
$I=I_{\rm max}-8, I_{\rm max}-16 \cdots $ g.s. for $n=4$, 
  $G_J =-1$  with  $J=2j-3$ ($j > 13/2)$ produces
an $I=I_{\rm max}-12$ g.s. for $n=5$, and  $G_J=-1$ with 
  $J=2j-3$ ($j > 15/2)$  produces  an  
$I=I_{\rm max}-20$ g.s. for $n=6$.  No understanding
of these features are available.

Now we  exemplify the applications of Table VI for 
four fermions in a single-$j$ shell by the case of $j=9/2$,
where $N=j+1/2=5$. 
Here one finds according to the second row of Table VI 
that ${\cal N}_0 = 3$, and
${\cal N}_4={\cal N}_{I_{\rm max}} = 1$.
The predicted values of $P(I)$'s are 
$P(0)=60\%$ and $P(4)=P(I_{\rm max})=20\%$ while            
all other predicted $P(I)$'s are zero. 
For 1000 runs of the TBRE Hamiltonian \cite{Zhao2}, the  
$P(0)=66.4\%$, $P(4)=11.8\%$ and $P(I_{\rm max})=17.9\%$  
while  all  other $P(I)$'s are close to zero. 
The agreement between the  predicted $P(I)$'s by the empirical
method of Refs. \cite{Zhao2,Zhao4} and those 
obtained by diagonalizing the TBRE Hamiltonian is thus very good.

This empirical method can be also applied to predict 
$P(I)$'s of odd numbers of fermion systems. For example,
for five fermions in a    $j=9/2$ shell, 
one finds from the first row of Table IV in Ref. \cite{Zhao4} 
that ${\cal N}_j=2$, ${\cal N}_{3/2} = {\cal N}_{5/2}
= {\cal N}_{I_{\rm max}}=1$. The predicted values of
$P(I)$'s are thus $P(3/2)= P(5/2)= P({I_{\rm max}}) = 20 \%$ 
and $P(j) = 40\%$ while all other $P(I)$'s are predicted to be zero.
$P(I)$'s obtained by 1000 runs of diagonalizing
the TBRE Hamiltonian are: 
$P(3/2)= 20.5\%$, $P(5/2)=15.5\%$,  $P({I_{\rm max}}) = 18.4 \%$,
and $P(j) = 33.9 \%$ while all other $P(I)$'s are close to zero.
Good agreement is also obtained between the
predicted values of $P(I)$'s and those obtained
by using the TBRE Hamiltonian.

 Figure 9 gives a comparison between the predicted 
$P(0)$'s (open squares), which are
obtained by using Table VI of this paper and
Table V of Ref. \cite{Zhao4}, 
  and those obtained by diagonalizing the 
 TBRE Hamiltonian  (solid squares) for $n=4$ and 6. 
It can  be seen that a good    
agreement   is  obtained  for   fermions in    a 
single-$j$ shell  for both small and large $j$. 
 The predicted 0 g.s. probabilities 
exhibit  a  similar staggering  as  those obtained by 
diagonalizing the  TBRE Hamiltonian.

 It is interesting to note  
 that  the  $P(0)$'s of four and six fermions 
 can also be fitted by empirical formulas:
\begin{eqnarray}
&& {\rm for} ~ n=4: 
P(0) = \frac{ \left[ (2j+1)/6 \right] + k }{j+\frac{1}{2}}
\times 100 \%, ~~ k =
\left\{
\begin{array}{ll}
1       &  {\rm if}~ 2j = 3m   \\
0       &  {\rm if}~ 2j+1 = 3m   \\
-1      &  {\rm if}~ 2j-1 = 3m   
\end{array}  \right.;
\nonumber  \\   
&& {\rm for} ~ n=6: 
 P(0) = \frac{ 2 \left[ (2j+3)/6 \right] -1}{j-\frac{1}{2}}  
\times 100 \%,    \label{test}
\end{eqnarray} 
where the $``\left[ ~~ \right]"$ means to take  the largest integer not 
exceeding the value inside. 
These empirical formulas are  
interesting because they present scenarios  
for very large-$j$ cases where  it would be not possible to 
diagonalize the TBRE Hamiltonian.  A comparison
between the values predicted by these formulae and those
 obtained by diagonalizing the TBRE Hamiltonian  
is also given in Fig. 9.

A correlation can be found in this figure 
 between  $P(j)$ for an odd number of fermions
in a single-$j$ shell and  $P(0)$ for a neighboring even number of fermions
in a single-$j$ shell. This correlation originates  from an attractive
monopole pairing interaction:
 $G_0=-1$ gives the 0 g.s. for fermions with an
even number  $n$ of particles 
and $j$ g.s. for those with the neighboring odd $n$. 
Namely, at a large probability,   random interactions with 
attractive $G_0$  give  
0 g.s. for the case of  even $n$ and  $j$ g.s. for the case
of odd $n$. If one  switches off $G_0$ term, there will be  
in general much fewer cases 
for which the same sets of random interactions 
give 0 g.s. for the case of   even $n$ and  $j$ 
g.s. for the case of the  neighboring  odd $n$.

\subsection{Fermions in  many-$j$ shells,  $sd$- and $sdg$-boson systems}

Now we discuss $P(I)$'s 
 of fermions in   many-$j$ shells and bosons with 
many  spin $l$'s by using  the empirical approach of 
  Eq. (\ref{empirical-II}). This formula 
remains the same here except that the $G_J$ should be   
replaced by the general two-body matrix elements  
$G_{JT}(j_1 j_2, j_3 j_4)$.

Let us first  exemplify
a two-$j$ ($j=\frac{7}{2}, \frac{5}{2}$) shell with 
$n$=4 to 7. 
A comparison of the predicted $P(I)$'s by using the  
empirical formula of Eq. (\ref{empirical-II}) with those obtained by 
diagonalizing the TBRE Hamiltonian  of fermions in   
a two-$j$ ($j=\frac{7}{2}, \frac{5}{2}$) shell with  
$n$=4 to 7 is shown in Fig. 10. One sees a  
reasonable agreement    
\footnote{For fermions in   many-$j$ shells, the number of two-body matrix elements
is usually large.  In such cases, 
especially  for odd-fermion systems, there are ``quasi-degeneracy" problems 
in counting ${\cal N}_I$:
 sometimes  the lowest eigenvalue is  quite close to the second lowest
one when one uses $ G_J(j_1 j_2, j_3 j_4) =-1$ and others 0. 
 For such two-body matrix elements,
one should actually introduce an additional ``rule"
in order to have a more reliable prediction.  
Namely, it is not appropriate to count ${\cal N}_I$ in
the most naive way. 
In order to avoid confusions, however,
we do not modify the  way to count ${\cal N}_I$ in  such 
cases  throughout this article.  It is noted that
  $I=\frac{7}{2}$ in Fig. 10(b) and $I=\frac{3}{2}$ in Fig. 5(d) 
are cases of ``quasi-degeneracy". Improvement 
of agreement between the predicted $P(I)$'s 
and those obtained by diagonalizing the TBRE Hamiltonian can be achieved
by appropriately considering  the above ``quasi-degeneracy".}.

Now let us study $sd$-boson systems similarly.  
Table VII presents spins $I$  which appear in the lowest states 
when one of the above $sd$-boson parameters is set to be $-1$ and others 0.
We predict, according to Table VII and the procedure of the
empirical method, that
only $I=0$, 2, and $2n$  g.s. probabilities are sizable, while  other 
 $I$  g.s. probabilities  are close to zero. 
Fig. 11 shows a comparison of the predicted 
$P(I)$'s and those obtained by diagonalizing the TBRE Hamiltonian 
of $sd$-boson systems, with boson numbers ranging from
 6 to 16. One can see a reasonable agreement between 
 $P(I)$'s obtained by using the TBRE Hamiltonian and those
 predicted by the empirical formula of Eq. (\ref{empirical-II}).

Because the empirical formula of Eq. (\ref{empirical-II})
reasonably predicts  $P(I)$'s of both simple and complicated systems,
it is also called an empirical rule of $P(I)$'s in this paper.

\subsection{Simple argument of the 0 g.s. dominance  
 for four fermions in a single-$j$ shell and
four bosons with   spin  $l$}

A simple argument of 0 g.s. dominance
for four fermions in a single-$j$ shell and four bosons with 
spin $l$ was recently discussed in Ref. \cite{Zhaoym}.  The essential point
is that there is only one $I=0$ state with non-zero eigenvalue
among many   $I=0$ states of these systems, while  
those of $I \neq 0$ are scattered with smaller values in
magnitude when only one of the two-body matrix elements is taken
to be $-1$ and others are zero. 
  This feature can be proved 
 by constructing the $I=0$ states by using
 a pair basis \cite{Zhao8}. We shall discuss below 
  for the case of  four fermions in a single-$j$ shell. 
According to the empirical rule, 
the $I=0$ states have a large probability
to be the ground states for the TBRE Hamiltonian. 

The single-$j$  Hamiltonian  
for which one of the $G_J$'s is $-1$ and others are zero
 is defined by 
\begin{equation}
H= H_J  = -\sum_M A^{\dagger (J)}_M A^{(J)}_M. 
\end{equation}  
We define
\begin{equation}
|r_1 r_2: I\rangle \rangle =
\frac{1}{ \sqrt{N_{r_1 r_2 I}} }
\left(A^{(r_1)\dagger}   A^{(r_2) \dagger}  \right)^{(I)}
| 0 \rangle ~, \nonumber
\end{equation}
where  $| \rangle \rangle$ 
means  that the basis is normalized,
$A^{(r_i)\dagger}$  is  defined by Eq. (\ref{Clebsch}) without
isospin degree of freedom,  and 
$N_{r_1 r_2 I}$  is 
given by
\begin{equation}
N_{r_1 r_2 I} = 1 + \delta_{r_1 r_2} - 4  (2r_1+1) (2r_2+1)
 \left\{ \begin{array}{ccc}
 j    & j  & r_1 \\
 j    & j  & r_2 \\
 r_1  & r_2 & I  \end{array} \right\} ~. \nonumber   \label{e5.13}  
\end{equation}
The matrix elements of $H_J$
  are as follows  (see Ref. \cite{Yoshinaga2}), 
\begin{equation}
  \langle \langle r'_1 r'_2: I |
H_J  |
r_1 r_2: I\rangle \rangle 
= \frac{-1}{\sqrt{ N_{r_1 r_2 I} N_{r'_1 r'_2 I}} } \sum_{R={\rm even}}
 U_{r_1 r_2 J R}   U_{r'_1 r'_2 J R} ~,
 \label{matrix1}
\end{equation}
with
\begin{equation}
U_{r_1 r_2 J R} = \delta_{J r_1} \delta_{R r_2}
                + (-)^I \delta_{J r_2} \delta_{R r_1}
                - 4 \hat{L}_1  \hat{L}_2  \hat{J}  \hat{R}
 \left\{ \begin{array}{ccc}
 j    & j  & r_1 \\
 j    & j  & r_2 \\
 J  & R & I  \end{array} \right\}  \nonumber,    \label{e5.12}  
\end{equation}
where $\hat{r}_1$ is a short hand notation of  $\sqrt{2r_1+1}$.

Let $|\Phi_J \rangle \equiv | JJ: 0 \rangle \rangle $,
 $|\Phi_K \rangle \equiv | KK: 0 \rangle \rangle -  \langle \langle KK: 0|JJ:0 \rangle \rangle 
  |JJ:0 \rangle \rangle $ ($K \neq J$).
The  new basis constructed by $|\Phi_K \rangle$'s are 
orthogonalized with respect to  {\it only} $|\Phi_J \rangle$,  
and  not normalized except $|\Phi_J \rangle$. 
Similar to Eqs. (\ref{e5.12}) and (\ref{e5.13}), one has
\begin{equation}
  \langle \langle KK: 0 | H_J  |K'K':0 \rangle \rangle
=  \langle \langle KK: 0 |JJ:0 \rangle \rangle 
~ \langle  \langle K'K': 0 |JJ:0 \rangle \rangle N_{JJ0},  
\end{equation}
where $K'=0, 2, \cdots, 2j-1$.

By using this formula, one easily confirms that all matrix elements
$  \langle  \Phi_K | H_J  | \Phi_{K'}   \rangle$ are zero
except when $K=K'=J$. 
Namely, $ \langle \langle JJ: 0| H_J | JJ: 0 \rangle \rangle=N_{JJ0} =
E_I^{J(j)} $ 
is the {\it only} non-zero eigenvalue for the
$I=0$ states of four fermions interacting by  $H_J$ 
(This  procedure is also applicable for  
four bosons with  spin $l$).  
From the sum  rule of diagonal matrix elements \cite{Yoshinaga1} 
one obtains that $\sum_J E_I^{J(j)} = -\frac{1}{2} n(n-1) D_I^{(j)}
= -6 D_I^{(j)}$, 
where  $n$ is the number of 
fermions in the system ($n=4$ here), and
$D_I^{(j)}$is the  number of  states with spin  $I$. 
  For $I \neq 0$ states the
eigenvalues for $G_J = -\delta_{JJ'}$ are scattered while
for $I=0$ there is only one non-zero eigenvalue which
is {\it on average} larger in magnitude  than eigenvalues 
for  states with other  $I$'s   except
for very small $j$ \footnote{This does not mean that
the $I=0$ state with nonzero eigenvalue
are always lower in energy than other $I$ states 
for $G_J = -\delta_{JJ'}$. There are exceptions where
other $I$ states appear lower than $I=0$. 
For example, the $I_{\rm max}$ state is the lowest
for $J={\rm 2j-1}$. }.
This suggests that ${\cal N}_0$ is larger than  
${\cal N}_I$ ($I\neq 0$). 
According to the empirical rule, the   
0 g.s.  probability  is larger 
than that  of all other $I$  states.

One sees that the increase of $P(0)$  ``coincides" with that 
 the  number of $I=0$ states for four 
fermions in a single-$j$ shell or that
 for (denoted as $D_0^{(l)}$) four bosons 
with  spin $l$, respectively. $D_0^{(j)}$  and $D_0^{(l)}$
take their values  1, 1,   1, 2, 2, 2, 3, 3, 3,
4, 4, 4, $\cdots$ for $2j$=3, 5, 7, 9, 11, $\cdots$ or 
$l$= 0, 1, 2, 3, 4, $\cdots$, etc. 
 Although it is still difficult to  prove  this correlation, below we give 
an argument based on the empirical method of Refs. \cite{Zhao2,Zhao4}.
Taking a system of
four fermions  as an example, the relation 
 $\sum_J E_I^{J(j)} = -6 D_I^{(j)}$ 
 means that the magnitude of the sum of $E_0^{J(j)}$ 
increases with $D_0^{(j)}$,  suggesting  
that ${\cal N}_0$ increases  with 
$D_0^{(j)}$ simultaneously.
Because $P(0) = {\cal N}_0 / N$, 
a regular increase of $D_0^{(j)}$ 
of four fermions   \cite{Ginocchio1,Zhaoym2}   produces 
a regular staggering of  the $P(0)$'s. It is noted that
a similar ``coincidence" of the    staggering of the $P(0)$'s 
 with increasing number of $I$=0 states  was   observed 
in Refs. \cite{Mulhall2,Zhao1} for four fermions in a 
single-$j$ shell, but without an explanation or argument.

\subsection{Schematic interpretation of the empirical approach}

The   rationale of the above empirical approach can be seen from the following analysis.
Although the  relation between the eigenvalues and 
two-body interactions is complicated, the   
eigenvalues are always linear in terms of two-body interactions 
in a ``local" space (explained below) of the TBRE.  Therefore,
  instead of studying the effects of all two-body matrix
elements simultaneously, we can decompose the problem into $N$ parts. 
In each part, we focus on only one interaction matrix element, i.e.,
the entanglements between the two-body matrix elements are
neglected.  Let us take a certain $G_J=-1$ and   $G_{J'}=0$ ($J' \neq J$), and 
diagonalize the  Hamiltonian of, e.g., fermions in a single-$j$
shell. Suppose that $E_{I \beta}^J$ is the lowest eigenvalue. 
The   wave function corresponding to $E_{I \beta}^J$ is 
\begin{equation}
\Phi (j^n, I \beta J) =  \sum_{K J'\gamma}
\langle j^{n-2} (K \gamma), j^2 (J') | \}
j^n I \beta J \rangle
\left[ \Phi \left(j^{n-2} (K \gamma) \right)
\Phi \left(j^2 (J') \right) \right]^{(I)}, 
\label{linear}
\end{equation}
where $\langle j^{n-2} (K \gamma), j^2 (J') | \}
j^n I \beta J \rangle $'s are the
two-body coefficients of fractional parentage (cfp's), 
 which have been widely used in   shell model calculations. 
$\beta$ ($\gamma$) refers to   
additional quantum numbers needed 
to define a state of $n$  (or $n-2$)  
fermions with total angular momentum $I$ (or $K)$ uniquely.

Now we introduce a small perturbation by  
adding $\epsilon  \{G_J'\}$ to $G_J=-1$.  $G_J=-1$ and  
$\epsilon  \{G_J'\}$ define the ``local" space  
of our TBRE. Let us call it the $J$ subspace.   The  
  new eigenvalue is approximated in the first order by   
\begin{equation}
\left( E_{I \beta}^J \right)' = E_{I\beta}^J
+ \epsilon \frac{n(n-1)}{2} \sum_{K J'\gamma}
\left[ \langle j^{n-2} (K \gamma), j^2 (J') | \}
j^n I \beta J \rangle \right]^2 G_{J'}.  
\end{equation}
This means that the  $E_{I \beta}^J$ is 
linear in terms of $\{G_{J'} \}$ in the $J$ subspace.
Because $E_{I\beta}^J$ 
gives the lowest eigenvalue
for the case with $G_J=-1$ and others zero, 
this angular momentum $I$ continues to  
give the lowest eigenvalue
in this   local space ($\epsilon$ can be 0.2-0.3 in most cases  according to 
our numerical experiments).  The full space of the  TBRE Hamiltonian 
can be covered mostly by the  $N$ subspaces defined above.

Thus the empirical method of Refs. \cite{Zhao2,Zhao4}
implicitly assumes that a very large  part of the full space of the 
TBRE Hamiltonian can be covered  by 
the $N$ local subspaces, which are defined by  
  introducing a small perturbation   
 $\{\epsilon G_{J'}\}$ ($\epsilon$ is small)
to a fixed  $G_J=-1$ ($J \neq J')$. For 
 fermions in a single-$j$ shell, for instance,  $G_J=-1$ and
$\{\epsilon G_{J'}\}$ define the $(J/2+1)$-th local subspace 
of two-body matrix elements. The local subspace with $J=0$ is the  
first subspace, and that with $J=2$ the second, and so on.

This philosophy can also be shown from the following numerical experiments:  
Let us take four fermions in a
single-$j$ shell with $j=\frac{17}{2}$. In Fig. 12(a) 
$G_{J_{\rm max}} (J_{\rm max} = 16)=-1$  and  all the other
two-body matrix elements
$G_{J'}$ ($J' \neq J_{\rm max}$) are set
to be the TBRE, but with a factor $\epsilon$ multiplied. 
One sees that almost all cases of  the  g.s. belong  to
$I=I_{\rm max}=4j-6$=28 when  $\epsilon$ is small (say, 0.4).
If one uses $G_{J_{\rm max}} (J_{\rm max} = 16)= 1$, then 
$P(I_{\rm max}) \sim 0$, which  means that the cases
of the TBRE with  $G_{J_{\rm max}} < 0$
produce almost all the $I_{\rm max}$ g.s. in a single-$j$ shell.   
 Fig. 12(b) shows the  results for the same system  with 
$G_0$ being $-1$ and other $G_J$'s being the TBRE
multiplied by   $\epsilon$. It is seen  that  the 0 g.s.
is overwhelming for small $\epsilon$. 
When   $\epsilon=0$ and $G_0=+1$,  the 
0 g.s. probability is also sizable because of the contributions
from $J=6$ and $J=12$ (refer to Table  VI).

The above numerical experiments are not trivial.
By this procedure one  can  find which interactions, 
not only monopole pairing, are important in obtaining  the 0 g.s. dominance. 
Taking four fermions in a single-$j$ ($j=31/2$) shell as an example, 
the 0 g.s. probability is $\sim$0.2$\%$ if we
delete all  two-body interactions which
produce $I=0$ g.s. ($J=0, 6, 8, 12,$ and 22, refer to the last row
of Table VI). 
This means that the 0 g.s. dominance comes
essentially from those five interactions. 
Previously, Johnson {\it et al.} noticed that the robustness of the 0 g.s. 
dominance is  more or less independent
of monopole pairing \cite{Johnson1,Johnson2}.
It was not known, however,  whether  a certain two-body
matrix element is essential or partly responsible, and
how to find which interactions are essential, 
in producing the 0 g.s. dominance for a given system.

A shortcoming of the  empirical approach of Refs. \cite{Zhao2,Zhao4} 
is as follows:
One takes one  of  $G_J$'s is set to be $-1$ (other $G_J$'s are zero) 
in each numerical experiment, and  one finds the angular momentum
of the lowest state.  The case with $G_J=1$ is  excluded in  
the numerical experiments because  
one would likely obtain 
degenerate levels with the lowest energy for this case.  
 Thus the good  consistence between the predicted $I$ g.s.   
probabilities by the  empirical approach of Refs. \cite{Zhao2,Zhao4}  
and those obtained by diagonalizing
the TBRE Hamiltonian indicates that  
the properties of local spaces
 defined by $\{G_J=-1 + \epsilon G_{J'}  (J' \neq J) \}$ 
  more or less  represent the  features of the full space 
\footnote{However, some $\{G_J=1 + \epsilon G_{J'}  (J' \neq J) \}$
local subspaces also contribute to  $I$ g.s. probabilities. 
 In some  cases we 
find that for an even number of fermions 
the 0 g.s. probability may   not be very small  even 
if one deletes all the terms which give $I=0$ g.s. with
only one of $G_J=-1$ and other $G_{J'}$ switched off.
For example, for $n=4$ and $j=\frac{15}{2}$ the
$P(0)$=50.2$\%$ if we use the full TBRE Hamiltonian, and 
$P(0)$=23.1$\%$ if we delete $G_J$'s with $J=0, 4, 8, 10$
which present $I=0$ g.s. if these $G_J=-1$ (refer to Table VI). 
If we deleted $G_{12}$ then the  
$P(0)$ would be 14.2$\%$.  Nevertheless, 
 numerical experiments provide a very simple recipe to find   
what interactions are essential for a certain $I$ g.s. probability. }.

\subsection{Spin $I_{\rm max}$ g.s. probabilities}

For fermions in a  single-$j$ shell,  
the state with the maximum angular momentum (denoted as 
$I_{\rm max}$)   was found to have a sizable probability to
be the  g.s. \cite{Mulhall1,Zhao1}. 
This phenomenon can be  explained  by the observation that 
 ${\cal N}_{I_{\rm max}} =1$ always   
\footnote{In Ref. \cite{Zhao4},   an argument was given why 
 $E_{I_{\rm max}}^{J_{\rm max}}$ 
is the lowest eigenvalue when $G_{J_{\rm max}}=-1$
and other $G_J$'s are zero. }. 
The predicted $I_{\rm max}$ g.s. probabilities 
 of fermions in  a single-$j$ shell  are 
$\frac{1}{N} = \frac{1}{j+1/2} \times 100 \%$, independent of 
all particle numbers (even or odd), which is shown in Fig. 13(a).

The above argument on  $I_{\rm max}$g.s. probabilities for single-$j$
shells can be readily generalized to many-$j$ shells. Consider,
for example, two shells with angular momenta $j_1$ and $j_2$.
Following the same logic as was used for a single-$j$ shell, we
predict that the two angular momenta $I_{\rm max}'=I_{\rm max}(j_1^n)$ and 
$I_{\rm max}(j_2^n)$ have  g.s. probabilities which are at least as
large as $1/N \times 100\%$. Here, $I_{\rm max}(j^n_i)$ ~ $(i=1$ or 2)
is the largest  angular momentum of a state constructed from the $j^n_i$
configuration, and $N$ is the number
of independent  two-body interactions
in the ($j_1, j_2$) shells. 
In other words, we can predict in this way the
lower limit for these $I_{\rm max}'$g.s. probabilities.

Figure 13(b) presents
 $I_{\rm max}'=I_{\rm max}(j_1^n)$ and $I_{\rm max}(j_2^n)$ 
g.s. probabilities. They are compared with a simple $1/N$ plot.
Indeed, the  predicted lower limit of $1/N$ for  $I_{\rm max}'$g.s.
probabilities works very well. It should also be noted that 
$I$g.s. probabilities with $I$ very near $I_{\rm max}'$ are extremely
small (less than 1$\%$) in all these examples.

Figure 13(c) shows the $P(I_{\rm max})$'s vs. spin $l$ of bosons, with $n$ 
ranging from three to six.
When   $l$  is small 
the agreement is   good; when $l$ becomes larger,
deviations between  the values 
calculated by diagonalizing the TBRE Hamiltonian 
(denoted as $P^{\rm TBRE}(I_{\rm max},l)$) 
and those predicted by the above $1/N=1/(l+1)$ relation appear.  
  The  $P^{\rm TBRE}(I_{\rm max},l)$'s are systematically larger  
than $1/N$ and increase with $n$.

An argument why the  behavior of the
$P^{\rm TBRE}(I_{\rm max},l)$'s is different for 
bosons and fermions, given  in Ref. \cite{Zhaoym}, is as follows. 
As discussed above, 
the $P(I_{\rm max})$ comes essentially
from a gap produced by the pairing 
interaction  $G_{L_{\rm max}}$ for bosons with  spin $l$,
or the pairing interaction $G_{J_{\rm max}}$  for fermions in a single-$j$ shell, where
$L_{\rm max}=2l$ and $J_{\rm max}=2j-1$, respectively.
One can trace the ``anomaly" of $P^{\rm TBRE}(I_{\rm max},l)$  
back to the gap associated with $G_{J_{\rm max}}$ or $G_{L_{\rm max}}$ by 
using  analytical formulas of $(E_{I_{\rm max}-2} -  E_{I_{\rm max}})$. 
Here the state with $I=I_{\rm max}$ ~ ($I_{\rm max}-2$) 
is found to be the ground (first
excited) state if   $G_{J_{\rm max}}$ or $G_{L_{\rm max}}$ is $-1$. 
We obtain $(E_{I_{\rm max}-2} -  E_{I_{\rm max}})$ as follows, 
\begin{eqnarray}
  {\rm boson ~~ systems}: &&
\frac{2ln-1}{4l-1};
\nonumber  \\
  {\rm fermion ~~ systems, } ~ n=4: &&
\frac{3}{8} + \frac{105}{128(4j-7)} + \frac{135}{64(4j-5)}
\nonumber  \\
&& 
- \frac{63}{128(4j-3)};
\nonumber  \\
  {\rm fermion ~~systems, } ~ n=5: &&
\frac{35}{128} + \frac{2205}{2048(4j-9)}
+ \frac{5145}{2048(4j-7)}
\nonumber  \\
&& 
- \frac{1785}{2048(4j-5)}
-  \frac{189}{2048(4j-3)};
\nonumber  \\
  {\rm fermion~~ systems, } ~ n=6: &&
\frac{27}{128} + \frac{10395}{8192(4j-11)} + \frac{2835}{1024(4j-9)}
\nonumber  \\
&& 
- \frac{4725}{4096(4j-7)} - \frac{45}{256(4j-5)}
-  \frac{297}{8192(4j-3)}.
\nonumber  \\
\label{final1}
\end{eqnarray}
One easily sees that the gap for  bosons
with  spin $l$   increases  regularly with $n$ 
at an interval$\sim 1/2$  if $l$ is large, while that for  
fermions in a single-$j$ shell is much smaller
(almost one order) in magnitude 
and comparable for different $n$ and $j$.
For instance,  the gap is 0.47, 0.39,  0.35  for
$n=$4, 5 and 6 fermions in a $j=15/2$ shell, respectively, while 
the gap is 2.03, 2.56, 3.07
for $n=4$, 5, 6 bosons with  spin $l=7$, respectively.
According to the empirical rule discussed above, 
a relatively larger gap makes  the corresponding 
$P(I)$  larger: the larger the gap is,
the larger the corresponding $P(I)$ is.
It is pointed out again,  however, that the 
  the $P^{\rm TBRE}(I_{\rm max},l)$ follows the $1/N$ prediction 
when the spins of bosons in the system are small.

For  $sd$ bosons,  it was found in 
Refs. \cite{Bijker2,Bijker3}   that   $I_{\rm max}$ g.s. probabilities 
are  large, which  can   be  explained in the same way: 
 Among the  two-body matrix elements,
 interactions with  $c_4=-1$ and others being 0 produce 
the lowest eigenvalue for the $I_{\rm max}=2n$ state.
The predicted $I=2n$  g.s. probability is 1/$N$=1/6=16.7$\%$, 
consistent with  that obtained by
diagonalizing the TBRE Hamiltonian ($\sim 15\%$) 
\footnote{Note that the  term
$ (s^{\dagger} d^{\dagger}) (s d)$ gives degenerate 
lowest eigenvalues   for many $I$ states when 
$e_{sdsd}$ is set to be $-1$ and others are 0. Therefore,
we use six (instead of seven)  as the number of 
independent two-body matrix elements, $N$.
The difference due to this minor modification is very  small
($\sim 2 \%$), though.}. In $sdg$-boson systems, 
the predicted $I_{\rm max}=4n$ g.s probabilities is  
$1/N \sim 3.2\%$, where $N=32$. The $I_{\rm max}$ g.s. probability that we 
obtain  by diagonalizing the TBRE Hamiltonian  are 
$3.3\%$, $4.2\%$, $3.3\%$ for $n=$4, 5, 6, respectively.

Therefore,
for fermions in a single-$j$
shell the $1/N$ relation works very well for both small and large $j$,
and for both small and large $n$;
for fermions in   many-$j$ shells, the $1/N$ relation predicts very well 
the lower limit for   $I=I'_{\rm max}=I_{\rm max} (j^n_i)$ g.s. 
probabilities. 
The $P^{\rm TBRE}(I_{\rm max},l)$'s 
of bosons follow the $1/N$ prediction when the spins  ($l$)  
of the bosons in the system are small. Large
deviation from the $1/N=1/(l+1)$ 
relation appears when $l$ is large.

\subsection{Argument based on width}  

Two definitions of width were used in the literature.
One is defined by $g_I = \sqrt{ \langle H^2 \rangle /D_I}$,
where $D_I$ is the number of states with angular momentum $I$. 
The width $g_{I(v)}$'s in Table IV are calculated by using
this definition. The other is defined by
$\sigma_I = \sqrt{ \langle (H-\overline{E_I})^2 \rangle /  D_I }$.
In this subsection and Sec. 6.1, the
overhead line above a certain quantity  (such as $E_I$) means to take the 
average of this quantity over all the states with spin $I$ 
 obtained by {\it one set} of the TBRE Hamiltonian, and 
$\langle   ~ \rangle$ outside  
 means to take the TBRE average (in our case averaging over 1000 runs).

As discussed in Sec. 4.1, the argument based on
 $g_I$ is not 
applicable to cases of fermions in a single-$j$ shell. 
For example, although the $I=I_{\rm max}$ state for 
 fermions in a single-$j$ shell has the
 the largest width $g_I$, $P(I_{\rm max})$ is smaller than
 $P(0)$  for all $j > 7/2$. Here $P(I_{\rm max})$ was shown
to be $1/(j+\frac{1}{2})$ in Sec. 5.5.  The width $g_I$ does not
have a clear relation or correlation with $P(I)$ for fermions in a 
single-$j$ shell. In Ref. \cite{Bijker3}, Bijker and Frank found
a similar result for $sd$ bosons. 
                                   
Very recently,  Papenbrock and Weidenmueller
refined the width ($g_I$)  argument in Ref. \cite{Papen_new}.
They  derived
the distribution of and the correlation between the 
$g_I$'s.  They empirically obtained that 
the maximum of $|E_{I\beta}| \simeq r_I g_I$ ($I$ fixed), where 
  the value $r_I$ is a constant and 
is called   ``scaling factor".  For 
a single-$j$ shell with $n=6$ and $j=19/2$, $r_I$ 
$\sim 1.8$ for small and medium $I$, and $\sim 1$ for 
$I \sim I_{\rm max}$.  Interestingly, 
they found that the $P(I) \simeq $ probability 
for  $r_I g_I$ to be  maximal for six or eight fermions 
in a $j=19/2$ shell. It was speculated in Ref. \cite{Papen_new} that
similar considerations  would also apply to other  many-body systems.

In Refs. \cite{Yoshinaga1,Arima4}  we tried to understand  
   $P(I)$'s by combining the property of width $\sigma_I$  and the 
probability for $\overline{E_I}$ to be the lowest energy  (this probability  
is denoted as ${\cal P}(I)$ and will be   
be discussed in details in Sec. 6). The idea of Refs. 
\cite{Yoshinaga1,Arima4} is as follows: 
  ${\cal P}(I_{\rm min})$ and 
${\cal P}(I_{\rm max})$ are always large, which can be traced back to the 
quasi-randomness of two-body coefficients of fractional parentage
(as discussed in Ref. \cite{Zhao5}). 
 On the other hand, one should be aware that there is only one 
state with $I=I_{\rm max}$, while
the number of $I=I_{\rm min}$=0 states for an
even number of fermions  in a single-$j$ shell  is usually 
larger than one. Some  of these $I=0$ states are pushed down far from their 
average energy. 

We define
\begin{eqnarray}
&& {\alpha}_{I\beta \beta'}^J = \frac{n(n-1)}{2} \sum_{K \beta}  
\langle j^{n-2}(K\gamma) j^2(J)|\}j^n\beta I \rangle
\langle j^{n-2}(K\gamma) j^2(J)|\}j^n\beta' I \rangle ~,  \nonumber 
\end{eqnarray}
 and ${\alpha}_{I\beta}^J = {\alpha}_{I\beta \beta}^J $.
The coefficients $\langle j^{n-2}(K\gamma) j^2(J)|\}j^n\beta I \rangle$
are the two-body cfp's. The $\alpha_{I\beta}$ such defined
is consistent with that used in Sec. 3 and Sec. 4. 

For the case of fermions in a single-$j$ shell,
 Refs.  \cite{Yoshinaga1,Arima4} interpreted large $P(I_{\rm min})$ and
 $P(I_{\rm max})$ in terms of  the large
fluctuations of    $\overline{\alpha^J_{I}}= \sum_{\beta}
\alpha_{I\beta}^J /D_I$, and large  $\sigma_I$ 
for $I=0$ states 
\footnote{One should be aware that
this definition of $\sigma_I$ is different from $g_I$ in Table IV or 
that in Ref. \cite{Bijker1}, 
where the energy centroid is not taken into account.}
 in terms of the statistical point of view \cite{Yoshinaga1,Arima4}. 

One thus expects that the probability of a $0^+$ 
state to be the ground state is larger than that 
of the $I=I_{\rm max}$  state. This is a reasonable 
argument for the 0 g.s. dominance for a system with an even number 
of fermions, although it is at a qualitative level.

Along the line
of  Refs.  \cite{Yoshinaga1,Arima4}, 
the case of four fermions in a single-$j$ shell can be further 
elucidated by evaluating $\sigma_I$'s 
\footnote{ Here one needs to know the number of  
nonzero eigenvalues
$E_{I,i}^{J(j)}$ for states with  the angular momentum $I$  
when $G_J = -\delta_{JJ'}$. 
For $I=0$ there is always one nonzero eigenvalue
$E_{0}^{J(j)}$ corresponding 
to $G_J = -\delta_{JJ'}$ (refer to
Sec. 5.3). For the cases of $I=2$ and 4 we give 
the numbers of nonzero eigenvalues
$E_{I,i}^{J(j)}$ ($i$ is the index for the nonzero eigenvalues)
for $G_J = -\delta_{JJ'}$ without details:
\begin{eqnarray}
&I=2 &
 \left\{ \begin{array}{ll}
 1       & {\rm if ~} J=0 \\
 3    & {\rm if ~} 0< J < J_{\rm max}  \\
 2  & {\rm if ~} J=J_{\rm max} \end{array}  \right. ~, \nonumber   \\
 \nonumber \\
&I=4 &
 \left\{ \begin{array}{ll}
 1       & {\rm if ~} J=0 \\
 3    & {\rm if ~} J=2, J_{\rm max}  \\
 5    & {\rm if ~}  2< J < J_{\rm max}-2  \\
 4  & {\rm if ~} J=J_{\rm max}-2 
 \end{array} \right.  ~. \nonumber   
\end{eqnarray} }
for $I=0$, 2 and 4. These $\sigma$'s are found 
 empirically to take the  largest values. 
It was  shown in Ref. \cite{Yoshinaga1} that 
\begin{eqnarray}
&& \sigma^2_I = \langle \overline{ \left({E}_{I} \right)^2 } \rangle 
- \langle \left(\overline{E_I} \right)^2 \rangle 
\nonumber
\end{eqnarray}
with
\begin{eqnarray}
\langle \overline{ \left({E}_{I} \right)^2 } \rangle  =
  \sum_J \sum_{\beta, \beta'} \left(
\alpha_{I \beta \beta'} \right)^2 / \left(D_I \right)
\nonumber
\end{eqnarray}
and
\begin{eqnarray}
\langle \left( \overline{ E_I } \right)^2 \rangle  =
 \sum_J \left( \sum_{\beta} 
\alpha_{I \beta}^J \right)^2 / \left( D_I \right)^2. 
\nonumber 
\end{eqnarray}
The above expressions for 
$\langle \overline{ \left({E}_I \right)^2} \rangle $ and 
$ \langle \left(\overline{E_I} \right)^2 \rangle$ can be
further simplified in terms of  non-zero eigenvalues
 $E_{I,i}^{J(j)}$
($i$ is the index for the nonzero eigenvalues,
the number of $E_{I,i}^{J(j)}$ is usually much smaller than
$D_I$) for $G_{J'} = -\delta_{JJ'}$: 
\begin{eqnarray}
\langle \overline{ \left( {E}_I \right)^2} \rangle  =
  \sum_J \sum_{i} \left( 
E_{I,i}^{J(j)} \right)^2  / \left(D_I \right)  \label{y1}
\end{eqnarray}
and
\begin{eqnarray}
\langle \left( \overline{ {E}_I}  \right)^2 \rangle  =
  \sum_J \left( \sum_{i} 
E_{I,i}^{J(j)} \right)^2 /\left( D_I \right)^2 ~. \label{y2}
\end{eqnarray}
Although it is oversimplified to assume  
that all non-zero eigenvalues $E_{I,i}^{J(j)} $'s are equal,
it is very instructive   to estimate the 
$\sigma_I$'s by using  this assumption. One can then obtain
\begin{eqnarray}
&& \sigma_I^2 \sim (j+\frac{1}{2}) \sum_i \left( E_{I,i}^{J(j)} \right)^2   /D_I 
 - 36 /(j+\frac{1}{2}) ~. \nonumber 
\end{eqnarray}
By using the analytical expressions of 
the number of states with angular momentum $I$  for four fermions in a 
single-$j$ shell \cite{Zhao9}, one finally obtains 
\footnote{Let us take $I=0$ as an example. One has
 $E_{0}^{J(j)} \sim 6 D_0^{(j)}/(j+\frac{1}{2}) \sim 2 $, where
$D_0^{(j)} \sim j/3$. This gives $\sigma_0 \sim \sqrt{12}$ in the large $j$
limit.}
\begin{eqnarray}
&{\rm for} ~ I=0:& \sigma_I^2 \sim 12  - 36 /(j+\frac{1}{2}) ~ , \nonumber \\
&{\rm for} ~ I=2:& \sigma_I^2 \sim 8  - 36 /(j+\frac{1}{2}) ~ ,  \nonumber \\
&{\rm for} ~ I=4:& \sigma_I^2 \sim 7  - 36 /(j+\frac{1}{2})  ~.   \nonumber 
\end{eqnarray}
In the  large $j$ limit, these    $\sigma_I$ 
saturate at $\sim \sqrt{12}$, $\sqrt{8}$ and $\sqrt{7}$, 
respectively. For $j=\frac{31}{2}$, the   $\sigma_I$
obtained by 1000 runs of the TBRE Hamiltonian  
 is 3.52, 3.16 and 2.97 for $I=0$, 2 and 4, respectively, which are 
reasonably consistent with the above estimations, 
$\sqrt{12} \sim 3.46$, $\sqrt{8} \sim 2.82$, and $\sqrt{7} \sim 2.64$.

One therefore sees that the width $\sigma_I$ for $I=0$ 
states is larger than those of  states with  other $I$'s. Because 
${\cal P}(0)$ is   large, the $P(0)$ is expected
to be even larger  due to  the large $\sigma_0$.
This leads to  the spin   0 g. s. dominance for the cases of 
four fermions in a single-$j$ shell, and  similarly for the
case of four bosons with spin $l$. 

For   cases with very large dimensions (larger than 100) 
and large enough particle numbers,  Zuker and collaborators  presented  
a formula \cite{Zukerx} to obtain the  
lower bound for the energy of  states with angular momentum $I$: 
$\overline{E_I} - \sqrt{{\rm ln}{D_I}/{\rm ln}{2}} \sigma_I$. 
In Ref. \cite{Zuker} Vel\'azquez and Zuker made an effort 
to relate this  lower bound to  
the problem of the $0$ g.s. dominance in the presence 
of random interactions. 
One of the conclusions of Ref. \cite{Zuker} is that  
the width $\sigma_I$, rather than the  energy centroid
$\overline{E}_I$, plays a crucial role for the lower bound energy. 
However,  the  origin of  large $\sigma_0$ is 
not yet clear, except the case of four fermions
in a single-$j$ shell as discussed above.

\subsection{Relation between  
0 g.s. wavefunctions of systems with mass number
differing by two}

In Ref. \cite{Johnson2} it was reported  that 
the pairing phenomenon seems to be 
favored simply as a consequence of the two-body
nature of the interaction. The ``pairing" here means that
there is a large matrix element of the
$S$ pair annihilation operator between the ground states
of a $n$ fermion system and a $n+2$ fermion system in the
same shell. This seems to suggest that the spin 0
ground states obtained by using the TBRE Hamiltonian 
are, to a large extent, $S$-pair condensation. In order to test this 
hypothesis, Johnson {\it et al.} followed the example of generalized
seniority and considered the general pair creation 
and annihilation operators
$S^\dag = \sum_j \xi_j S^\dag_j$, and 
$S = \sum_j \xi_j S_j$
where 
$S^\dag_j= ( a^\dag_j a^\dag_j )^{(0)}$
and $S = (S^\dag)^\dag$. The coefficients $\xi_j$ is
given by  $\langle n | S_j| n+2 \rangle$ for   samplings  with
0 g.s. for both the $n$ and $n+2$ fermion systems.
(It is noted that the  $S$ pair such determined is different from that
determined in Sec. 6.2.2).  
The pair-transfer fractional collectivity from the 0 g.s.
of $n+2$ particles to that of $n$ particles is
defined as follows:
\begin{equation}
f_{\rm transfer} = \frac{ \left( \langle n, I=0 {\rm ~ g.s.}| S| n+2, I=0 ~ {\rm
g.s.}\rangle
\right)^2}{\langle n, I=0 {\rm ~ g.s.}| S^{\dagger} S| n, I=0 {\rm ~ g.s.} \rangle}. 
\end{equation}
Thus the $I=0$ ground states  are condensates of $S$ pairs
if $f_{\rm transfer} =1$.  They showed that 
 $f_{\rm transfer}$'s between 
$I=0$ ground state  of $n$ and that of $n+2$ are close to 
 $f_{\rm transfer} =1$ instead of 0, suggesting a
 correlation  of the pairing-like condensates. 

The pair-transfer fractional collectivity may  be
defined in another form:
\begin{equation}
f'_{\rm transfer} = \frac{ \left( \langle n+2, I=0 {\rm ~ g.s.} | S^{\dagger}|
n, I=0 {\rm ~ g.s.}  \rangle
\right)^2}{\langle n, I=0 {\rm ~ g.s.} | S S^{\dagger}| n, I=0 {\rm ~ g.s.} \rangle}. 
\end{equation}

In order to investigate the correlation between the 0 g.s.
of a system with $n$ fermions  and that with $n+2$
fermions in the same shell, 
the case of fermions in a single-$j$ shell
(where the seniority quantum
number $v$ is well defined), was 
checked in Ref. \cite{Zhao4}. 
Below  a few examples are discussed.

The simplest case is four and six fermions in 
the $j=11/2$ shell. The 0 g.s. probability for $n=4$ and 6
is 41.2$\%$ and  66.4$\%$, respectively.
Among 1000 sets of the  TBRE Hamiltonian, 
364 sets  give 0 g.s. both for  $n=4$ and 6 simultaneously.
Namely, 
 the TBRE Hamiltonian by which  the ground  
state has spin $I=0$ for $n=4$ has  an extremely large probability  
(around 90$\%$) to produce the $I=0$ ground states also for $n=6$. 
The $f'_{\rm transfer}$'s are 
in most cases around 0.8-0.9 and $f_{\rm transfer}$'s
are typically around 0.9-1.0. 
This means that the $S$   annihilation operator takes (approximately) 
the 0 g.s. of six fermions to that of four fermions.

Now let us take a larger value of $j$, i.e. $j=15/2$,  which 
is  good enough for our discussion. 
The 0 g.s. probability for  $n=4, 6$ and  8
is 50.2$\%$, 68.2$\%$ and 32.1$\%$, 
 respectively.  Among 1000 sets of the TBRE Hamiltonian, 
 We found  310 sets   which produce 
0 g.s. simultaneously for   $n=$4, 6 and 8. Considering this 
  31$\%$ and  $P(0) = 32.1\%$ which produce 0 g.s. 
for $n=8$, we can say that 
almost all those TBRE Hamiltonian which produces  0 g.s. 
for $n=8$, also produce 0 g.s. for $n=4$ and 6. 
It should be noted  that the difference 
of 0 g.s. probabilities among those for  $n=4$, 6, and 8   
are  large. As discussed above, 
the 0 g.s. probability of six fermions   is 68.2$\%$ while 
that of eight fermions is 32.1$\%$, which means that 
more than 50$\%$ of the 0 g.s. for $n=6$ are not related to the 
 chain in which the 0 g.s. of $n$ fermions can be obtained  
by   annihilating one $S$ pair from  that of  $(n+2)$ fermions. 

Figure 14 shows the distribution of seniority in the 
0 g.s. which are obtained by using the TBRE Hamiltonian, for 
 a few cases of four and six fermions in a single-$j$ shell. 
Low seniority components in the wavefunctions
of these 0 g.s. are not favored at all. This means that 
the contribution to the total 0 g.s.  beyond the 
 seniority chain described in Ref. \cite{Johnson2} is  
more important in the 0 g.s. of  these systems. 
                                   
Based on the above discussions, we conclude that
  a chain of angular momentum zero ground states, which 
were  suggested  in Ref. \cite{Johnson2} to be 
linked (approximately)  by the $S$ pair operator,   can be also seen  
frequently in systems with  even numbers of fermions in a single-$j$ shell  
for small $j$. However,  this chain covers  only a  
part of the 0 g.s.,  because the contribution  
beyond this chain can be more important, and this link  
becomes  weak for large $j$.

As for fermions in many-$j$ shells,  recent 
calculations   \cite{Zhaoym2} showed
that the  seniority distribution in the 0 g.s.  
is very complicated. In the $sd$ shell  
systems  low seniority states do not dominate  
in the spin zero ground states.

\subsection{Other results}

One alluring suggestion on
the origin of the 0 g.s. dominance is time reversal 
${\cal T}$ invariance. Because the time reversal  
invariance plays a key role in the formation of   $0^+$ pairs  
  in the ground states of even-even nuclei, 
one    expects that this invariance may imply a 
built-in favoring of $I=0$ ground states in the
presence of the TBRE Hamiltonian. To see whether or not this  is true, 
Bijker, Frank and Pittel \cite{Bijker1} 
analyzed a system of identical
nucleons in the $sd$ shell. They took the Gaussian unitary ensemble
 for two-body interactions,  
for which the time-reversal invariance does not hold, 
rather than the TBRE.  The outcome
of their calculations  was that  
 the 0 g.s. dominance is even 
more pronounced, which suggests that the time-reversal 
invariance is not the origin of the 0 g.s. dominance
\footnote{In Ref. \cite{Zuker} Vel\'azquez and Zuker 
claimed that the time reversal invariance suggests 
the 0 g. s. dominance, though it does not imply it. Namely, 
 the 0 g.s. dominance must be associated to 
some general cause, and the time reversal invariance is a good  
candidate, whose influence can be detected through the abundance  
of self-conjugate ${\cal T}$ pairs. For fermions in a single-$j$ shell,  
 considerably large components of
0 g.s. are given by low  seniority     pairs,  
in particular for cases with $j$ not large. This 
alluring suggestion deserves further studies. }. 
Instead, these authors showed that for the cases which they
checked,   $I=0$ states have a larger width than
other $I\neq 0$ states. 

Another view was suggested  by Drozdz and Wojcik in Ref. 
\cite{Drozdz}. They found  that 
the non-zero off-diagonal matrix elements
of all $G_J$'s for $I=0$ states have a wider distribution 
on average than those of $I \neq 0$ states, and thus
the $I=0$ states are expected to spread over a broader energy interval 
even though the number of states is usually much smaller
than for the $I \neq 0 $ cases. However, 
one can not conclude that the 0 g.s. dominance
arises just from this  phenomenon. 
As pointed out  by the  same authors \cite{Drozdz}, the 0 g.s. dominance 
results from an interplay between
the diagonal and off-diagonal matrix elements.

In Ref. \cite{Kaplan1}, Kaplan and Papenbrock studied
the structure of eigenstates for many-body
fermion systems in the presence of  the TBRE Hamiltonian. 
They found that near the edge of the spectrum, wave function  
intensities of the TBRE Hamiltonian exhibit fluctuations
which deviate significantly from the expectations of the
random matrix  theory. A
simple formula was given which relates these
fluctuations to the fluctuations for the TBRE Hamiltonian. 
The possible connection between this deviation and
the 0 g.s. dominance in fermion systems with even numbers of particles   
is unclear.

Because large energy gaps
were found in Refs. \cite{Johnson1,Johnson2},
between the 0 g.s. and excited levels, one can ask 
whether there exist  certain universal features of these gaps. 
In Ref. \cite{Santos1}, Santos, Kusnezov and Jacquod defined the
energy gap as follows: For $sp$ ($sd$) boson systems,
it is defined by the energy difference between the
first $1^-$  ($2^+$) state and 
 the $I=0$ ground state energy;  
For an even number of fermions in a single-$j$ shell, it is
defined by the energy difference between
the first $1^+$ state and the $I=0$ ground state (after a re-scaling). 
They showed that  the distribution of the gap such defined 
 is robust and may  be 
helpful in understanding the $0$ g.s. dominance.

Spin-1/2 fermions (i.e., without orbital angular momentum) 
in the presence of random interactions were studied
recently in Refs. \cite{Jacquod,Kota1,Kaplan2}. 
Jacquod and Stone \cite{Jacquod} derived a formula for fixed-$I$ (total
spin of the system) widths for the  TBRE Hamiltonian  
by using a heuristic argument,   
while Kota and Kar \cite{Kota1}
used a group theoretical approach
coupled with a so-called binary correlation approximation. 
Kaplan {\it et al.} \cite{Kaplan2} showed that one has to go beyond the
width and consider 
a so-called excess parameter which gives deviations from
Gaussian distributions.   
Adding this   correction,  one sees the smallest $I$ to be 
lowest in energy for spin-1/2 fermions.

\subsection{Summary of this Section}

In this Section  we  have reviewed the  results for 
complicated systems.
We first explained and applied  an empirical rule  to predict 
 $P(I)$'s. 
This rule was found to work   for fermions 
in a single-$j$ shell or in  many-$j$ shells, 
with even numbers of particles or odd numbers of particles.
The same rule works for  bosons.

The empirical rule also means  that 
the 0 g.s. dominance is essentially given by two-body matrix elements 
which produce $I$=0 g.s. when one of two-body matrix elements
is $-1$ and all others are zero. For fermions in a single-$j$ shell or
bosons with spin $l$, 
 the origin of the $I_{\rm max}$ g.s. probability 
is  clarified:  The large $P(I_{\rm max})$ is essentially contributed 
by an attractive $G_{J_{\rm max}}$.  
The $P(I_{\rm max})$ would be close to zero if one
deleted $G_{J_{\rm max}}$ for fermions  or $G_{L_{\rm max}}$ for bosons. 
The simple relation   $P(I_{\rm max}) \sim 1/N$ works very 
well for fermions and also for bosons with small $l$.

The disadvantage of this approach is that one
must diagonalize the Hamiltonian under the requirement that
one of two-body matrix elements is $-1$. Therefore, this
interpretation is not so transparent.

We further reviewed our results presented in Refs.  \cite{Zhaoym,Zhao8},
where an argument for the 0 g.s. dominance and the regular
staggering of $P(0)$ versus $j$ was given 
for four fermions in a single-$j$ shell
and four bosons with spin $l$, based on  this empirical rule.  
This simple argument  was found to be restricted 
to $n=4$ with single-$j$ or single-$l$ shell
and to a large $j$ ($l$) shell with  $n=6$. It is  
difficult to ``generalize" to other cases.
The disadvantage of this argument is that it does not  
provide a quantitative evaluation of $P(I)$'s. 

There were a few interesting efforts to relate \cite{Zuker} the 0 g.s.  
dominance to the time reversal invariance of the Hamiltonian. None of these  
alluring arguments provides us a good description of $I$ g.s. probability. 
 In contrast, it was  found in Ref. \cite{Bijker1} that a Hamiltonian 
which breaks the time reversal symmetry enhances the $P(0)$.

There were also arguments \cite{Yoshinaga1,Arima4} of the 0 g.s. dominance
in terms of  the large  value of 
width $\sigma_0$ defined by $\sqrt{\langle (H-\overline{E_{I=0}})^2
\rangle /D_{I=0} }$. First, it was realized
\cite{Yoshinaga1,Arima4,Zhao5} that the probability ${\cal P}(0)$ for
energy centroids with $I \sim I_{\rm min}$
and $I\sim I_{\rm max}$ to be the lowest is large. Thus 
 $P(0)$ is large. Second, it can be shown  
 schematically, at least for the case of four fermions
in a single-$j$ shell or four bosons with spin $l$, that the $I=0$ 
states have the largest width $\sigma_0$, based on which one expects 
that the $P(0)$ should be even larger. This argument is   
interesting but it is  unknown yet how to relate 
$P(I)$ to ${\cal P}(I)$ and $\sigma_I$ in an explicit way.

We also reviewed a regularity of 
the so-called pair-transfer fractional collectivity $f_{\rm transfer}$, 
which is  defined by matrix elements (after normalization) 
of the $S$ pair operator between the 0 g.s. of systems 
with  fermion numbers differing by two.  
It was found in Ref.  \cite{Johnson2}   that  the $f_{\rm transfer}$
values are  large (larger
than 0.5 in most cases) for 
even numbers of valence nucleons  in  the $sd$ shell.
In Ref. \cite{Zhao4}  it was found that 
for the case of single-$j$ shell with small $j$ the  
$f_{\rm transfer}$ values are  almost 1.0. One should be 
aware, on the other hand,  that the chain of the 0 g.s.
for systems with mass number differing by two covers
only one part of the 0 g.s., and $f_{\rm transfer}$
is not large when $j$ is large.

The results in Refs. \cite{Drozdz,Kaplan1,Santos1} 
might be also helpful towards understanding of 
the 0 g.s. dominance. Interesting results
include  an observation for a robust energy gap
between the 0 g.s. and some specific 
excited states \cite{Santos1}, an observation 
of a wider distribution of off-diagonal
matrix elements for $I=0$ states \cite{Drozdz},
discussion of structure of eigenstates obtained by random interactions
\cite{Kaplan1}. However,  the results of Refs.
 \cite{Drozdz,Kaplan1,Santos1}    are far from  applications 
to  detailed analysis of $P(I)$'s. 
For instance, these authors were unable to give any predictions 
of $P(0)$'s for the systems that they studied.
The analysis in Refs. \cite{Jacquod,Kota1,Kaplan2} are restricted 
to systems with spin-$1/2$ systems.

\newpage

\section{Average energies, collectivity and yrast state spin} 

In previous Sections we discussed the regularities
of  the ground states    in the presence of the TBRE 
Hamiltonian.  In particular we focused on 
$P(I)$'s and the approaches to predict
them.

In this Section we shall 
the behavior of average energies,  collective motion and
normal ordering of yrast state spin 
of many-body systems in the presence of the TBRE Hamiltonian. 
  We shall also discuss 
some results on one of the original questions, 
how arbitrary an interaction can be in order to reproduce
the global regularities of atomic nuclei.

\subsection{Behavior of average energies}

There were  a few studies addressing  average energies. It should be
noted that the meaning of  ``average energies"
used  by different authors can be different,
and one should be careful about their definition. 

\subsubsection{Definitions of average energies}

Let us begin with      average energies defined in Refs. 
\cite{Johnson1,Velazquez}, where the authors averaged energies of 
 the yrast states with even values of  $I$. 
Ref. \cite{Johnson1} showed  an 
indication of  $I(I+1)$ behavior for these average energies, 
which was called ``non-collective" rotation in nuclear spectroscopy.  
The authors of Ref. \cite{Velazquez}
 investigated the transition from the realistic 
two-body interaction to   purely random ones
(also refer to  Sec. 6.3), and  
showed  that the relative ordering
among the energies of the yrast states averaged over the TBRE 
 survives with exceptions of low $I$, but their relative 
separations change significantly when two-body interactions 
change from realistic to random ones.

In Refs. \cite{Mulhall1,Mulhall2} Mulhall {\it et al.} chose a subset  
 which gives $I=0$ g.s. or $I=I_{\rm max}$ g.s.  among
 1000 runs  of the TBRE Hamiltonian. Then they  
 averaged the  energies of the yrast states
 for this  subset of the ensemble, and 
 found that these average energies of the yrast states
 such defined follow 
 a parabolic function of $I$ with a strong odd-even effect.

Let us change here the definition of average energies.   
From now on,  we take  the centroids (denoted as $\overline{E_I}$)  
 of all eigenvalues of  states with spin $I$ as the average energies,
 as  defined in Sec. 5.6. One sees that 
   $\overline{E_I}$ is a linear combination of  $G_J$'s:  
\begin{equation} 
\overline{E_I} = \sum_J \overline{\alpha^J_I} G_J,  \label{average}  
\end{equation} 
where $\overline{\alpha^J_I}$ is obtained by averaging
$\alpha_{I \beta }^J$ over all $\beta$'s. 
One can apply the empirical approach of  Sec. 4.2 
to predict the  probability (denoted as ${\cal P}(I)$) for 
$\overline{E_I}$  to be   the lowest energy. 

The motivation to investigate   $\overline{E_I}$  is that
its behavior is much  simpler than that
of the $E_{I \beta}$'s, and we expect a   more transparent 
explanation of its behavior. This explanation  might be 
very helpful to understand the original problems  
such as the 0 g.s. dominance  
of many-body systems in the presence of the TBRE Hamiltonian,
as we have discussed in Sec. 5.6.

\subsubsection{The probability for $\overline{E_I}$ to be the lowest energy } 

 One should be aware that
a displacement of the TBRE  produces only a constant shift to
all $\overline{E_I}$'s. This can be easily seen from Eq. (\ref{average}). 
A displacement TBRE with $c$  defined by Eq. (\ref{tbrex}) gives 
the same $\overline{E_I}$ plus a shift $\frac{1}{2} n(n-1) c$, 
because $\sum_J \overline{\alpha_I^J} = \frac{1}{2} n (n-1)$. 
Thus all regularities of $\overline{E_I}$'s, including   probabilities 
${\cal P}(I)$'s, 
are {\it robust} regardless of the displacement, while 
$P(I)$ is very sensitive to the displacement
(refer to Sec. 3.5).

In Fig. 15 we plot  ${\cal P}(I)$'s for a few   different 
systems: four fermions in a single-$j$ shell ($j=\frac{15}{2})$,
 six fermions   in   two-$j$  ($2j_1, 2j_2$)=(11,7) shells, 
       six $sdg$ bosons, and five  fermions in 
  a single-$j$ shell ($j=\frac{9}{2})$.
  They are  typical examples among the 
many cases that we have  checked:
four, five, and six valence fermions in a single-$j$ shell 
    up to $j=\frac{31}{2}$,  
both  even and odd numbers of  fermions ($n=4$ to 9) in 
   two-$j$ shells with  ($2j_1, 2j_2)$=(7,5), (11,3), (11, 5), (11,7), (11,9),
(13,9),  $d$-boson systems with $n_d$ changing from 3 to
45, $sd$-boson systems with $n$ changing from
4 to 17, and $sdg$-boson systems with 
$n=4$, 5,  and 6.  One sees that 
the ${\cal P}(I)$'s are   large
if $I \sim I_{\rm min}$ or $I_{\rm max}$, and  are
close to zero otherwise. This feature holds 
 for all the cases that we have checked.

Now  let us  describe the ${\cal P}(I)$'s by using the  
empirical approach in Sec. 4.2. As a specific example, we  
discuss   four fermions in   $j=\frac{9}{2}$ shell.
We   predict  the ${\cal P}(I)$'s by using  integrals similar to
Eq. (\ref{exact}), without diagonalizing  
the TBRE Hamiltonian. 
The predicted ${\cal P}(I)$'s for four fermions in    $j=\frac{9}{2}$ shell 
 are listed in the column ``pred1." in Table VIII. One sees that 
  the  ${\cal P}(I)$ is large  if $\overline{E_I}$  has one or more 
$\overline{\alpha^J_I}$ which are the 
largest (or the  smallest)  for different $I$'s. 
The ${\cal P}(I)$'s predicted  by   Eq. (\ref{empirical-I}) 
  are   listed in the
column ``pred2." of  Table VIII. In Eq. (\ref{empirical-I}) 
${\cal N}'_I$ is the number of times
for  $\overline{\alpha^J_I}$ to be 
either the smallest or the largest for each $I$, and   $N_m=2N-1=9$. 
The ${\cal P}(I)$'s obtained
by using 1000 sets of the TBRE Hamiltonian  are given 
in the column ``TBRE".  It is seen that
the two  predicted ${\cal P}(I)$'s are reasonably consistent with
those obtained by using the TBRE Hamiltonian.  

The origin of the fact   that ${\cal P}(I)$'s
are large only if $I\sim I_{\rm min}$ or $I\sim I_{\rm max}$
was  argued  in Refs. \cite{Mulhall1,Zhao5} based on 
the assumption   that  the two-particle cfp's  are  randomly  distributed,
and in Ref.   \cite{Kota1} based on group symmetries of the TBRE.
We shall discuss the formula of $\overline{E_I}$
derived  in Refs. \cite{Mulhall1,Kota1} in Sec. 6.1.4.

\subsubsection{The $I(I+1)$ behavior of $\overline{E_I}$}

Examining  the ordering  of the  average 
energies $\overline{E_I}$'s with respect to $I$, one sees  that, when 
 the spin $I$ of the lowest $\overline{E_I}$ 
  is $ \sim I_{\rm min}$ ($I_{\rm max}$) in {\it one} set 
 of the TBRE Hamiltonian,    $\overline{E_I}$ given by
 the same set of interactions increases  (decreases)
 with  $I$ in most cases, and is proportional  to $I(I+1)$ on average. 
 One should be aware that
 both even  and odd values of $I$ are included
here. Although the authors of Refs. \cite{Johnson1,Velazquez}
discussed the $I(I+1)$ behavior of average energies, their
definition of averaging differs from our  $\overline{E_I}$, and 
they were interested only in   even $I$ values (see Sec. 6.1.1 for
their definition). 
Also, one should not confuse the $\overline{E_I}$'s 
 with Bethe's expression for the level densities \cite{Bohr} which is 
 is based on the Fermi gas approach \footnote{ 
The $I(I+1)$ behavior of $\overline{E_I}$ in  Bethe's expression for the  
level densities   and   that of  $\langle \overline{E_I} \rangle_{\rm min}$ 
(or $\langle \overline{E_I} \rangle_{\rm max}$) discussed in
this paper  are   completely  different. 
For example,   ${\cal J}$ in Bethe's expression changes
with particle number,  but in the present context it
is not sensitive to the particle number $n$ but 
to the orbits of the shell, as shown in Ref. \cite{Zhao5}. 
Furthermore,   systems which show  the  $I(I+1)$ behavior of 
 $\langle \overline{E_I} \rangle_{\rm min}$ 
 can be very simple, and
those described by Bethe's expression require complexity
in energy levels so that one
needs a statistical approach. }.

Let $ \langle \overline{E_I}\rangle_{\rm min}$ 
($\langle \overline{E_I}\rangle_{\rm max}$) be a quantity obtained by  
averaging the   energies $\overline{E_I}$  over the cases of 
 $E_{I\sim I_{\rm min}}$ ($E_{I\sim I_{\rm max}}$) being the lowest.  
 We find that both
$\langle \overline{E_I}\rangle_{\rm min}$
and $\langle \overline{E_I}\rangle_{\rm max}$  
are proportional to    $I(I+1)$,   similar to
 ``rotational" spectra. 

In Fig.  16 we  show  $\langle \overline{E_I}\rangle_{\rm min}$ vs. $I(I+1)$ for
twenty  $d$ bosons,  ten  $sd$ bosons,   
  four fermions in a single-$j$ ($j$=17/2) shell, 
 and a  system with four fermions in   two-$j$ shells.  
For the sake of simplicity,   we introduce the ``moment of
inertia" ${\cal J}$, defined by the optimal  coefficient as 
 $\langle \overline{E_I}\rangle_{\rm min}$  $= \frac{1}{2 \cal J} I (I+1)$.
Below we do not discuss $\langle \overline{E_I}\rangle_{\rm max}$ but note 
that $\langle \overline{E_I}\rangle_{\rm max} \simeq 
\frac{1}{2 \cal J} \left[ I_{\rm max} (I_{\rm max}+1) - I (I+1) \right]$,
where the ${\cal J}$ for 
$\langle \overline{E_I}\rangle_{\rm max}$ and
that for $\langle \overline{E_I}\rangle_{\rm min}$ are approximately
 equal for all cases that we have checked.

 An empirical relation between  ${\cal J}$ and $j$ is
summarized in Fig. 17. For fermions in a single-$j$ shell, 
  $d$ boson systems and $sd$ boson  systems, ${\cal J}$
  is fitted by a trajectory of the form $\sqrt{{\cal J}} \simeq 1.42 j$ 
  (we take $j=2$ for $d$ boson and $sd$ boson
systems); for $sdg$ bosons and fermions in  two-$j$ shells, there seems  
a slight shift from the trajectory  $\sqrt{{\cal J}} \simeq1.42 j$,
where  $j^2 \equiv j_1^2 + j_2^2$.

Another feature of ${\cal P}(I)$'s is 
   that the ${\cal P}(I_{\rm max})$'s are always 
quite ``stable"   (about 28-35$\%$), while 
 the ${\cal P}(I_{\rm max}-2)$'s of fermions  in a 
  single-$j$ shell   and boson systems, and the 
  ${\cal P}(I_{\rm max}-1)$'s of fermions in  many-$j$ shells, 
    are drastically smaller 
   than the ${\cal P}(I_{\rm max})$'s, though  still
   sizable     ($\sim 5-15 \%$). Moreover,  
  there may be 2 or 3 sizable
${\cal P}(I)$'s for the cases of $I\sim I_{\rm min}$, and  
the ${\cal P}(I_{\rm min})$'s are not {\it always}   larger than the other
${\cal P}(I)$'s  (with $I \sim I_{\rm min}$).

In Ref. \cite{Zhao5}, an argument of 
 the above asymmetry for  ${\cal P}(I)$ was given in terms of 
 the fluctuations   of $\overline{E_I}$:  
The $\overline{E_I}$'s are   proportional to 
$I(I+1)$  with  some deviations in each set  of the TBRE Hamiltonian. 
Because   $ \left( \overline{E}_{I+1} - \overline{E_I} \right)$
is small if $I$ is small  and large if $I$ is large,
the possibilities to change the order of the 
 $\overline{E_I}$'s  due to the deviations  when   $I \sim I_{\rm min}$ 
are  much larger  than 
when $I \sim I_{\rm max}$. This explains why there are 2 or 3 sizable
and comparable ${\cal P}(I)$'s with $I\sim I_{\rm min}$ but
only one large ${\cal P}(I)$  with $I= I_{\rm max}$, 
${\cal P}(I_{\rm max}-1) \ll {\cal P}(I_{\rm max})$,
and ${\cal P}(I_{\rm max}-2) \ll {\cal P}(I_{\rm max})$.

The relation  
  $\langle \overline{E_I} \rangle_{\rm min} \sim I(I+1)$   was also discussed 
 in Ref.  \cite{Zhao5} by  
assuming  that the two-body cfp's are randomly distributed. The analysis
 in Ref.  \cite{Zhao5}  showed that such an assumption can explain the  
 $I(I+1)$ behavior of $\langle \overline{E_I} \rangle_{\rm min}$.

The   ${\cal J}$ for fermions in a single-$j$ shell 
can be evaluated in the following way.
Let us  assume that  
$\langle \overline{E_{I=I_{\rm max}}} \rangle_{\rm min}
\sim \alpha^{J_{\rm max}}_{I_{\rm max}}
\langle G_{J_{\rm max}} \rangle_{\rm min}$ 
by neglecting contributions to 
 $\overline{E_{I_{\rm max}}}  \equiv E_{I_{\rm max}} $ from other $G_J$'s. 
If one uses   $\langle G_{J_{\rm max}} \rangle=0.7$, as  
found empirically in  Refs. \cite{Mulhall2,Zhao5}   
 for four fermions in all  single-$j$ shells,  one  obtains  
\begin{equation}
\alpha^{J_{\rm max}}_{I_{\rm max}} \langle G_{J_{\rm max}} \rangle_{\rm min} \sim
0.7 \alpha^{J_{\rm max}}_{I_{\rm max}} \sim
\frac{1}{2 \cal J} I_{\rm max} (I_{\rm max}+1) = 
\frac{1}{2 \cal J} (4j-6)(4j-5) \sim \frac{8}{ \cal J}   j^2. \label{estimate}
\end{equation}
It was shown in  Ref. \cite{Zhao4}  that the   
 $\alpha^{J_{\rm max}}_{I_{\rm max}}$
 of $n=4$ saturates quickly at $\frac{29}{8}$ when $j$  increases. 
Then the left hand side $\sim \frac{29}{8} \times 0.7$ $\sim  2.54$. 
One finds that $\sqrt{\cal J}  \sim 1.77 j$, which is slightly larger than
 the $\sqrt{\cal J}$  obtained in Fig. 17
(where $\sqrt{{\cal J}} \simeq 1.42 j$).   
 The   ${\cal J}$
in Eq.~(\ref{estimate}) is over-estimated, 
 because    the contributions from  
$G_{J_{\rm max}-2}$  and $G_{J_{\rm max}-4}$ were neglected. 

\subsubsection{The formulas by Mulhall {\it et al.} and Kota {\it et al.}}

For fermions in a single-$j$ shell,
 Mulhall, Volya and   Zelevinsky  derived a 
formula in  Refs. \cite{Mulhall1,Mulhall2} 
for the  expectation value of energy of spin  $I$ states 
by minimizing the ground state energy and   assuming the
statistical point of view   
for the angular momentum couplings 
in an $n$-body system  (``geometric chaoticity" called by these authors). 
The energy such obtained is actually the centroid of eigenvalues 
of spin $I$ states
\footnote{The energy such obtained
was interpreted in Refs. \cite{Mulhall1,Mulhall2} to be
the lowest spin $I$ energy. However, it  was later pointed out
in Refs. \cite{Zhao4,Kota1,Papen_new,Zhao5} 
that it should be  $\overline{E_I}$, the energy centroid of
the $E_{I\beta}$'s. }, 
namely,  the average energy $\overline{E_I}$ 
defined by Eq. (\ref{average}). 
Their final result was written as follows, 
\begin{eqnarray}
&& \overline{E_I} = \sum_J (2J+1) G_J \left( \frac{n}{2j+1} \right)^2 \nonumber \\ 
& +& I (I+1) \sum_J (2J+1) \frac{3 (J^2 - 2 j(j+1)) }{2 j^2 (j+1)^2 (2j+1)^2}  G_J
+ O (I^2(I+1)^2), \label{chaotic}
\end{eqnarray}
where $O (I^2 (I+1)^2) $ refers to higher $I$ terms which seem to be 
negligible. The first term of this formula is a 
constant which is independent of $I$. 
 The  second term is proportional to 
$I(I+1)$.  However, this term does not  explain  
$\sqrt{{\cal J}} \simeq 1.42 j$ (refer to Fig. 17). 
Let us take   $J \sim 2j$ and $G_J \sim 0.7$ for 
the coefficient of this term,
then one sees that the second term $\sim \frac{2.1}{j^3} I(I+1)$.
This is different from our result:
$\overline{E_I} \simeq \overline{E_{I_{\rm min}}} + \frac{I(I+1)}{4j^2}$
(we used ${\cal J} \simeq 2j^2$ here).

In Ref.   \cite{Kota1}, Kota and Kar obtained Eq. (\ref{chaotic}) 
for  $\overline{E_I}$ by resorting to the group structure of
 $U(2j+1) \supset O(3)$   for  $n$ fermions in a single-$j$ shell. They also 
pointed out that  the use of the  
cranking approximation and the Fermi-Dirac occupancies of particles  in 
Refs. \cite{Mulhall1,Mulhall2} are equivalent to
the approach by using group symmetries of the TBRE as in Ref. \cite{Kota1}.

Although the approach of Refs. \cite{Mulhall1,Mulhall2} does not
produce the same results as those in Refs. \cite{Yoshinaga1,Arima4,Zhao5}, 
 the concept of geometric chaoticity introduced  
in Refs. \cite{Mulhall1,Mulhall2} is essentially  
the same as the randomness of two-body coefficients
of fractional parentage introduced
in Refs. \cite{Yoshinaga1,Arima4,Zhao5}. Both of  these two concepts 
are based on  the complexity of angular momentum couplings of 
$n$ particles.

\subsubsection{A short summary}

As a short summary of this subsection, let us first repeat 
two   robust regularities
of many-body systems interacting  by  the  TBRE Hamiltonian:

1. The   ${\cal P}(I)$'s   are
large if and only if  $I \sim I_{\rm min}$ or $I_{\rm max}$. 

2. The $I$ dependence of the 
$\langle \overline{E_I}\rangle_{\rm min}$ and $\langle \overline{E_I}\rangle_{\rm max}$
is roughly  $I(I+1)$. 

These regularities have been argued in terms of  
the statistical distribution  (geometric chaoticity)
of the two-particle cfp's.

We  emphasize here that the ${\cal P}(I)$'s, 
  which are related to the randomness of two-particle cfp's 
 or  another appellation of ``geometric 
chaoticity" for the angular momentum couplings, 
and the $P(I)$'s discussed in the last Section, 
are  different quantities.
For examples, $P(j)$ is always sizable for an odd number of particles,  
but ${\cal P}(j)$ is close to zero in this case;
$P(I_{\rm max})$ is not large $(\sim 1/(j+\frac{1}{2})$)  
 for fermions in a large single-$j$ shell,    
 but ${\cal P}(I_{\rm max})$  $\sim 38\%$.   
In particular,
${\cal P}(I)$'s are independent of the displacement $c$ in the
displaced TBRE, but $P(I)$'s are sensitive to $c$. 
We should not mix up these two probabilities.

This  shows  that  
    $I$ g.s. probabilities (and 0 g.s. dominance) 
cannot be explained   by the geometric  chaoticity. The  
role of variances or the width of states with  spin $I$ is 
important and complicated.  
This was  stressed in Refs. \cite{Yoshinaga1,Arima4,Zhao5,Kota1}.

\subsection{Collective motion in the presence of random interactions}

The structure of levels of many-body systems in the presence
of the TBRE Hamiltonian  is another interesting topic.
In Refs. \cite{Bijker2,Bijker3} it was shown that  both vibrational and
rotational features arise in the   IBM 
\cite{Iachello1} in the presence of the TBRE Hamiltonian. In
contrast, as pointed out by many authors,  rotational behavior does not  
generically arise in fermion systems when their interactions are random. 
It was suggested, therefore, that some constraints should be imposed on 
 the random Hamiltonian   to obtain a generic rotational
behavior in fermion systems \cite{Johnson3}.

\subsubsection{Vibration and rotation in the vibron model and the IBM}

The occurrence of both vibrational and  rotational
band structure within the frameworks of
the $sd$ IBM and the vibron model was elegantly
demonstrated  in Refs. \cite{Bijker2,Bijker3}.

The following ratio 
\begin{equation}
   R=(E_{4_1^+}-E_{0_1^+}) / (E_{2_1^+} - E_{0_1^+}) ~   \label{R-value}
\end{equation}
has been used as  collective indicator, and 
 extensively investigated  for medium-heavy
  even-even nuclei \cite{Casten-review}.
 $R$ has the characteristic value of 2 for vibrational systems, 
  and $10/3$ for rotational systems.

For  the vibron model \cite{vibron}, the collective
indicator $R$ is given by
\begin{equation}
   R=(E_{2_1^+}-E_{0_1^+}) / (E_{1_1^+} - E_{0_1^+}) ~, 
\end{equation}
which is equal to 2 in the vibrational limit and 3 in the rotational limit. 
Because the results for $sp$ bosons and those for $sd$ bosons are  
similar,  we give below only the results for $sd$ bosons.

The $R$ values were calculated by using
the shell model by Johnson \cite{Johnson3}. 
He found that the $R \sim 1.5$ with a broad distribution 
for $^{22}$O ranging from 0 to $\sim 3$
by diagonalizing the TBRE Hamiltonian. This means that
the rotational band structure does not appear
for fermions interacting by two-body random forces.

In Ref. \cite{Bijker2}  Bijker and Frank calculated 
low-lying states  within the $sd$ IBM by using the TBRE Hamiltonian.  
They obtained that  about  60$\%$ among 1000 sets of the TBRE produce 
  $I=0$ ground states.  The same sets, which produce  
$0^+$ ground states, were used to calculate 
$E_{2_1^+}$ and $E_{4_1^+}$. These $E_{2_1^+}$ and $E_{4_1^+}$ as well
as $E_{0_1^+}$ determine  $R$ values in Eq. (\ref{R-value}). 
The $R$'s such obtained are larger than 1 for most of them. 

Figure 18  plots the  distribution of $R$ with  
boson number $n$=3, 6, 10, and 16 (taken from Ref. \cite{Bijker2}).  
One sees that two sharp peaks emerge at $R\sim 2.0$ and 3.3 when
 the boson number $n$ increases from 3 to 10.

The E2 transition rates also take  characteristic values 
for vibrational and rotational modes. For example,
$\frac{ {\rm B(E2,} 4_1^+ \rightarrow 2_1^+) }{{\rm B(E2,} 2_1^+ 
\rightarrow 0_1^+) }$ take the value 
of  $ \frac{2 n}{n-1}$ in the vibrational
limit and the value of
$ \frac{10 (n-1) (n+5)}{7n(n+4)}$ in the rotational limit 
of the IBM \cite{Arima1,Arima2,Arima3,Iachello1}. 
When boson number $n$ is large,
these characteristic  values in the IBM  
are equal to those in the Bohr-Mottelson model: 
2 for vibration and $|\frac{ (4 0, 20 | 20)}{ (20, 20| 00)} |^2 =
\frac{10}{7}$ for rotation.

A correlation between $R$ and  
$\frac{ {\rm B(E2,} 4_1^+ \rightarrow 2_1^+) }{{\rm B(E2,} 2_1^+ 
\rightarrow 0_1^+) }$  for $n=16$ is shown in Fig. 19 (taken from
Ref. \cite{Bijker2}), where 
one-body parameter $e_d$ in Eq. (\ref{sd-boson1}) was also included as
a random parameter together with the TBRE.

The remarkable rotational peak obtained in  Refs. \cite{Bijker2,Bijker3}   
suggested that the key for obtaining a rotational peak from 
a random shell model Hamiltonian might be to restrict the space to a collective
subspace built  from the lowest $S$ and $D$ pairs, because these
$SD$ pairs are the objects which are approximated
by   $s$ and $d$ bosons  in the IBM \cite{Iachello1}. 
To see  whether this is  true,
calculations in the truncated $SD$-pair space were carried out
in Ref. \cite{Zhao3} by using 
 a general TBRE Hamiltonian defined by  Eq. (\ref{V}).

The calculations of Ref. \cite{Zhao3} 
 were performed  in the following procedures:
One first selects only  random interactions  which 
  produce    $0^+$ ground states and
  $2^+$ first excited states in 
  two-particle systems. Then it is assumed that the
 collective $S$ and $D$ pairs are given by these 
$0^+$ and $2^+$ wavefunctions of the two-body systems.
One proceeds to  calculate the 
spectra for six-particle systems by using 
the same random two-body interactions.

Figure 20  shows the distribution of $R$ values  in  
the $SD$-pair truncated shell model space for  six identical  
nucleons in the $sd$, $pf$ and $sdg$ fermion shells, respectively.  One sees 
that the distribution of the $R$ values in the $sd$ shell within an 
$SD$-pair subspace is similar to that obtained in the full shell model 
space \cite{Johnson3} -- a broad distribution extending to $R \sim 
1.3$. When one goes to larger shells, the distributions become  
sharper, and shift to the right from the $sd$ shell ($R \sim 1.3$)
to the $sdg$ shell ($R\sim 1.91$). Nevertheless, no sharp peak at 
$R\sim 3.33$ appears. 
 From this it is  concluded that 
rotational motion does not seem to emerge from $SD$ truncated
shell model calculations.  Statistically, the full 
shell model space and the $SD$ truncated truncated shell model space
defined here give essentially the same results for a general
two-body interaction.

\subsubsection{Rotational spectra in the $SD$-pair subspace}

Because rotational motion does not arise from the $SD$-pair truncated
shell model space with the TBRE, let us investigate whether 
it might appear when we use a more restrictive Hamiltonian  defined by Eq. 
(\ref{sep1}) and take
the  $SD$-pair truncation as  well. 
Here we fix  $H_1$ (single-particle term) to be  zero, and  keep only  
monopole pairing, quadrupole pairing and quadrupole-quadrupole
interaction, with their strength parameters 
(i.e., $G_0, G_2 $ and $\kappa$)  generated randomly. Each  of them 
follows   a  Gaussian distribution with an average being zero and
a  width being  one.

The calculations   were carried out in Ref. \cite{Zhao3}
in the same procedure as the $SD$-pair truncated shell model calculations 
by using the  Hamiltonian of Eq. (\ref{V}), except that
the $G_J(j_1 j_2 j_3 j_4)$'s are replaced by the strength parameters
$G_0, G_2 $ and $\kappa$.  Fig.  21  shows the distribution  of   $R$
 thus  calculated for six identical nucleons in the $sd$ shell.
When all three   strengths are  
treated on the same footing, one arrives at the distribution of $R$ 
  shown in Fig. 21(a). In this case, no sharp rotational peak is 
observed. Instead, a peak appears around $R\sim 1.3$, with a long 
tail extending to $R\sim 3.1$.  If   the $\sum_M Q_M  Q_M$ strength
parameter $\kappa$ is  artificially enhanced by a factor $\epsilon$
($>1$), one arrives at the results shown in Figs. 21b-d. As $\epsilon$ is 
increased, i.e., as the quadrupole-quadrupole strength is enhanced, 
a peak at $R\sim 3.1$ gradually appears. On the other hand,
the probability of $R>3.1$ remains very small.

As the size of the shell is progressively increased, the peak at 
$R\sim 1.3 $ gradually disappears and another peak at  
$R\sim 3.3$ emerges. This is illustrated in Fig. 22 
for six identical nucleons in the $pf$, $sdg$, $pfh$, and $sdgi$
shells even with $\epsilon=1.0$. For a large shell, the peak
at $R \sim 3.3$ becomes very well pronounced.

As  fingerprints of occurrence of  rotational motion,
ratios of the E2 transition rates of six nucleons in the $sdgi$ shell
(cf. Fig. 22 (d) )  are examined.  
According to the Elliott Model \cite{Elliott}, the  ratio 
$\frac{ {\rm B(E2,} 4_1^+ \rightarrow 2_1^+) }{{\rm B(E2,} 2_1^+ 
\rightarrow 0_1^+) }$ is 1.35 for the $sdgi$ shell
(this ratio is 1.16 in the IBM for three $sd$ bosons), and  
$\frac{ {\rm B(E2,} 2_2^+ \rightarrow 2_1^+) }{{\rm B(E2,} 2_1^+ 
\rightarrow 0_1^+) }= $
$\frac{ {\rm B(E2,} 0_2^+ \rightarrow 2_1^+) }{{\rm B(E2,} 2_1^+ 
\rightarrow 0_1^+) } = $
$\frac{ {\rm B(E2,} 2_2^+ \rightarrow 0_1^+) }{{\rm B(E2,} 2_1^+ 
\rightarrow 0_1^+) }=0$.

It is seen in Fig. 23 that there is a strong correlation between
  $R$ and the ratios of these
E2 transition rates. One   sees   
a concentration of   points near the coordinates 
$(R,  \frac{ {\rm B(E2,} 2_2^+ \rightarrow 2_1^+) }{{\rm B(E2,} 2_1^+ 
\rightarrow 0_1^+) } )$=$(3.3, 0)$, 
$(R,  \frac{ {\rm B(E2,} 0_2^+ \rightarrow 2_1^+) }{{\rm B(E2,} 2_1^+ 
\rightarrow 0_1^+) } )$=$(3.3, 0)$,   
$(R,  \frac{ {\rm B(E2,} 2_2^+ \rightarrow 0_1^+) }{{\rm B(E2,} 2_1^+ 
\rightarrow 0_1^+) } )$=$(3.3, 0)$, and 
$(R,  \frac{ {\rm B(E2,} 4_1^+ \rightarrow 2_1^+) }{{\rm B(E2,} 2_1^+ 
\rightarrow 0_1^+) } )$=$(3.3, 1.3)$, respectively.

Based on the above results  it was concluded
in Ref. \cite{Zhao3} that the rotational 
motion is related closely to the form of  two-body interactions. 
In particular, for systems of identical nucleons there must be a strong
quadrupole-quadrupole component in  interactions for the 
rotational motion  to occur. 

\subsubsection{Rotations based on  displaced random interactions }

The philosophy that   the occurrence of
 rotational band structure  requires the  
Hamiltonian to have some specific features was also discussed
by Vel\'azquez and Zuker in Refs. \cite{Zuker,Velazquez}.
Since the B(E2) enhancement   
is not produced by the TBRE Hamiltonian, they 
took a displaced TBRE which is centered 
at a negative constant  $c$, namely, the displaced  
TBRE of  Vel\'azquez and Zuker is attractive on the average. 
This idea can be traced  back  \cite{VZ} to the fact that the 
realistic interaction in the $pf$ shell, such as the KB3 \cite{KB3}, 
are mostly attractive.
In Ref. \cite{Zuker}, Vel\'azquez and Zuker
found a gradual  buildup of B(E2) values as well as a gradual increase of 
$P(0)$  in the ($f_{7/2} p_{3/2}$) space as $ c $ decreases from 0 to $-3$.
 $R=3.3$ was found  to be also clearly  favored. 
  
One criticism   \cite{Johnson-comment}
was that the magnitude of displacement used 
in Ref. \cite{Zuker} is too large. The 
width of the KB3 matrix elements is close to their   
average value  in magnitude.  However,  
the value of $|c|$ used in Ref. \cite{Zuker} 
is about five times larger than 
  the average value of the KB3 matrix elements. 
When the magnitude of $c$ is so large, the 
B(E2) values are dominated by the large value of $c$,
according to Ref. \cite{Johnson-comment}.

Vel\'azquez and Zuker  suggested in Ref.  \cite{Zuker}
another  possibility to obtain rotational
band structure for fermions. 
Their suggestion is based on the observation that
the nucleus $^{20}$Ne (with
four valence neutrons) has a rotational spectrum
while the nucleus $^{36}$Ar (with four neutron holes 
 and two proton holes) has a vibrational spectrum
 although the same realistic  two-body matrix elements are used for these
two nuclei in  the $sd$ shell.  
This difference comes possibly from  the changes of the mean field.  
They suggested  that one would obtain both 
  vibration and rotation in one nucleus by randomizing 
the single-particle energies with the two-body matrix elements fixed.

\subsubsection{A short summary}

To summarize this subsection,  we first reviewed the discovery of  
generic vibration and rotation within the $sd$ IBM. Similar results
can be seen for the $sp$ bosons.
It was found   that in  a truncated $SD$-pair subspace vibrations   
arise   for the general TBRE Hamiltonian but rotations do 
not. With a restricted Hamiltonian defined by Eq. (\ref{sep1})    
collective rotations appear.  
Not surprisingly, the quadrupole-quadrupole interaction seems to play
a key role in obtaining a peak at $R\sim 3.33$. 

According to  Refs.  \cite{Zuker,Velazquez}, 
a negatively displaced TBRE is also able to produce 
a  rotational band structure as well as an enhanced B(E2) transition rate.
One criticism to this suggestion was that
the displacement used in Refs.  \cite{Zuker,Velazquez} is
too  large.

It is interesting to discuss why the IBM with
the TBRE Hamiltonian is able to give rise to rotations, 
while the shell model truncated to $SD$ pairs cannot.
As generally believed,  the $sd$ IBM 
 is   a consequence of quadrupole and pairing correlations contained
 in the realistic interaction. Then random interactions among $s$ and
 $d$ bosons  already absorb some parts of quadrupole correlations among
 the realistic interaction.   There is no
inconsistency, therefore, between the results of Refs. 
\cite{Bijker2,Bijker3} by using the TBRE within 
the IBM, those of Ref. \cite{Johnson3} by using the TBRE within the
shell model,  and those of Ref. \cite{Zhao3} by using
both the TBRE and the restricted Hamiltonian defined
in Eq. (\ref{sep1})  within the $SD$ nucleon pair
approximation \cite{Chen,unified}.

\subsection{Normal ordering of spin in the yrast band}

The first study of   normal ordering of spin $I$ in the presence of the
TBRE Hamiltonian was  done  by Cortes, Haq, and Zuker \cite{Cortes} 
more than twenty years ago.  These authors took the case of 
 $^{20}$Ne nucleus and used
 a Hamiltonian consisting of the Elliott SU(3) component \cite{Elliott}  and 
the TBRE component. Namely, the Hamiltonian of Ref. \cite{Cortes} was defined by
\begin{equation}
H = (1-b) H_{\rm Elliott} + b H_{\rm TBRE}, \nonumber 
\end{equation}
where $H_{\rm Elliott}$ is the Elliott $Q$-$Q$ force and
$H_{\rm TBRE}$ refers to the TBRE Hamiltonian.
The parameter $b$ runs from 0 to 1
in order to study the effect on the normal ordering
in the spectrum from the noise, the $H_{\rm TBRE}$ part.

It should be noted interestingly that the calculations of
Ref.  \cite{Cortes} 
almost observed the 0 g.s. dominance but the large 
variances therein prevented the authors from drawing this conclusion. 
The Fig. 2 (8) in Ref. \cite{Cortes} already suggested 
that the yrast spin zero state is lower than those 
for spin two, four, etc. on average, even when $b=1$ 
(the pure TBRE Hamiltonian).

In Ref. \cite{Velazquez} Vel\'azquez and collaborators applied
the similar idea of Ref.  \cite{Cortes} 
to  $^{24}$Mg,  $^{44}$Ti and  $^{48}$Cr. 
The motivation was to investigate the transition from  realistic 
two-body interaction  to   purely random ones. The realistic 
interaction  which they used is the Kuo-Brown (KB3) interaction \cite{KB3} 
for the $fp$ shell and the Wildenthal interaction \cite{Wildenthal} for the 
$sd$ shell. Random interactions were   taken as the TBRE. 
It was found in Ref. \cite{Velazquez} that 
the average energies of yrast
 states with different angular momenta $I$ keep the  
ordering of the band when the Hamiltonian changes from
the realistic interaction to the TBRE interactions when $b$ 
is small, and that the probability that the yrast states keep the 
ordering is quite large ($\sim 30\%$) even for  purely random interactions,
suggesting a strong correlation  between these states.

Another relevant result was given in Ref. \cite{Johnson1}, 
where it was  shown that there is an 
indication of a so-called ``non-collective"
rotation in the spectrum of the nucleus $^{46}$Ca, i.e.,  
 $I(I+1)$ behavior of  the yrast energies  obtained by averaging over 
  the 0 g.s. subset of the TBRE, where $I$ is even.

\subsection{Constraints on random interactions in nuclei}

As discussed earlier, within both  the shell model and the IBM,
 the TBRE Hamiltonian reproduces some features of nuclear
 properties. However, the realistic interactions between nucleons are
of course not random.  This leads to
essential differences between   calculated results by using
the realistic interaction and those by using the TBRE Hamiltonian.
For example,  the observed $P(0)$ of even-even  
nuclei is 100$\%$ without exceptions,  
while for the TBRE Hamiltonian it is typically
$30-70\%$. Therefore, although some characteristic 
properties survive as the interaction changes   
from realistic to random, it would be interesting to study,
 as was asked in Ref. \cite{Johnson3}, how arbitrary  a set of
interactions can be in order to reproduce realistic nuclear properties.
In this subsection we shall review results in this context.

The IBM is a very proper tool to investigate this
context due to its simplicity.
Although the IBM with the TBRE can explain some typical
features of quadrupole type collectivity, there are certain quantities
which cannot be simply explained by this approach. 
In Ref. \cite{Kusnezov1},  Kusnezov {\it et al.}   
investigated this question   within the $sd$ IBM.  
 One of the features discerned  
in Ref.  \cite{Kusnezov1} is that for all nuclei throughout the nuclear
chart with $6\sim18$ valence nucleons outside the
doubly closed shell  
  the   experimental values of $R$ show a peak at 
2.3 while the IBM with random interactions tends to give a
peak at $\sim 2$. This indicates that we need to restrict
our random  interactions. Instead of using
the general Hamiltonian for  the $sd$ bosons
in Eq. (\ref{sd-boson1}), these authors used a schematic $sd$ boson 
Hamiltonian, $H = e_d n_d - \kappa
\sum_M {\cal Q}_M  {\cal Q}_M$,
which is called  the extended consistent ${\cal Q}$
formalism proposed in Ref.  \cite{Q-formalism}.  
Here  ${\cal Q}_M= (-)^M s^{\dag} d_M + d^{\dag}_M s 
+ \chi \sum_{m_1 m_2} (-)^{m_2}
(2 m_1 2 m_2|2M) d^{\dag}_{m_1} d_{-m_2}$. 
They obtained statistical ranges of 
$e_d / \kappa$ and $\chi$ values which 
give $R$ in the range of 2.2-2.4 for seven $sd$ bosons.

A similar study was performed by Zhang and collaborators  
in Ref.  \cite{Zhang1}  within  the geometric collective
model (GCM) \cite{Greiner}.   
Their calculations demonstrated that the experimental values 
of $R$ constrain  some ratios between the parameters in  
the GCM potential.

This problem  was also  studied  preliminarily in the shell model. 
In Ref. \cite{Horoi1}, Horoi {\it et al.} 
studied a system of four protons and four neutrons
in the $sd$ shell  (corresponding
to the $^{24}$Mg nucleus)  by using  
random two-body interactions which  distribute uniformly between $-1$ and 1. 
They found that the overlap  between the $I=T=0$ ground 
state wavefunctions obtained by random interactions  
 and those obtained by  the 
realistic effective interaction is about 
0.02  on the average. 
The B(E2, $2_1^+ \rightarrow 0_1^+$) values 
obtained by using  random interactions are typically 
one order of magnitude smaller than those obtained by the realistic 
interaction, indicating  that collectivity
of those states calculated by using random
two-body random interactions is  not as strong as that obtained  by using 
the realistic two-body interaction. 

More  extensive comparison between the collectivity arising from
the TBRE shell model Hamiltonian  and experimental data  is necessary
but  difficult  because of the huge dimensionality of the shell model.
Due to this difficulty, little has been known about the constraints
on random interactions of the shell model  Hamiltonian so far.

\subsection{Summary of this Section}

In this Section we first discussed
 the behavior of average energies, and  the occurrence of vibrational and   
rotational band structure, for many-body systems in the presence of
random two-body interactions.
Then we discussed
normal ordering of yrast spins
calculated by using the TBRE Hamiltonian.  
We also reviewed the results of constraints on
the random Hamiltonian in order to reproduce the global properties (such
as the distribution of $R$)  of realistic nuclei.

Energy centroids $\overline{E_I}$   were discussed in  many papers 
\cite{Mulhall1,Mulhall2,Zuker,Kota1,Yoshinaga1,Arima4}. 
 It was found that the probability for $\overline{E_I}$   
   to be the lowest is large if $I\sim I_{\rm min}$ or $\sim I_{\rm max}$. 
We thus divide the TBRE into two subsets, one of which gives
  $\overline{E_{I \sim I_{\rm min}}}$  the lowest energy,
  and the other of which gives  $\overline{E_{I  \sim I_{\rm max}}}$
 the  lowest energy.
The  $\langle \overline{E_I} \rangle_{\rm min}$  
 ($\langle \overline{E_I} \rangle_{\rm min}$), obtained by averaging
 the $ \overline{E_I}$ over the $\overline{E_{I \sim I_{\rm min}}}$
 ($\overline{E_{I \sim I_{\rm max}}}$) subset, 
the $I(I+1)$ behavior.
 These features can be explained in terms of   the quasi-randomness of  
two-body coefficients of fractional parentage.

The occurrence of vibrational and rotational
structure for the $sp$- and $sd$-boson systems with the TBRE  
was discovered and discussed in Refs. \cite{Bijker2,Bijker3}. 
However, the rotational  motion does not arise  in  
fermion systems  if one takes the general TBRE Hamiltonian. 
Additional requirements are thus necessary to obtain 
a rotational band structure for fermion systems. Two kinds of random
Hamiltonians  have been discussed so far to obtain
rotational band structure  for fermion systems:
a Hamiltonian by taking a displaced TBRE with   
an attractive average  \cite{Zuker}, 
and a restricted Hamiltonian   with quadrupole-quadrupole 
correlation \cite{Zhao3}.

In  Ref. \cite{Velazquez}, 
It was   found  that  to a very large extent  normal ordering  
 (i.e.,the sequence 0, 2, 4, etc.) 
of $I$ in the yrast band can be kept  when one changes the    
Hamiltonian  from the realistic effective interaction   
to the TBRE Hamiltonian. 
This  regularity  is very interesting, but its origin  has been  
discussed.

Constraints on  random interactions were studied by 
Kusnezov {\it et al.} \cite{Kusnezov1} within the framework
of the IBM  and by Zhang {\it et al.} \cite{Zhang1} within 
the geometric collective model. These works 
obtained some constraints on  random interaction 
parameters of these models in order to produce global properties  
 exhibited in the low-lying  states of atomic nuclei. 
Within the framework of  the shell model 
 Horoi {\it et al.} found  \cite{Horoi1} that the B(E2) value  
  obtained by random two-body  interactions is  too small 
in comparison to those obtained by the  realistic effective interaction. 
It is  difficult, however,  to obtain the  
constraints on the TBRE  in the shell model for heavy nuclei 
because of its huge dimensionality.

\newpage

\section{Summary}

The present subject  was  
stimulated by the discovery of spin 0  
 ground state (0 g.s.) dominance in the presence of random two-body
 interactions in 1998 \cite{Johnson1}. Because this discovery  is  
both surprising and interesting, it has sparked off a sudden 
interest in the origin of the 0 g.s. dominance 
in the ground states of even-even nuclei. 
It  also led to a number of other  discoveries, for example, the 
generic vibrational and rotational band
structure  within the frameworks of the vibron model and
the IBM. 

In Sec. 2 we established the notations and conventions. 
We   reviewed very briefly the   models of nuclear structure before we 
 defined   Hamiltonians for different systems involved 
in  this paper. Then we presented an introduction 
to    Monte Carlo samplings, 
and  defined the two-body random ensemble (TBRE).
It was noted that the statistical  
patterns  obtained by using other random two-body
ensembles  with an average being zero   (such as the so-called  
random quasiparticle ensemble)  
 are   similar to those obtained by using the TBRE.

In Sec. 3 we presented typical results of   $I$ g.s. probabilities,  
$P(I)$,  for various systems 
including fermions in a single-$j$ shell,   many-$j$ shells, and 
$d$-, $sp$-, $sd$- and $sdg$-boson systems. 
In systems with an even number of fermions  the 0 g.s. probability   
is usually dominant with few exceptions, while for an 
odd number of fermions, where   
  no $I=0$ state exists,  
the $I=j$ ($j$ is one of the  
angular momenta of single-particle states) g.s.  probability 
is large. For systems with an even number  of bosons  the  
0 g.s. probability is large, while for an odd number  
of bosons the probability for 
$I=l$ ($l$ is one of the spins of the bosons) g.s. 
is larger than for $I=0$ g.s. in many cases, although 
there may be $I=0$ states in these systems.
The pattern of $P(I)$'s is sensitive to the displacement of the 
TBRE,   except for  fermions in a single-$j$ shell and  
 bosons with spin $l$.

The parity distribution
in the ground states of the TBRE Hamiltonian   was  
found to be similar to that  of realistic nuclei with 
mass number $A$  larger than 120.  
The single-particle levels of these nuclei involve 
of both positive parity  and negative parity. 
The observed parity of the ground states of these nuclei       
is always positive for even-even nuclei, and  otherwise it is positive
or negative with  about 50$\%$ for each.  We  showed that   
in the presence of the TBRE Hamiltonian,  
  positive parity dominates   in the ground states 
of even-even nuclei,  despite the fact  that   
the numbers of states with positive   and   negative parity  
are very close to each other.  
Because parity is a quantity which is much easier to handle  than spin,   
we expect a sound understanding of parity distribution in the
ground states obtained by using the TBRE Hamiltonian in the 
near future.
                       
In Sec. 3, we also showed that the 
odd-even staggering of binding energies  
 arises from   random two-body interactions. 
As for the effect of random interactions of   higher rank,
the    discussion  
restricted to  the $sd$ bosons.  According to Ref. \cite{Bijker3}, 
the features   obtained from random Hamiltonians  including
three-body interactions do not change very much  
if  the number of bosons is much larger than the rank of
Hamiltonian.

In Sec. 4  we discussed some simple systems in  
which either the eigenvalues are linear in   the 
two-body matrix elements or one can classify the two-body
matrix elements in a simple way.  
Three techniques have been developed: the first 
technique is based on the geometry of eigenvalues
\cite{Chau}, and is
applicable to  systems in which the eigenvalues depend
 linearly on the two-body interactions. Such examples  include  
 $d$ bosons and  fermions in a $j\le 7/2$ shell. 
 The second technique is called the mean-field approach
 \cite{Bijker6,Bijker4}, and
 is applicable to the cases where  one can classify  
two-body interactions according to different geometric shapes of the
systems.  Such examples include  $sp$ bosons and $sd$ bosons.  
The discussion of  $sp$ bosons   based on random polynomials
was  also presented along this line. 
The  third technique is called the  empirical approach
\cite{Zhao2,Zhao4}.
Here one needs to know the lowest state and the highest state 
  when  one particular  two-body  
matrix element is $-1$ and all others are  zero. This
approach can be applied to all simple systems discussed above.

In Sec. 5  we   discussed  $P(I)$'s of 
  complicated systems,    for which
one cannot classify the two-body matrix elements as done
for the $sp$ and $sd$ bosons and the eigenvalues
are not linear in the  two-body interactions.
In this case the empirical approach 
 was found to     predict the  $P(I)$'s  reasonably well.  
Here one needs  the number of times for 
$I$ to be  ground state spin  when   one   
  two-body matrix element is set to be $-1$ and others zero, and 
  the same procedure is repeated for all the two-body matrix elements. 
It was also demonstrated  that the 0 g.s. dominance
in the presence of the TBRE Hamiltonian 
arises essentially from the two-body matrix elements 
which give $I=0$ ground states in this process. 
Differences between $P(I_{\rm max})$'s of fermions and those of bosons were  
found and understood by using this empirical approach. 

For some systems such as four fermions in a single-$j$ shell and  
four bosons with spin $l$, 
an argument was given that the 0 g.s. dominance is partly due to
the fact that  there is only one non-zero eigenvalue for  $I=0$ states    
when only one of the two-body matrix elements is switched on.

In Sec. 5, we also reviewed
   an  alluring but controversial argument of the 0 g.s. dominance 
 based on  time reversal invariance of the Hamiltonian, 
 the efforts to understand the 0 g.s. dominance
 based on the large width of the distribution   
 of eigenvalues for $I=0$  states,   and
an observation of a   large pair-transfer fractional collectivity 
for 0 g.s. of systems with  particle numbers differing by two.

Some by-products were obtained. For example, 
 the ground state spin $I$ of $n$ fermions in a single-$j$ shell  
 was found  to be equal to $n$ when $G_2=-1$ and  
 others zero ($n$ is even).

In Sec. 6, we first reviewed   
 energy centroids of states with spin $I$, denoted
by $\overline{E_I}$. The probability for $\overline{E_I}$ to be the
lowest energy is large only when $I\simeq I_{\rm min}$ or  $I\simeq
I_{\rm max}$.
One thus can approximately  divide the TBRE
into two subsets, one  which produces 
  $\overline{E_{I \sim I_{\rm min}}}$ as the lowest energy,
  and the other   which produces  $\overline{E_{I  \sim I_{\rm max}}}$  
as the  lowest energy.
The $\overline{E_I}$ averaged over each subset  of the TBRE
follows the $I(I+1)$ relation approximately. These  features were   
explained in terms of  the chaoticity of 
two-body coefficients of fractional parentage. 
The geometric chaoticity  related to the regularities 
of  energy centroids was also discussed   by Zelevinsky  
and Volya  in Ref. \cite{Zele-review}.

In Sec. 6, we also showed that 
generic vibration  and rotation  arise
in  $sp$- and $sd$-boson systems with TBRE
Hamiltonians.  As for fermion systems,  
 one needs additional requirements 
in order to contrive an ensemble that  
exhibits rotational behavior.
To this end, 
two Hamiltonians have been suggested so far: 
One involves a displaced TBRE with   
an attractive average, the other involves 
the quadrupole-quadrupole correlation.

In the third part of Sec. 6, we have seen that
there is a large probability that the yrast states keep the
``correct" ordering (i.e., 0, 2, 4 $\cdots$) even for
purely random interactions.  In the fourth part, we discussed 
 constraints on the TBRE  
  Hamiltonian for the $sd$ IBM \cite{Iachello1} 
 and geometric collective model \cite{Greiner} in order to reproduce the
global properties exhibited by atomic nuclei.

Parallel to the studies in the context of  nuclear structure discussed  
in this paper, interesting features of randomly interacting
quantum systems have been  discussed, such as  metallic clusters  
\cite{Papenbrock}, quantum dots \cite{Weiden} and etc. \cite{Jacquod,Rotter}.
These works suggests that quantum systems or the dynamics of
many-body systems in the presence of random interactions
is  an interdisciplinary and new field in theoretical physics.

In conclusion, there are many interesting  
 regularities for many-body systems in the   
presence of the two-body random ensemble, as we extensively
discussed in this paper.  Many of these regularities have been
well explained theoretically. 
The 0 g.s. dominance has been well confirmed by a large
amount of numerical experiments. Many efforts  
have been devoted to understand why the 0 g.s. dominance occurs.
However, the underlying physical origin of the 0 g.s. dominance is still to be
explained in future.

{\bf Acknowledgement}
We are also grateful to Drs. W. Bentz, G. Bertsch, R. Bijker,
 N. D. Dang, J. N. Ginocchio, 
 V. K. B. Kota, and A. P. Zuker  for their reading of this manuscript. 
We gratefully acknowledge  interesting discussions 
with Drs. R. Bijker, R. F. Casten,   S. Drozdz, 
A. Frank, J. N. Ginocchio, P. Van Isacker, 
C. W. Johnson, V. K. B. Kota, B. R. Mottelson, 
S. Pittel, O. Scholten, N. Shimizu,  
I. Talmi, V. Zelevinsky, and A. P. Zuker. 
The authors would like to thank 
Dr. R. Bijker for his allowing us to use
Figures  18  and 19, which were taken from his papers.

\newpage

\newpage

{TABLE I.  0 g.s. probabilities for different random
ensembles, as compared to the percentage of all states
in the model spaces that have these quantum numbers.
Data are taken  from Table I of Ref. \cite{Johnson2}. }

\vspace{0.1in}

\begin{tabular}{cccccc} \hline \hline
Nucleus  & TBRE &  RQE  & RQE-NP & RQE-SPE & $I=0$ \\
 &  &  &  &  & (total space) \\ \hline
$^{20}$O   & 50$\%$   &  68$\%$   &  50$\%$ & 49$\%$  & 11.1 $\%$  \\ 
$^{22}$O   & 71$\%$   &  72$\%$   &  68$\%$ & 77$\%$  & 9.8$\%$ \\ 
$^{24}$O   & 55$\%$   &  66$\%$   &  51$\%$ & 78$\%$  & 11.1$\%$ \\
$^{44}$Ca   & 41$\%$   &  70$\%$   &  46$\%$ & 70$\%$  & 5.0$\%$ \\
$^{46}$Ca   & 56$\%$   &  76$\%$   &  59$\%$ & 74$\%$  & 3.5$\%$ \\
$^{48}$Ca   & 58$\%$   &  72$\%$   &  53$\%$ & 71$\%$  & 2.9$\%$ \\ \hline \hline
\end{tabular}

\vspace{0.4in}

{TABLE II. Observed parity distribution in the ground states  
of atomic nuclei.  We included all available data
with mass number $A$ larger than 120.  
The single-particle levels of the shell model space   
for these nuclei involve both positive and negative parity. 
The statistics here is based on the compilation in 
 Ref. \cite{Firestone}. }

\vspace{0.1in}

\begin{tabular}{cccc} \hline \hline
counts & even-even & odd-$A$  & odd-odd  \\  \hline
verified (+)   & 361   & 182  & 68  \\
verified (-)   & 0     & 164  & 78  \\  \hline 
tentative (+) & 0     & 109  &  53 \\
tentative (-) & 0     &  109 &  53 \\ \hline \hline 
\end{tabular}

\vspace{0.3in}

\newpage

{TABLE III.  Probabilities for the ground states (in \%)
calculated by using the TBRE Hamiltonian 
to have positive parity. In brackets the  
number of neutrons and protons $(N_p, N_n)$ 
is given for each basis.}

\vspace{0.1in}

\begin{tabular}{cccccccc} \hline \hline 
\multicolumn{2}{c}{basis A}\\
$(0,4)$& $(0,6)$& $(2,2)$& $(2,4)$& $(2, 6)$\\
 $86.8\%$& $86.2\%$& $93.1\%$&  $81.8\%$& $88.8\%$\\ \hline
 $(2,3)$&  $(1,4)$& $(1,3)$& $(0,5)$& $(1,5)$& $(6,1)$& $(2, 1)$\\
 $42.8\%$& $38.6\%$& $77.1\%$& $45.0\%$&    $69.8\%$& 38.4$\%$& $31.2\%$ \\\hline
\multicolumn{2}{c}{basis B}\\
$(2,2)$& $(2,4)$& $(4,2)$  \\
 $72.7\%$& $80.5\%$& $81.0\%$ \\\hline
 $(3,4)$& $(3,3)$& $(2,3)$&$(5,1)$& $(3,2$)& $(4,1)$& $(1,4)$& $(5,0)$\\
 $42.5\%$& $74.9\%$&$72.4\%$& $42.9\%$& $39.1\%$& $75.1\%$&  $26.4\%$ & $44.1\%$  \\\hline
\multicolumn{2}{c}{basis C}\\
 $(2,2)$& $(2,4)$& $(4,0)$& $(6,0)$ \\
 $92.2\%$& $81.1\%$& $80.9\%$&  $82.4\%$ \\\hline
 $(1,3)$& $(1,5)$&$(2,3)$& $(5,0)$&  $(4,1)$\\
 $73.0\%$& $64.4\%$& $52.0\%$& $42.6\%$& $56.5\%$ \\\hline
\multicolumn{2}{c}{basis D}\\
$(2,2)$& $(4,2)$& $(2,4)$& $(0,6)$ \\
 $67.2\%$& $76.1\%$& $74.6\%$&  $83.0\%$  \\ \hline
 $(3,3)$& $(3,2)$& $(2,3)$& $(0,5)$\\
 $54.5\%$& $54.2\%$&$54.0\%$&  $45.9\%$\\    \hline \hline
\end{tabular}

\newpage

{TABLE IV. Probability for each state to be  the ground  
state, and the distribution width of each eigenvalue 
for a $j=7/2$ shell with 
four fermions. Each of the states is 
labeled uniquely by its  angular momentum $I$ and seniority
number $v$. 
Probabilities of the row ``TBRE" are obtained by 1000 runs 
 of the  TBRE Hamiltonian, and those of  ``pred1." 
 are obtained by calculating integrals such as  
 Eq.~(\ref{exact}) for $0^+$ state of $n=4, j=\frac{7}{2}$ case.
 The row ``pred2." is obtained by using the empirical approach of 
  Eq. (\ref{empirical-I}) in Sec. 4.2. 
 The row ``exact" is obtained by using the geometry method proposed
 by Chau {\it et al.} \cite{Chau}. 
 The  distribution  width,   $g_{I(v)}$, of each eigenvalue,
 is listed in the last row.  }

\vspace{0.1in}

\begin{tabular}{ccccccccc} \hline \hline
$I(v)$ & 0(0) & 2(2) & 2(4) & 4(2) & 4(4) & 5(4) & 6(2) & 8(4) \\  \hline
TBRE & $19.9\%$ & $1.2\%$ & $31.7\%$ & $0.0\%$ & $25.0\%$ &
$0.0\%$ & 0.0$\%$  & 22.2$\%$ \\
pred1. & 18.19$\%$ & 0.89$\%$ & 33.25$\%$ & 0.00$\%$ &
22.96$\%$ & 0.00$\%$ & 0.02$\%$ & 24.15$\%$ \\
pred2. & 14.3$\%$ & 0$\%$ & 28.6$\%$ & 0$\%$ &
28.6$\%$ & 0 & 0$\%$ & 28.6$\%$ \\
exact & 18.33$\%$ & 1.06$\%$ & 33.22$\%$ & 0$\%$ &
23.17$\%$ & 0 & 0.05$\%$ & 24.16$\%$ \\
$g_{I(v)}$ & 3.14 & 3.25& 4.12 & 3.45 & 3.68 & 3.62 & 3.64 &4.22  \\ \hline \hline
\end{tabular}

\vspace{0.3in}

{TABLE V. Angular momenta which give the largest (smallest) eigenvalues
when $c_l=-1$  and other $c_{l'}$ ($l' \neq l$) parameters are   0 for 
$d$ boson systems.  Here $N_m=5$,
and ${\cal N}'_0 = $3, 0, 1, 2, 1, 0, 
for $6 \kappa$, $6 \kappa+1$, $6 \kappa+2$, $\cdots$,
 $6 \kappa+5$, respectively. $\kappa$ is a natural number. 
${\cal N}'_{I_{\rm max}} = 2$ for all $n$, and 
${\cal N}'_2 =$ 5- ${\cal N}'_0$.  This suggests periodical
$P(I)$'s versus $n$ for $d$ bosons, according to Eq.
(\ref{empirical-I}) in Sec. 4.2.
This table is obtained based on  the reduction rule for 
$U(5) \rightarrow O(3)$. 
}

\vspace{0.1in}

\begin{tabular}{cccccc} \hline  \hline
$n$   &  $c_0$(min) &  $c_2$(min) &  $c_2$(max) &  $c_4$(min) &  $c_4$(max)   \\  \hline
6$\kappa$  &      0      &  0  & $I_{\rm max}$ & $I_{\rm max}$ &0 \\
6$\kappa$+1&      2      &  2  & $I_{\rm max}$ & $I_{\rm max}$ &2 \\
6$\kappa$+2&      0      &  2  & $I_{\rm max}$ & $I_{\rm max}$ &2 \\
6$\kappa$+3&      2      &  0  & $I_{\rm max}$ & $I_{\rm max}$ &0 \\
6$\kappa$+4&      0      &  2  & $I_{\rm max}$ & $I_{\rm max}$ &2 \\
6$\kappa$+5&      2      &  2  & $I_{\rm max}$ & $I_{\rm max}$ &2 \\
    \hline  \hline
\end{tabular}

\vspace{0.4in}

\newpage

{TABLE VI. Angular momenta which give the lowest eigenvalues  
when  $G_J=-1$  and all other two-body matrix elements
are zero for four fermions in a 
single-$j$ shell.   }

\vspace{0.1in}

\begin{footnotesize}
\begin{tabular}{ccccccccccccccccc} \hline  \hline
$2j$ &  $G_0$ &  $G_2$ &  $G_4$ &  $G_6$ &  $G_8$ &  $G_{10}$ &  $G_{12}$
&  $G_{14}$ &  $G_{16}$ &  $G_{18}$ &  $G_{20}$ &  $G_{22}$ &  $G_{24}$
&  $G_{26}$ &  $G_{28}$ &  $G_{30}$   \\  \hline
7  & 0 &4 &2 &8 &   &   & & & &  & & & & & & \\
9  & 0 &4 &0 &0 &12 &   & & & &  & & & & & & \\
11 & 0 &4 &0 &4 &8  &16 & & & &  & & & & & & \\
13 & 0 &4 &0 &2 &2  &12 &20 & & &  & & & & & & \\
15 & 0 &4 &0 &2 &0  &0  &16 &24 & &  & & & & & & \\
17 & 0 &4 &6 &0 &4  &2  &0  &20 &28 &  & & & & & & \\
19 & 0 &4 &8 &0 &2  &8  &2  &16 &24 &32 & & & & & & \\
21 & 0 &4 &8 &0 &2  &0  &0  &0  &20 &28 &36 & & & & & \\
23 & 0 &4 &8 &0 &2  &0  &10 &2  &0  &24 &32 &40 & & & & \\
25 & 0 &4 &8 &0 &2  &4  &8  &10 &6  &0  &28 &36 &44 & & & \\
27 & 0 &4 &8 &0 &2  &4  &2  &0  &0  &4  &20 &32 &40 &48 & & \\
29 & 0 &4 &8 &0 &0  &2  &6  &8  &12 &8  &0  &24 &36 &44 &52 & \\
31 & 0 &4 &8 &0 &0  &2  &0  &8  &14 &16 &6  &0  &32 &40 &48 & 56 \\
\\   \hline  \hline
\end{tabular}
\end{footnotesize}

\vspace{0.4in}

{TABLE VII. Same as Table V for $sd$-boson systems.
The angular momentum $I$  corresponding to $e_{sdsd}=-1$ is omitted because 
it always presents  degenerate levels for  states of many $I$'s.
The one-body parameter $e_d$ is omitted because we are
interested in  $P(I)$'s in the presence of
the TBRE Hamiltonian. }

\vspace{0.1in}

\begin{tabular}{ccccccc} \hline  \hline
$n$ &  $e_{ssss} $ &  $e_{sddd}$ &  $e_{ssdd}$ &   $c_0$ &  $c_2$ &  $c_4$
  \\  \hline
6 & 0 &0 &0 &0 & 0  & $I_{\rm max}$    \\
7 & 0 &0 &0 &2 & 2  & $I_{\rm max}$    \\
8 & 0 &0 &0 &0 & 2  & $I_{\rm max}$    \\
9 & 0 &0 &0 &2 & 0  & $I_{\rm max}$    \\
10& 0 &0 &0 &0 & 2  & $I_{\rm max}$    \\
11& 0 &0 &0 &2 & 2  & $I_{\rm max}$    \\
12& 0 &0 &0 &0 & 0  & $I_{\rm max}$    \\
13& 0 &0 &0 &2 & 2  & $I_{\rm max}$    \\
14& 0 &0 &0 &0 & 2  & $I_{\rm max}$    \\
15& 0 &0 &0 &2 & 0  & $I_{\rm max}$    \\
16& 0 &0 &0 &0 & 2  & $I_{\rm max}$    \\
   \hline  \hline
\end{tabular}

\vspace{0.4in}

\newpage

{TABLE  VIII. Coefficients $\overline{\alpha_I^J}$ and ${\cal P}(I)$ 
for four fermions in a $j=\frac{9}{2}$ shell. 
 Bold font is used   for the largest
 $\overline{\alpha^J_I}$,  and italic for
 the smallest  $\overline{\alpha^J_I}$ for a given $J$.  Probabilities
 in the column ``pred1." are obtained by  
 integrals similar to Eq. (7) in Ref. \cite{Zhao1}, and those
 in the column ``pred2." are obtained by
 the empirical formula given in Eq. (\ref{empirical-I}). 
 The ${\cal P}(I)$'s in the last column ``TBRE" (in $\%$) 
  are  obtained by diagonalizing
the TBRE Hamiltonian for 1000 runs. We take
 both the smallest and the largest $\overline{\alpha^J_I}$ 
 when counting ${\cal N}'_I$. }

\vspace{0.1in}

\begin{tabular}{ccccccccc} \hline  \hline
$I$   &  $G_0$ &  $G_2$ &  $G_4$ &  $G_6$ &  $G_8$ & pred1.($\%$) & pred2.($\%$) & TBRE  \\  \hline 
0 &  {\bf 0.80} & 0.35 & 1.74 & 2.11 & 1.01 & 11.97 & 11.1 & 10.2 \\
2 &  0.30 & {\bf 1.39} & 1.45 & {\it 1.29} & 1.56 & 14.51 & 22.2  & 15.4  \\
3 &  0.00  & 0.36 & {\bf 2.28} & {\bf 2.63} & {\it 0.71} &28.17 & 33.3 & 28.9 \\ 
4 &  0.20 & 1.07 & 1.38 & 1.91 & 1.44 & 1.74 &  0 &  1.7 \\
5 &   0.00  & 1.00 & 1.59 & 1.84 & 1.57 & 0.30 &  06 &  0.6 \\ 
6 &  0.20 & 0.79 & 1.50 & 1.58 & 1.93 & 0.22 &  0 &  0.3 \\  
7 &   0.00  & 1.20 & 1.09 & 1.40 & 2.31 & 3.44 &  0 &  3.2 \\  
8 &  0.30 & 0.48 & 1.05 & 1.82 & 2.36 & 0.03 &  0 &  0 \\  
9 &   0.00  & 0.17 & 1.33 & 2.12 & 2.38 & 0.01 &  0 &  0 \\ 
10&   0.00  & 0.70 & 0.69 & 1.41 & 3.21 & 6.76 & 0 &  8.7 \\ 
12&   0.00  & {\it 0.00} & {\it 0.52} & 1.69 & {\bf 3.78} & 32.64 & 33.3 & 31.0  \\   \hline  \hline
\end{tabular}

\newpage

{\bf Figure captions}: 

FIG. 1 ~~ Probabilities of $I^+$ ground states for 
different $j$ shells with four fermions.  
All probabilities are obtained from 
1000 runs of the TBRE Hamiltonian.   
One sees that  $P(0)$ periodically  staggers with the  
value of $j$ at an interval $\delta_j=3$. 

\vspace{0.4in}

FIG. 2 ~~ Same as Fig. 1 for $n$=5.

\vspace{0.4in}
 
FIG.~3 ~~     $P(0)$'s of $n=4, 6$ and $P(j)$'s of
$n=5, 7$ fermions   in a single-$j$ shell.
They stagger  synchronously at  $\delta_j=3$ when $j$ is small and  
seem to saturate when $j$ becomes large.

\vspace{0.4in}

FIG.~4 ~~ $I$ g.s. probabilities vs. $l$ of four bosons with  spin $l$.
The results are obtained by 1000 runs of the TBRE Hamiltonian.
The $P(0)$ staggers with spin $l$   
 at an interval $\delta_l=3$  (similar 
to the $P(0)$ of four fermions in a single-$j$ shell, refer to Fig. 1), 
and that the $P(I_{\rm max})$ is very large.

\vspace{0.4in}

FIG. 5 ~~  $I=0$ and $I=l$ g.s. probabilities 
versus $n$ for $l= 4$ and 6. 
One sees that $P(0)$ is usually smaller than the  
corresponding $P(l)$ when $n$ is an odd number, indicating that
the 0 g.s. dominance might be associated with 
an odd-even effect of boson number. 

\vspace{0.4in}

FIG. 6 ~~   $I$ g.s. probabilities for $d$ bosons, with boson number $n$  
ranging from  4 to 44. Only states with  $I=$0, 2, and $I_{\rm max}=2n$
are possible as   ground states.   0  g.s.,
2  g.s. and $I_{\rm max}=2n$  g.s.
probabilities are periodically close to 0, 20$\%$, 40$\%$ or 60$\%$.  
 $P(0) \sim 0$  when ${n_d} = 6 \kappa \pm 1$.  
 The predicted $P(I)$'s (open squares) are well
 consistent with those (solid squares)
 obtained by using the TBRE Hamiltonian.

\vspace{0.4in}

FIG. 7 ~~ Polygons corresponding to  a system 
of five $d$ bosons. Each state is represented by a dot. 
The dots inside the polygon  never come to
the ground, and the g.s.  probability of each state of the vertex
is determined by Eq. (\ref{vertex}). Here $\theta_1$ (corresponding
to the $I(v)=2(1)$ state) and $\theta_3$ (corresponding to
the $I(v)=2(5)$ state) 
lead $P(2) \sim 60\%$, $\theta_4$ leads to $P(I_{\rm max}) \sim 40 \%$,
and $\theta_2$ (corresponding to the
$I(v)=0(3)$ state) leads to $P(0) \sim 4\%$.

\vspace{0.4in}

FIG. 8 ~~  $P(0)$, $P(1)$ and $P(n)$ for $sp$ bosons. 
The results are obtained by 1000 runs of the TBRE Hamiltonian.

\vspace{0.4in}

FIG. 9 ~~   $P(0)$'s of fermions in a    
 single-$j$ shell. Solid squares 
 are obtained by 1000 runs of the TBRE Hamiltonian.  
The   open squares  are predicted $P(0)$'s. ~  
a)  $n=4$; ~ b)   $n=6$. 
Solid triangles  are obtained from  the empirical 
formula of Eq. (\ref{test}). 

\vspace{0.4in}

FIG. 10 ~~
Comparison of $P(I)$'s obtained by diagonalizing the TBRE
Hamiltonian with those predicted by the empirical rule
of Eq. (\ref{empirical-II}). Here we show  
fermions in   two-$j$ shells with  ($j_1, j_2$) 
= $(\frac{7}{2}, \frac{5}{2})$. $n=$4, 5, 6, 7
in a), b), c) and d), respectively. Solid squares are
obtained by 1000 runs of the TBRE Hamiltonian and open squares are
the  predicted values.

\vspace{0.4in}

FIG. 11 ~~  The $P(0)$, $P(2)$ and $P(I_{\rm max}$) of $sd$-boson systems.
 Solid symbols  are $P(I)$'s obtained from 1000 runs of the TBRE Hamiltonian.  
Open symbols  are $P(I)$'s predicted by
the empirical approach introduced in Eq. (\ref{empirical-II}). 
Only $I=0$, 2, $I_{\rm max}$  g.s. probabilities are
included. All other $P(I)$'s obtained by  
diagonalizing the TBRE Hamiltonian are close to zero, and
the predicted $P(I)$'s  are also zero.

\vspace{0.4in}

FIG. 12 ~~
a)  ~The $I_{\rm max}$ g.s. probability obtaining by fixing
 $G_{16}$=$\pm 1$ ($J_{\rm max}$=16) and  
all other $G_J$ being  a  TBRE multiplied by $\epsilon$; 
b)  ~  $0$ g.s. probabilities obtained by
fixing  $G_{0}$=$\pm 1$  and all other $G_J$ being the
TBRE Hamiltonian multiplied by $\epsilon$. 
In this figure,  $j=\frac{17}{2}$ and $n=4$. ~   

\vspace{0.4in}

FIG. 13 ~~  Regularities of   $I_{\rm max}$g.s. probabilities.  
The squares are predicted by the $1/N$ relation  
 whereas all other results were obtained by  
diagonalizing  the TBRE Hamiltonian. (a) Fermions in a single-$j$ shell;
(b) $I'_{\rm max}$ g.s.
probabilities for fermions in two-$j$ shells; (c)   bosons with spin $l$.
The $I'_{\rm max}$ is defined by $I_{\rm max} (j_1^n)$
and $I_{\rm max} (j_2^n)$. 
One sees that the $1/N$ relation applies well to the fermions in a
single-$j$ shell but underestimates the $P(I_{\rm max})$ for
bosons with spin $l$ when  $l$ becomes large. 
The $1/N$ relation predicts reasonably the lower limit
of $P(I'_{\rm max})$ for fermions in  two-$j$ shells. Refer to
the text for details.

\vspace{0.4in}

FIG.~14 ~~ Seniority distribution in the angular momentum $I=0$  
ground states. No bias of low seniority is observed in these systems with 
four  and six fermions in a single-$j$ shell.
 
\vspace{0.4in}

FIG.~15 ~~ Typical results for ${\cal P}(I)$'s:    
a)~ single-$j$ ($j=\frac{15}{2}$) with four fermions,
b)~ two-$j$ shell ($2j_1,2j_2)=(11,7)$ with six fermions,
c)~ 6-$sdg$ bosons,  
d)~ single-$j$ ($j=\frac{9}{2}$) with five fermions. 

\vspace{0.4in}

FIG.~16 ~~    $\langle \overline{E_I}\rangle_{\rm min}$ vs. $I(I+1)$. 
   $\langle \overline{E_I}\rangle_{\rm min}$'s are  obtained
by averaging over   $\overline{E_I}$  
  with the requirement that
 $\overline{E_{I \sim I_{\rm min}}}$ be the lowest energy.
The results are obtained by applying 1000 sets of the TBRE Hamiltonian. 
a). twenty $d$ bosons; b) ten $sd$ bosons;
c) single-$j$ shell with $j$=17 and $n=4$,
and d)  a two-$j$ shell with $j_1=5/2, j_2 =7/2$, and  $n=$4.  
The quantity  $\langle \overline{E_I}\rangle_{\rm min}$
has the same behavior except for a difference of sign. Refer to the
text for details.

\vspace{0.4in}

FIG.~17 ~~Correlation between $\sqrt{\cal J}$ and $j=\sqrt{ \sum_i j^2_i}$. 
$\sqrt{\cal J} \simeq 1.42   \sqrt{ \sum_i j^2_i}$ 
 for $d$, $sd$ bosons, and fermions 
 in a single-$j$ shell. This correlation  
 is shifted  very slightly to the right  for fermions in many-$j$ shells 
 and $sdg$ bosons.

\vspace{0.4in}

FIG. 18 ~ Distribution of $R$ of $sd$ bosons in the
presence of random two-body interactions with $n=$3 (dash-dotted),
6 (dotted), 10(dashed) and 16 (solid). 

\vspace{0.4in}

FIG. 19 ~ Correlation between $R$
and $\frac{ {\rm B(E2,} 4_1^+ \rightarrow 2_1^+) }{{\rm B(E2,} 2_1^+
\rightarrow 0_1^+) }$ for sixteen $sd$ bosons with one-body and two-body
random interactions. It is seen that 
$(R,  \frac{ {\rm B(E2,} 4_1^+ \rightarrow 2_1^+) }{{\rm B(E2,} 2_1^+ 
\rightarrow 0_1^+) } )$  concentrate at  two points:
(2.0, 2.0) which is characteristic  for
vibrational motion and (3.3, $\frac{10}{7}$) which
is characteristic   for rotational motion in the large $n$ limit
of the IBM.

\vspace{0.4in}

FIG. 20 ~ The distribution of   $R$ values for six identical
nucleons interacting by the  TBRE Hamiltonian in a) the $sd$
shell, ~ b) the $pf$ shell, and c) the $sdg$ shell.
The calculations were done within the $SD$-pair truncated
subspace. 
        
\vspace{0.4in}

FIG. 21 ~ The distribution of  $R$ values for six identical
nucleons in the $sd$ shell interacting by a random Hamiltonian
containing monopole pairing, quadrupole pairing, and
quadrupole-quadrupole forces.  The strength $\kappa$ of the
quadrupole-quadrupole interaction is multiplied by a factor
$\epsilon$ to assess the importance of the quadrupole-quadrupole
interaction. The calculations were done within the $SD$-pair truncated
subspace.

\vspace{0.4in}

FIG. 22 ~ The distribution of  $R$ values for six identical
nucleons interacting by a random Hamiltonian containing monopole
pairing, quadrupole pairing, and quadrupole-quadrupole forces for
a)  the $pf$ shell, b)  the $sdg$ shell, c)  the $pfh$ shell, and
d) the $sdgi$ shell.   
The calculations were done within the $SD$-pair truncated
subspace, and $\epsilon$ is fixed to be 1.0.

\vspace{0.4in}

FIG. 23 ~ Correlation between ratios of E2 transition rates and $R$
for the same calculations as in Fig. 22d . The inserts in panels 
a), b) and d) focus on  critical regions.

\newpage

Appendix  ~~List  of mathematical notations used in this paper. 

\begin{eqnarray}                                                 
&A  &  {\rm mass ~ number ~ of ~ an ~ atomic ~ nucleus} \nonumber \\
&{\rm Z}  &  {\rm proton ~ number ~ of ~ an ~ atomic ~ nucleus} \nonumber \\
&{\rm N}  &  {\rm neutron ~ number ~ of ~ an ~  atomic ~ nucleus} \nonumber \\
&N_p ~ (N_n)  &  {\rm valuence ~ proton ~ (neutron) ~ number ~ outside
~ a  ~  closed  ~ shell} \nonumber \\
& n  & {\rm number ~ of ~ valence ~ particles } \nonumber \\
&j  &  {\rm angular ~ momentum ~ of ~ a ~  single-particle ~ state}  \nonumber  \\
&l  &  {\rm intrinsic ~ spin ~ for ~ a ~ boson ~ or ~ the ~ orbital ~ component ~
 of ~ } j \nonumber \\
& m  & z-{\rm component ~ of } ~ j \nonumber \\
& t  & {\rm isospin ~ for ~ a ~ nucleon} \nonumber \\ 
& m_t  & z-{\rm component ~ of } ~ t \nonumber \\
& T  & {\rm total ~ isospin ~ of ~ two ~ or ~ more ~ nucleons} \nonumber \\
& M_T & z-{\rm component ~ of } ~ T \nonumber \\ 
& J  & {\rm total ~ angular ~ momentum ~ of ~ two ~ fermions ~
 or ~ two ~ bosons} \nonumber \\
& M_J  & z-{\rm component ~ of } ~ J \nonumber \\ 
& \rho(x) & {\rm distribution ~ function ~ of ~ } x \nonumber \\
& e_{j m_t} & {\rm single-particle ~ energy ~ of ~
valence ~ protons ~ or ~ neutrons} ~ (m_t = \mp \frac{1}{2}) \nonumber \\ 
& e_d ~ (e_p) & {\rm single  ~ } d ~ (p) ~ {\rm boson ~ energy} \nonumber \\ 
& G_{JT} (j_1 j_2, j_3 j_4) & {\rm  
  two-body ~ matrix ~ elements ~ of ~ fermions } \nonumber \\
& G_J  & {\rm abbreviation ~ of} ~ G_{JT} (j_1 j_2, j_3 j_4) {\rm ~ for 
 ~ single-}j {\rm ~ fermions} \nonumber \\
& G_L & {\rm abbreviation ~ of} ~ G_{LT} (l_1 l_2, l_3 l_4) {\rm ~ for 
  ~  spin}-l  ~ {\rm bosons} \nonumber \\
& {\rm TBRE} & {\rm abbreviation ~  for ~ ``two-body ~ random ~ ensemble" } 
\nonumber \\
& N & {\rm number ~ of ~ independent ~ two-body ~ matrix ~ elements}
\nonumber \\
& N_m & 2N-1 \nonumber \\
& {\cal N}'_I & {\rm number ~ of ~ times ~ that ~ spin ~}  I ~
{\rm states ~ appear ~ either ~ as  ~ the  ~ ground  } 
  \nonumber \\
& & {\rm ~ state ~ or ~ the ~ highest ~state ~ when ~ one ~ of } ~
G_{JT}(j_1 j_2 j_3 j_4) {\rm 's} ~ {\rm is ~-1} \nonumber \\
& & {\rm ~ and ~ others ~ are ~ zero} \nonumber \\
& {\cal N}_I & {\rm number ~ of ~ times ~ that ~ spin ~}
  I ~ {\rm states ~ appear ~ as ~   the ~ ground ~ state ~ 
~ for }   \nonumber \\
& & {\rm  one ~ of ~} G_{JT}(j_1 j_2 j_3 j_4) {\rm 's}
~ {\rm to ~ be ~-1 ~ and ~ others ~ to ~ be ~ zero} \nonumber \\
& \kappa & {\rm parameter ~ for ~ quadrupole-quadrupole ~ interaction}
\nonumber \\
&I  &  {\rm total ~ angular ~  momentum  ~ for ~ a ~ state ~ of ~
many-body ~ systems} \nonumber \\
&I_{\rm max} & {\rm maximum ~ of} ~ I \nonumber \\
&I_{\rm min} & {\rm minimum ~ of} ~ I \nonumber \\
& v    & {\rm seniority ~ number ~ for ~ fermions ~ in ~ a ~ single-}j ~ {\rm shell}  \nonumber \\
& &   {\rm or ~ number ~ of } ~ d ~  {\rm bosons ~ not ~
paired ~ to ~ spin ~  zero } \nonumber \\
& \beta & {\rm additional ~ quantum ~ numbers ~ (except} ~ I) ~ {\rm
to ~ specify ~ a ~ state} \nonumber \\
& S ~ (D)   &  {\rm fermion ~ pairs ~ with ~ spin ~ zero ~ (two)} \nonumber \\
& P(I) & {\rm probability ~ that ~ the ~ ground ~ state ~ has ~
angular ~ momentum} ~ I \nonumber \\
&\overline{E_I} & {\rm energy ~ centroid ~ of ~ spin} ~ I 
~  {\rm states}  \nonumber \\
& {\cal P}(I) & {\rm probability ~ that} ~  \overline{E_I} ~ {\rm ~ is ~ the
 ~ lowest ~ among ~ all} ~ \overline{E}_{I'} \nonumber \\
& D_I & {\rm number  ~ of ~ the ~ angular ~ momentum} ~ I 
~  {\rm states}  \nonumber \\
& D_I^{(j)} & D_I
~  {\rm for ~ fermions ~ in ~ a ~ single-}j {\rm ~ shell}  \nonumber \\ 
& D_I^{(l)} & D_I
~  {\rm for ~ bosons ~ with ~ spin ~} l \nonumber \\
& E^{L(l)}_0 & {\rm the ~ non-zero ~ eigenvalue ~ for} ~ I=0 {\rm ~ 
of ~ four ~ bosons ~ with ~ spin ~} l  \nonumber \\ 
& & {\rm ~ and} ~  G_L=-\delta_{LL'}  \nonumber \\
& E^{J(j)}_0 & {\rm the ~ non-zero ~ eigenvalue ~ for} ~ I=0 {\rm ~ 
of ~ four ~ fermions ~ in} \nonumber \\ 
& &  {\rm ~ a ~ single-}j ~ {\rm shell} {\rm ~ and} ~  G_J=-\delta_{JJ'}  \nonumber \\
& \sigma_I & {\rm width ~ of ~ spin ~ } I ~ {\rm states, ~
defined ~ by ~ } \sigma_I^2 = \langle (H-\overline{E_I})^2 \rangle
/ D_I \nonumber \\
& g_{I(v)} & {\rm width ~ of ~ the ~} I(v) {\rm state, 
~ defined ~ by ~  } \nonumber \\
&  &  g_{I(v)}^2=  \sum_J \left(\alpha_{I(v)}^J
\right)^2, {\rm for ~ four ~ fermions ~ in ~} j \le \frac{7}{2} ~ {\rm shell} 
 \nonumber \\
& g_{I} & {\rm width ~ of   ~ spin ~ } I ~ {\rm states~, 
defined ~ by ~ } g_I^2 = \langle H^2 \rangle
/ D_I \nonumber \\ 
& {\alpha}_{I\beta \beta'}^J & \frac{n(n-1)}{2} \sum_{K \gamma}
\langle j^{n-2}(K\gamma) j^2(J)|\}j^n\beta I \rangle
\langle j^{n-2}(K\gamma) j^2(J)|\}j^n\beta' I \rangle  \nonumber \\
& {\alpha}_{I\beta}^J & \frac{n(n-1)}{2} \sum_{K \gamma} \left(
\langle j^{n-2}(K\gamma) j^2(J)|\}j^n\beta I \rangle \right)^2 = 
{\alpha}_{I\beta \beta}^J \nonumber \\ 
& \overline{{\alpha}_{I}^J} & \ \sum_{\beta} {\alpha}_{I\beta}^J /D_I \nonumber \\
& \langle \overline{E_I} \rangle_{\rm min} & {\rm ensemble ~ average ~ of ~}
\overline{E_I} ~ {\rm with ~ } \overline{E}_{I \sim I_{\rm min} } 
  {\rm being ~ the ~ lowest}  \nonumber \\
& \langle \overline{E_I} \rangle_{\rm max} & {\rm ensemble ~ average ~ of ~}
\overline{E_I} ~ {\rm with ~ } \overline{E}_{I \sim I_{\rm max} } 
  {\rm being ~ the ~ lowest}  \nonumber \\
& {\cal J} & {\rm coefficient ~ obtained ~ by} ~ 
\langle \overline{E_I} \rangle_{\rm min}=
\frac{I(I+1)}{2 {\cal J}} \nonumber \\
& &  {\rm  or ~ by }~
\langle \overline{E_I} \rangle_{\rm max}  = E_0 -\frac{I(I+1)}{2 {\cal J}} \nonumber \\
& | ~ \rangle \rangle & {\rm normalized ~ state ~ of ~ } | ~ \rangle \nonumber \\
& E_{I_1^+} & {\rm excited ~ energy  ~
of ~ the ~ first ~} I^+  
{\rm ~ state ~ of ~ an ~ even-even ~ nucleus} \nonumber \\
& R & E_{4_1^+}/ E_{4_1^+} \nonumber  \\
& f_{\rm transfer} & {\rm pair-transfer ~ fractional ~ collectivity}  \nonumber 
\end{eqnarray}

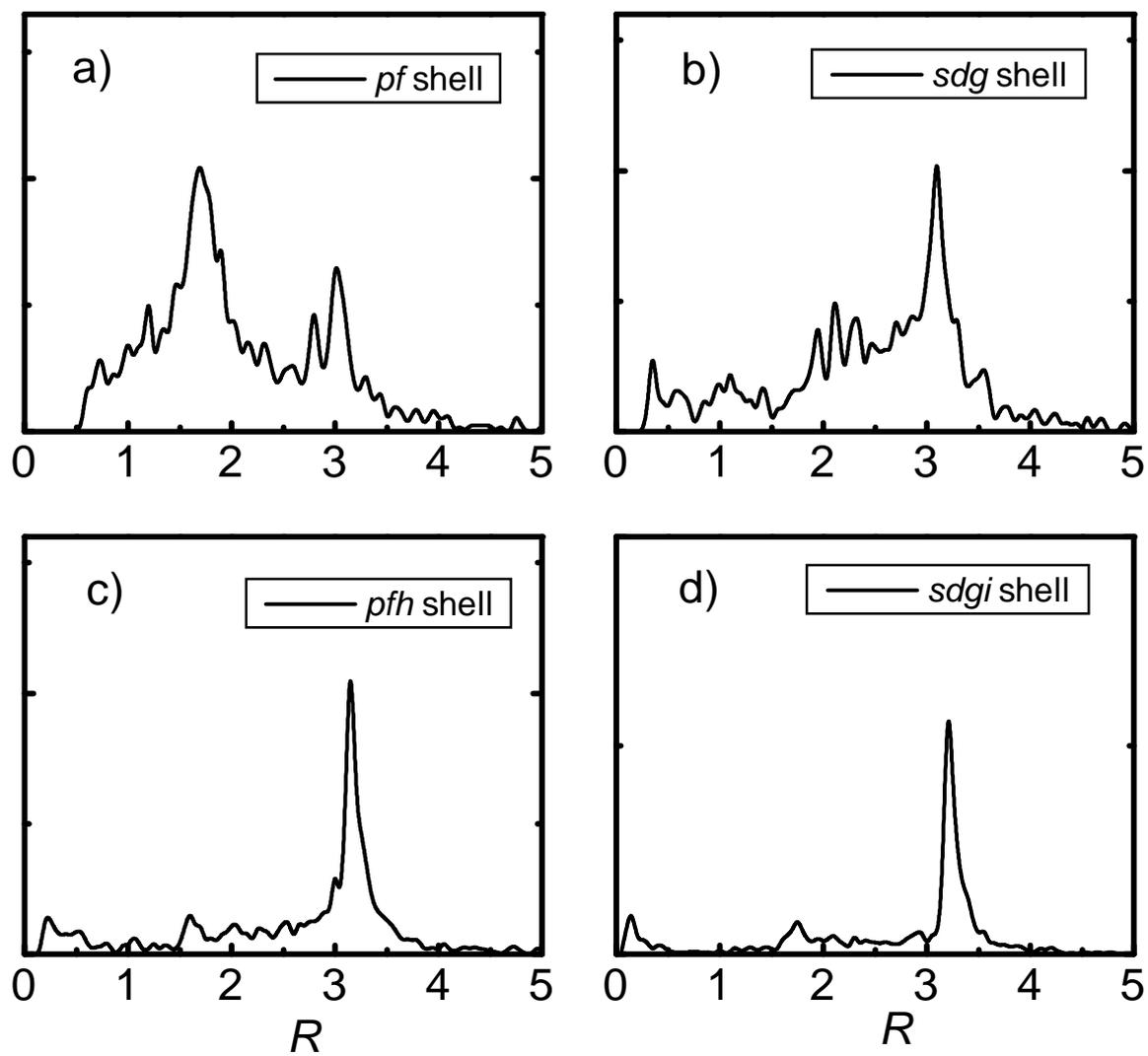
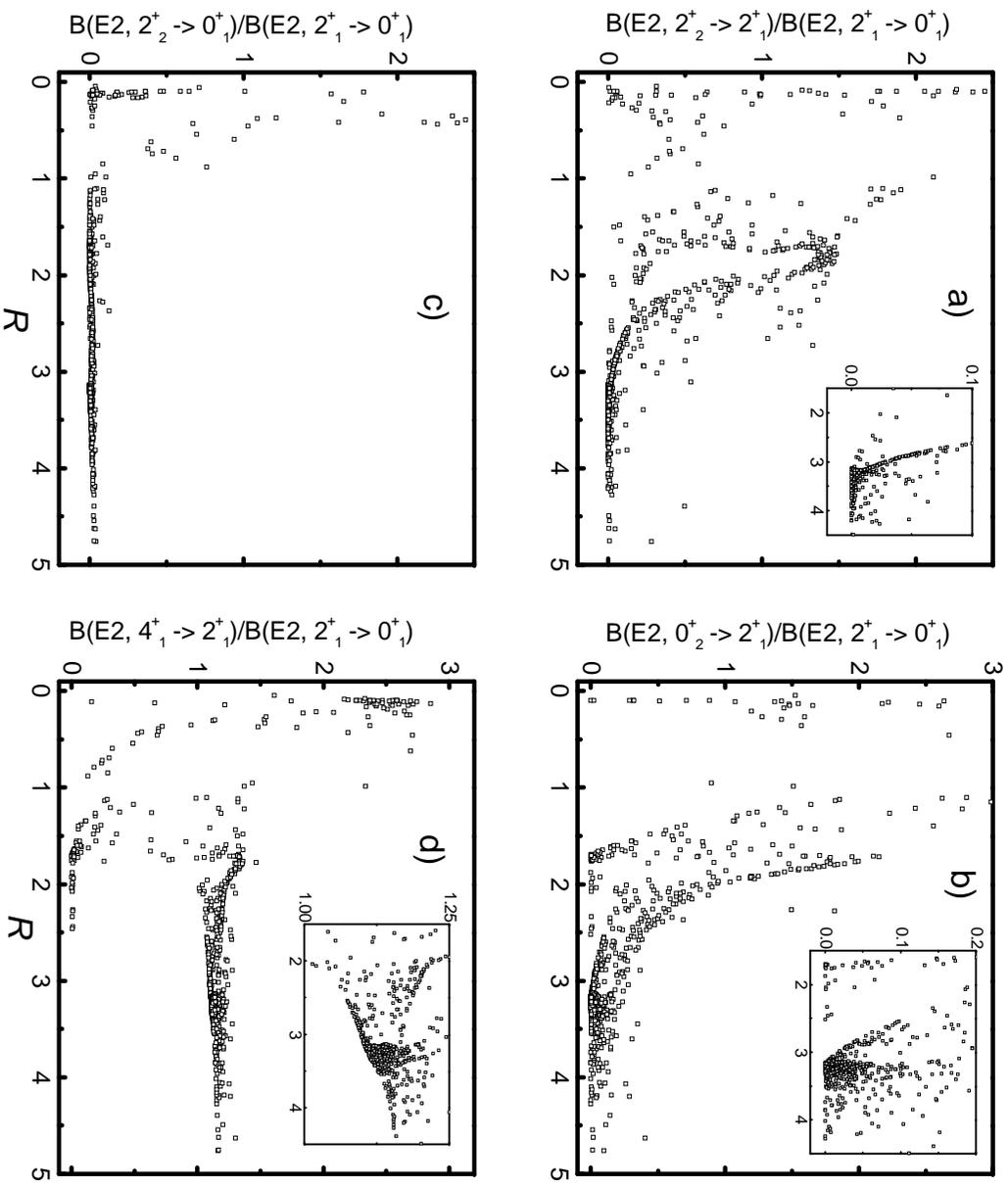
\end{document}